\documentclass[11pt, letterpaper]{article}
\usepackage[authoryear]{natbib}
\usepackage{graphicx}
\usepackage{amsmath,amssymb,amsfonts,amsthm}
\usepackage{setspace}
\usepackage{subfig}
\usepackage{booktabs}
\usepackage{multirow}
\usepackage{hyperref}
\usepackage{float}
\usepackage{color}
\usepackage{xargs}                      
\usepackage[pdftex,dvipsnames]{xcolor}  
\usepackage[colorinlistoftodos,prependcaption,textsize=tiny, disable]{todonotes}
\usepackage{tikz}
\usepackage{verbatim}
\usetikzlibrary{arrows}
\usepackage[flushleft]{threeparttable}

\usepackage{etoolbox}
\usepackage{xr}
\allowdisplaybreaks

\patchcmd{\abstract}{\small}{}{}{}

\hypersetup{
    colorlinks,%
    citecolor=black,%
    filecolor=black,%
    linkcolor=black,%
    urlcolor=blue
}

\newcommandx{\todoY}[2][1=]{\todo[linecolor=OliveGreen,backgroundcolor=OliveGreen!25,bordercolor=OliveGreen,#1]{#2}}

\newcommand{\ap}{\alpha}
\newcommand{\bt}{\beta}

\newcommand{\dt}{\delta}
\newcommand{\sig}{\sigma}
\newcommand{\sg}{\sigma}

\newcommand{\trace}{\text{tr}}

\newcommand{\ld}{\lambda}
\newcommand{\gm}{\gamma}

\newcommand{\eps}[0]{\ensuremath{\varepsilon}}

\newcommand{\lh}{\ld' H^{-1}}
\newcommand{\hl}{H^{-1} \ld }
\newcommand{\tlh}{\wtd{\ld}' \wtd{H}^{-1}}
\newcommand{\thl}{\wtd{H}^{-1} \wtd{\ld} }
\newcommand{\e}{e_f^k}
\newcommand{\te}{\wtd{e}_f^k}
\newcommand{\xik}{\xi_f^k}
\newcommand{\txik}{\wtd{\xi}_f^k}

\newcommand{\lt}{\left}
\newcommand{\rt}{\right}
\newcommand{\what}{\widehat}
\newcommand{\wtd}{\widetilde}
\newcommand{\opr}{o_p(\rho_{k,N})}

\newcommand{\Sg}{\Sigma}

\newcommand{\ben}{\begin{enumerate}}
\newcommand{\een}{\end{enumerate}}
\newcommand{\bit}{\begin{itemize}}
\newcommand{\eit}{\end{itemize}}

\newcommand{\parrow}[0]{\ensuremath{\stackrel{p}{\rightarrow}}}

\newcommand{\eq}[1]{\begin{align}#1\end{align}}
\newcommand{\eqs}[1]{\begin{align*}#1\end{align*}}

\usepackage[top=1 in, bottom=1 in, left=1 in , right=1 in]{geometry}

\usepackage{enumitem}
\setdescription{leftmargin=\parindent,labelindent=\parindent}

\newtheorem{theorem}{Theorem}
\newtheorem{lemma}{Lemma}

\newtheorem{corollary}{Corollary}
\newtheorem{assumption}{Assumption}
\newenvironment{definition}[1][Definition]{\begin{trivlist}
        \item[\hskip \labelsep {\bfseries #1}]}{\end{trivlist}}
\newenvironment{example}[1][Example]{\begin{trivlist}
        \item[\hskip \labelsep {\bfseries #1}]}{\end{trivlist}}

\DeclareMathOperator*{\argmin}{arg\,min}

\begin{document}
	
	\newpage
	
	\author{
		Seojeong Lee\thanks{\footnotesize
			Address: Level 4 Business School building,
			UNSW Sydney, NSW 2052,
			Australia. E-mail: \texttt{jay.lee@unsw.edu.au}} \\ \footnotesize University of New South Wales  \and
		Youngki  Shin\thanks{\footnotesize Address: 1280 Main St.\ W.,\ Hamilton, ON L8S 4L8, Canada. Email:
			\texttt{shiny11@mcmaster.ca}}\\  \footnotesize   McMaster University
	}

	\title{Complete Subset Averaging with Many Instruments
		\thanks{\footnotesize 	We are grateful to the Co-editor Victor Chernozhukov and the anonymous referee for helpful comments that improve the paper substantially. We also thank Graham Elliott, Ryo Okui, Marine Carrasco, and the seminar participants at Seoul National University, University of Melbourne, ISNPS 2018, CESG 2018, MEG 2018, BK21plus International Conference on
			Econometrics for helpful comments. Julius Owusu provided excellent research assistance. Shin gratefully acknowledges financial support received from
			the Social Sciences and Humanities Research Council of Canada (435-2018-0275) and from Australia Research Council (DP170100987). This work was made possible by the facilities of the Shared Hierarchical Academic Research Computing Network and Compute/Calcul Canada.
		}
	}
	\date{August 21, 2020}
	\maketitle

	\begin{abstract}
		
		We propose a two-stage least squares (2SLS) estimator whose first stage is the equal-weighted average over a complete subset with $k$ instruments among $K$ available, which we call the \textit{complete subset averaging (CSA) 2SLS}. The approximate mean squared error (MSE) is derived as a function of the subset size $k$ by the \citet{nagar1959bias} expansion. The subset size is chosen by minimizing the sample counterpart of the approximate MSE. We show that this method achieves the asymptotic optimality among the class of estimators with different subset sizes. To deal with averaging over a growing set of irrelevant instruments, we generalize the approximate MSE to find that the optimal $k$ is larger than otherwise. An extensive simulation experiment shows that the CSA-2SLS estimator outperforms the alternative estimators when instruments are correlated. As an empirical illustration, we estimate the logistic demand function in \citet*{berry1995automobile} and find the CSA-2SLS estimate is better supported by economic theory than the alternative estimates. 
		
	\end{abstract}
	
	\section{Introduction}
	Instrumental variables (IV) estimators are commonly used to estimate the parameters associated with endogenous variables in economic models. The two-stage least squares (2SLS) is the most popular IV estimator for linear regression models with endogenous regressors. While the 2SLS estimator is usually applied to just-identified models where the number of instruments is the same as the number of endogenous variables, there are many applications where more instruments are used than the number of endogenous variables (over-identified). In particular, when the instrument set is a large set of dummy variables or is constructed by interacting the original instruments with exogenous variables, the total number of instruments can be quite large. For example, \citet{angrist1991does} use as many as 180 instruments for one endogenous variable in one of their specifications to get a tighter confidence interval for the structural parameter than using a smaller set of instruments.
	
	Although using a large set of instruments can improve efficiency, there is a trade-off in terms of increased bias in the point estimate, e.g. \citet{kunitomo1980asymptotic}, \citet{morimune1983approximate}, and \citet{bekker1994alternative}.  Motivated by the trade-off relationship, \citet{donald2001choosing} propose to select the number of instruments by minimizing the approximate mean squared error (MSE) of IV estimators. \citet{kuersteiner2010constructing} propose an IV estimator that applies the model averaging approach of \citet{hansen2007least} in the first stage and show that the selected weights attain optimality in the sense of \citet{li1987asymptotic}. \citet{okui2011instrumental} proposes to average the first stage using shrinkage to obtain the shrinkage IV estimators. When the number of instruments can be much larger than the sample size, \citet*{belloni2012sparse} propose to select instruments using Lasso in the first stage under approximate sparsity and \citet{carrasco2012regularization} proposes a regularized IV estimator using all the instruments.
	
	In this paper, we propose a 2SLS estimator whose first stage is the equal-weighted average over a complete subset with $k$ instruments among the total of $K$ instruments. We call this estimator the \textit{complete subset averaging (CSA) 2SLS} estimator. Our approach differs from the important existing work in the many instruments literature:  Unlike \citet{donald2001choosing}, the CSA-2SLS is based on model averaging in the first-stage rather than model selection and does not require ordering of the instruments; Unlike \citet{kuersteiner2010constructing}, it does not require weight estimation; Unlike \citet{okui2011instrumental}, it does not require to specify the main set of instruments a priori.
	
	The main theoretical contribution of this paper is three-fold. First, we derive the approximate MSE for our CSA-2SLS estimator by the \citet{nagar1959bias} expansion. It is technically challenging because the average of non-nested projection matrices is not idempotent. In contrast, the existing literature usually assumes nested models. The derived formula shows the bias-variance trade-off and some interesting features which will be discussed in detail in the following sections. Second, we generalize the approximate MSE formula when irrelevant instruments exist whose number grows as the sample size increases. A penalty term decreasing with the subset size $k$ appears in the generalized formula, which suggests that the choice of a larger $k$ than otherwise would be desirable in the presence of irrelevant instruments.  Third, we prove that the CSA-2SLS estimator with the subset size minimizing the sample approximate MSE is asymptotically optimal in the sense that it attains the lowest possible MSE among the class of the CSA-2SLS estimators with different subset sizes. Our optimality proof is based on \citet{li1987asymptotic} and \citet{whittle1960bounds}.
	
\begin{figure}[t]
		\centering
		\subfloat[\citet*{angrist1991does}\label{subfig_AK}]{%
			\includegraphics[width=0.32\textwidth]{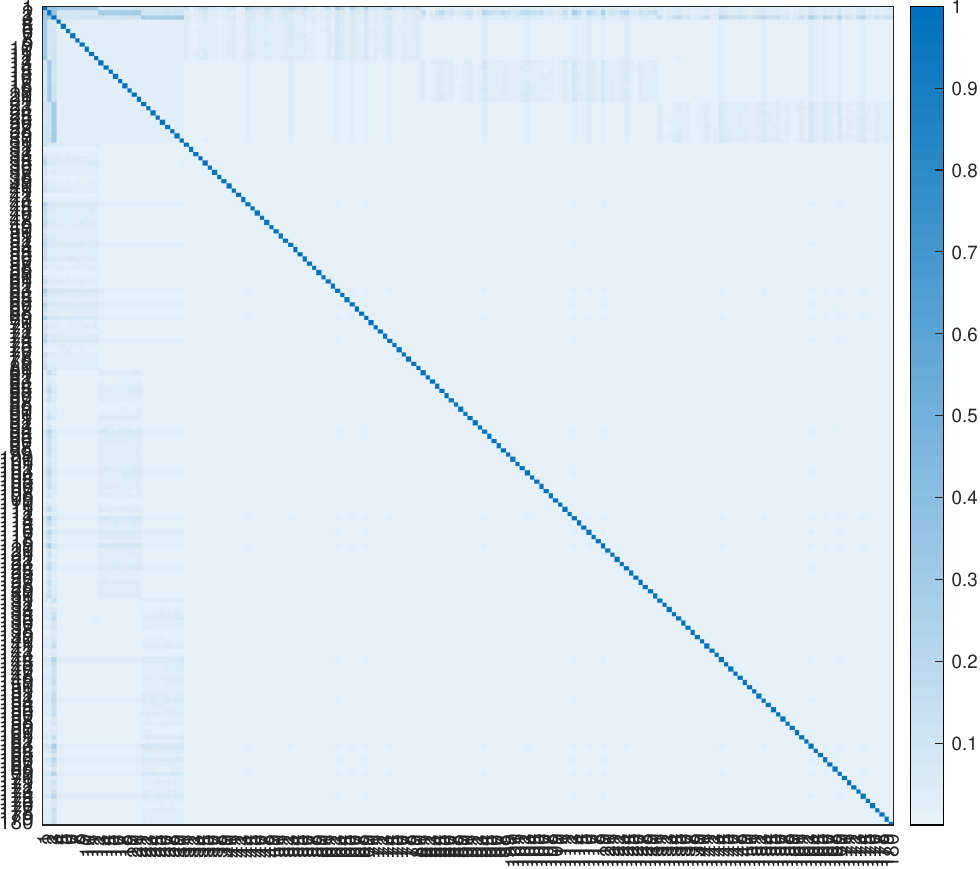}
		}
		\subfloat[\citet{berry1995automobile}\label{subfig_BLP}]{%
			\includegraphics[width=0.32\textwidth]{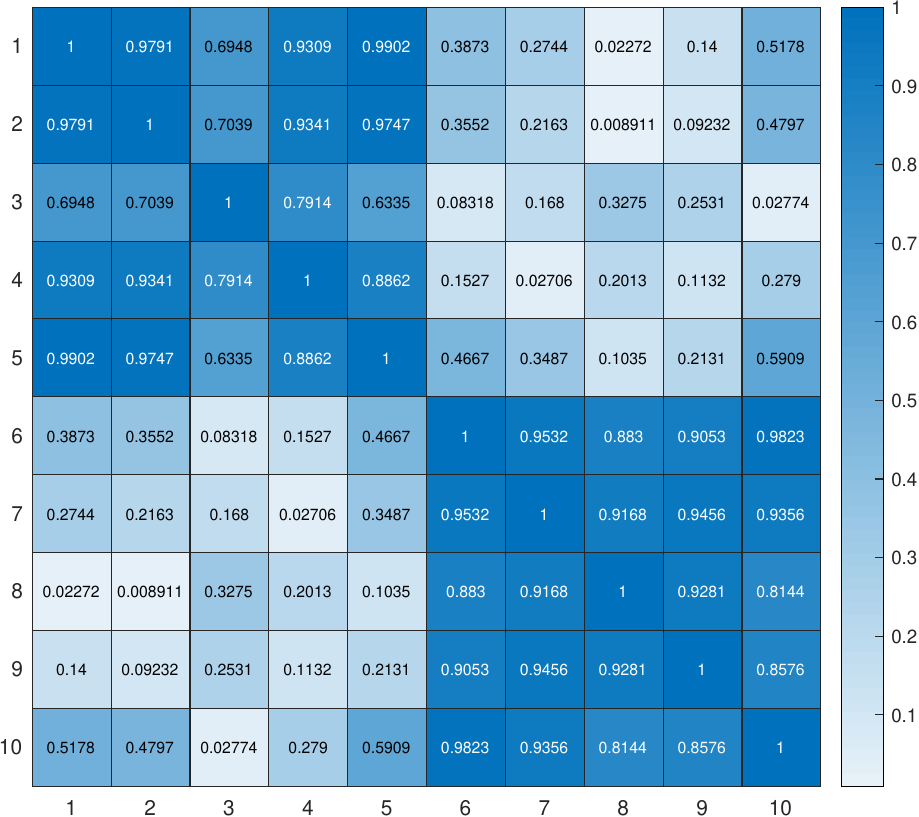}
		}
		\subfloat[\citet{acemoglu2008income}\label{subfig_AJRY}]{%
			\includegraphics[width=0.32\textwidth]{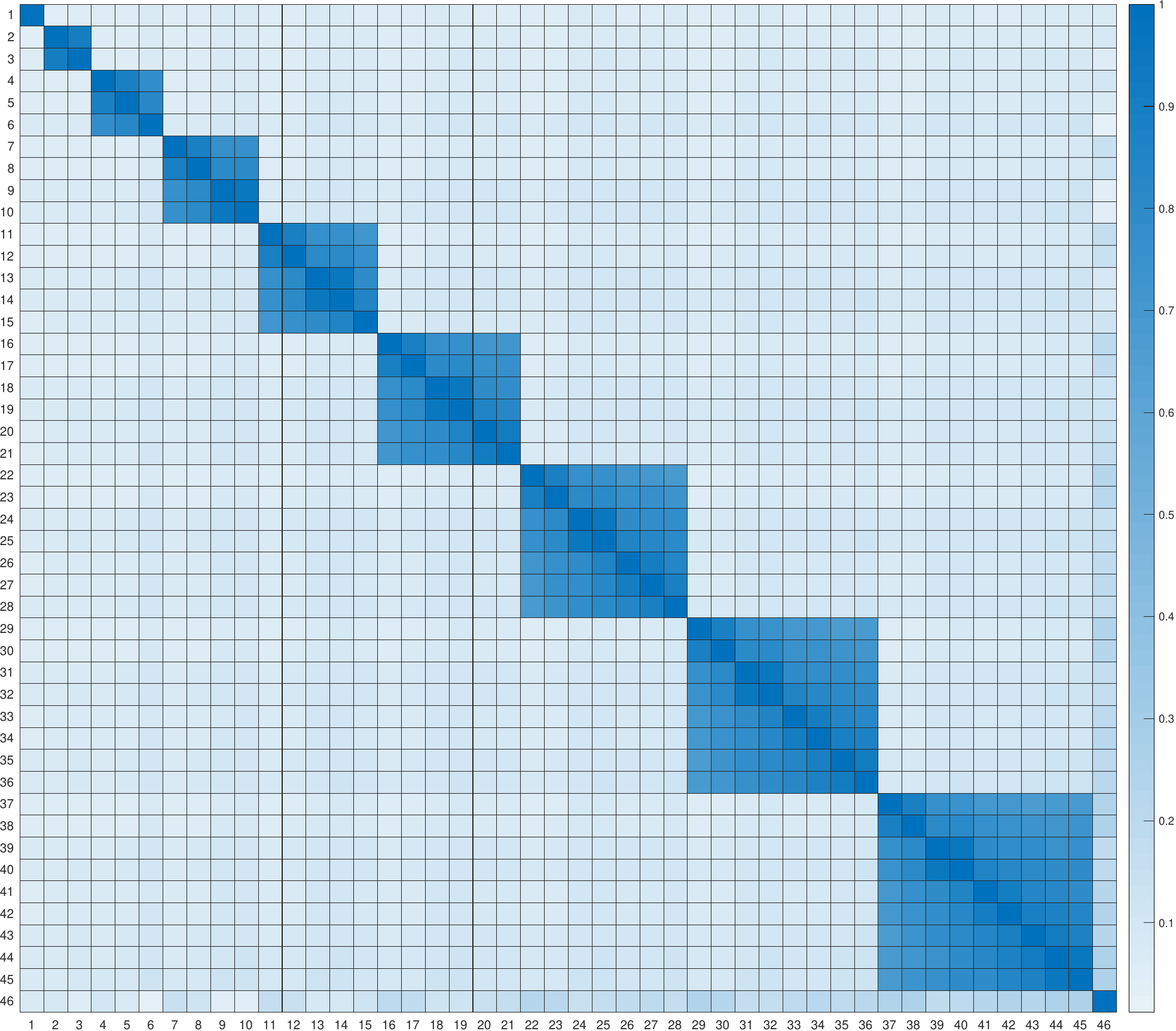}
		}
		\caption{Instruments Correlation Heatmap. Darker blue for larger absolute correlation}
		\label{fig:corr}
	\end{figure}

	Our approach is motivated by the following observations. First, a set of instruments in economic applications can be correlated with each other, often by construction.
	In this case, the model averaging approach in the first-stage is more appropriate. We borrow the intuition from the high-dimensional prediction literature. The model selection method, such as Lasso breaks down when predictors are highly correlated. Alternatively, ridge regression works better for highly correlated predictors and \citet*{elliott2013complete} show the connection between the complete subset averaging and ridge regression. Therefore, we believe that the averaging approach can work better with correlated instruments.
	Figure \ref{fig:corr} shows the correlation structure of instruments in three important studies: 180 instruments (interactions of dummy variables) in \citet*{angrist1991does}, 10 instruments (average characteristics of similar products) in \citet*{berry1995automobile}, and 46 instruments (lagged dependent variables) in \citet*{acemoglu2008income}. The instruments of the first study do not exhibit high correlation, but those of the other two show high correlation. Therefore, it would be empirically relevant to consider the averaging approach suggested in this paper.
	
	Our second motivation is that, in model averaging, estimating weights can cause finite sample efficiency loss, especially when the weight vector's dimension is large. In the forecasting literature it is not surprising to see that equal-weighted averaging outperforms other sophisticated optimal weighting schemes, e.g., see \citet{clemen1989combining}, \citet{stock2004combination}, and \citet{smith2009simple}. Bootstrap aggregating (known as bagging; \citet{breiman1996bagging}) is another example of equal-weight model averaging and is a popular method in the machine learning literature.
	
	It is worth emphasizing the important work by \citet*{elliott2013complete, elliott2015complete} who propose the equal-weight complete subset regressions in the forecasting context. They demonstrate the complete subset regression's excellent performance relative to competing methods such as ridge regression, Lasso, Elastic Net, bagging, and Bayesian model averaging. We build on their idea of the complete subset regression to provide a formal theoretical justification of the CSA-2SLS estimator with extensive Monte Carlo simulations. Indeed, we find that the CSA-2SLS exhibits potentially huge gains in terms of the bias and the MSE relative to existing methods especially when the instruments are correlated and there is large endogeneity.
	
	There are two limitations to be noted.
	First, conditionally homoskedastic errors are assumed in the derivation of the approximate MSE, similarly in \citet{donald2001choosing}, \citet{kuersteiner2010constructing}, \citet{hansen2007least}, \citet{okui2011instrumental}, and \citet*{carrasco2012regularization}. This is required to obtain the explicit order of the higher-order terms in the expansion of the MSE, which allows one-to-one comparison with the existing literature. \citet*{donald2009choosing} derive the approximate MSE under heteroskedasticity for the efficient generalized method of moment (GMM) and the generalized empirical likelihood (GEL) estimators at the expense of more complicated expressions. Also, note that the 2SLS is no longer efficient under heteroskedasticity.
	
	Second, we focus on the 2SLS estimator because of its utmost popularity among applied researchers. The limited information maximum likelihood (LIML) estimator has gained considerable attention recently because of its theoretical advantage over the 2SLS having a smaller bias under the many instruments asymptotics (\citet{donald2001choosing}). 
	To maintain our focus on the new averaging method, however, we defer any extension to the $k$-class estimators such as LIML and bias-corrected 2SLS to future research.

	Finally, we summarize related literature. The model averaging approach becomes prevalent in the econometrics literature. \citet{hansen2007least} shows that the weight choice based on the Mallows criterion achieves optimality. \citet{hansen2012jackknife} propose the jackknife model averaging, which allows heteroskedasticity. \citet*{ando2014model} present a model averaging for high-dimensional regression. \citet{ando2017weight} and \citet*{zhang2016optimal} consider the class of generalized linear models to show the optimality under the Kullback-Leibler loss function. \citet{zhang2018spatial} propose a model averaging in the spatial autoregressive models. \citet{kitagawa2016model} propose the propensity score model averaging estimator for the average treatment effects for treated. \citet{lee2015averaged} propose a model averaging approach over complete subsets in the second stage IV regression.
	
	Our approach is different from the many weak instruments asymptotics in \citet{chao2005consistent}, \citet{stock2005asymptotic}, \citet{han2006gmm}, \citet{andrews2007testing}, and \citet*{hansen2008estimation}, where the concentration parameter can grow slower than the sample size. Alternative estimators under heteroskedasticity and many instruments are proposed by \citet*{hausman2011properties} and \citet*{hausman2012instrumental}. \citet{kuersteiner2012kernel} extends the instrument selection criteria of \citet{donald2001choosing} to the time series setting and proposes a GMM estimator using lags as instruments. \citet{kang2018higher} derives the approximate MSE of IV estimators with locally invalid instruments. \citet*{antoine2014conditional} and \citet{escanciano2017simple} propose estimators free from choice variables by adopting the continuum of unconditional moment condition in the first stage.
	
	The remainder of the paper is organized as follows. Section \ref{sc:model} describes the model and proposes the CSA-2SLS estimator.
	Section \ref{sc:mse} derives the approximate MSE formula of the estimator and investigates its properties. We also show the asymptotic optimality result as well as the implementation procedure. Section \ref{sc:irrelevant} extends the result by allowing an increasing number of irrelevant instruments. Section \ref{sc:orthogonal-IV} investigates the relationship between our approximate MSE and that of \citet*{donald2001choosing} when the instruments are orthogonal. 
	Section \ref{sc:simulation} studies the finite sample properties via Monte Carlo simulations.
	Section \ref{sc:empirical} provides an empirical illustration.
	The online supplement contains proofs and additional simulation results.

	\section{Model and Estimator}\label{sc:model}
	
	We follow the setup of \citet{donald2001choosing} and \citet{kuersteiner2010constructing}. The model is
	\begin{eqnarray}
	y_{i} &=& Y_{i}'\beta_{y} + x_{1i}'\beta_{x} + \eps_{i} = X_{i}'\beta + \eps_{i},\\
	X_{i} &=& \left(\begin{array}{c}
	Y_{i} \\
	x_{1i}
	\end{array}\right) =  f(z_{i}) + u_{i}=\left(\begin{array}{c}
	E[Y_{i}|z_{i}] \\ x_{1i}
	\end{array}\right) + \left(\begin{array}{c}
	\eta_{i} \\ 0
	\end{array}\right) ,~~i=1,\ldots,N,
	\end{eqnarray}
	where $y_{i}$ is a scalar outcome variable, $Y_{i}$ is a $d_{1}\times 1$ vector of endogenous variables, $x_{1i}$ is a $d_{2}\times 1$ vector of included exogenous variables, $z_{i}$ is a vector of exogenous variables (including $x_{1i}$), $\eps_{i}$ and $u_{i}$ are unobserved random variables with finite second moments which do not depend on $z_{i}$, and $f(\cdot)$ is an unknown function of $z$. Let $f_{i} = f(z_{i})$ and $d=d_{1}+d_{2}$. The second equation represents a nonparametric reduced form relationship between $Y_{i}$ and the exogenous variables $z_{i}$, with $E[\eta_{i}|z_{i}]=0$ by construction. Define the $N\times1$ vectors $y = (y_{1},\ldots,y_{N})'$, $\eps = (\eps_{1},\ldots,\eps_{N})'$, and the $N\times d$ matrices $X=(X_{1},\ldots,X_{N})'$, $f=(f_{1},\ldots,f_{N})'$, and $u=(u_{1},\ldots,u_{N})'$.
	
	The set of instruments has the form $Z_{K,i}\equiv (\psi_{1}(z_{i}),\ldots,\psi_{K}(z_{i}),x_{1i})'$, where $\psi_{k}$'s are functions of $z_{i}$ such that $Z_{K,i}$ is a $(K(N)+d_{2})\times1$ vector of instruments. The total number of instruments $K(N)$ increases as $N\rightarrow\infty$ but we suppress the dependency on $N$ and write $K$ unless we need to express the dependence of $K$ on $N$ explicitly.
	Define its matrix version as $Z_{K}=(Z_{K,1},\ldots,Z_{K,N})'$. To define the complete subset averaging estimator, consider a subset of $k$ excluded instruments. Note that the included exogenous variables $x_{1i}$ are always included in the instruments set. To avoid confusion, we will use ``instruments'' to refer to excluded instruments throughout the paper. The number of subsets with $k$ instruments is
	\[\binom{K}{k}=\frac{K!}{k!(K-k)!}.\]
	A complete subset with size $k$ is the collection of all these subsets. Let $M(K,k) = \binom{K}{k}$, which is the number of models given $K$ and $k$. For brevity, the dependence of $M$ on $K$ and $k$ will be suppressed unless it is necessary. For any model $m$ with $k$ instruments, let $Z_{m,i}^{k}$ be an $(k+d_{2})\times 1$ vector of subset instruments including $x_{1i}$ where $d_{1}\leq k\leq K$ and $Z_{m}^{k}=(Z_{m,1}^{k},\ldots,Z_{m,N}^{k})'$ be an $N\times (k+d_{2})$ matrix for $m=1,\ldots, M$. For each $m$, the first stage equation can be rewritten as
	\begin{equation}
	X_{i} = \Pi_{m}^{k'}Z_{m,i}^{k} + u_{m,i}^{k},~~i=1,\ldots,N,
	\end{equation}
	or equivalently,
	\begin{equation}
	X = Z_{m}^{k}\Pi_{m}^{k} + u_{m}^{k},
	\end{equation}
	where $\Pi_{m}^{k}$ is the $(k+d_{2})\times d$ dimensional projection coefficient matrix for model $m$ with $k$ instruments, $u_{m,i}^{k}$ is the projection error, and $u_{m}^{k}=(u_{m,1}^{k},\ldots,u_{m,N}^{k})'$. The projection coefficient matrix is estimated by
	\begin{equation}
	\widehat{\Pi}_{m}^{k} = (Z_{m}^{k'}Z_{m}^{k})^{-1}Z_{m}^{k'}X, \label{eq:pi.hat}
	\end{equation}
	and the average fitted value of $X$ over complete subset $k$ becomes
	\begin{equation}
	\widehat{X} = \frac{1}{M}\sum_{m=1}^{M}Z_{m}^{k}\widehat{\Pi}_{m}^{k} = \frac{1}{M}\sum_{m=1}^{M}Z_{m}^{k}(Z_{m}^{k'}Z_{m}^{k})^{-1}Z_{m}^{k'}X \equiv \frac{1}{M}\sum_{m=1}^{M}P_{m}^{k}X  \equiv P^{k}X,
	\end{equation}
	where $P_{m}^{k} = Z_{m}^{k}(Z_{m}^{k'}Z_{m}^{k})^{-1}Z_{m}^{k'}$ and $P^{k} = \frac{1}{M}\sum_{m=1}^{M}P_{m}^{k}$. The complete subset averaging (CSA) 2SLS estimator with a given $k$ is now defined as
	\begin{equation}
	\widehat{\beta} = (\widehat{X}'X)^{-1}\widehat{X}'y = (X'P^{k}X)^{-1}X'P^{k}y.
	\label{aiv}
	\end{equation}
	The matrix $P^{k}$ is the average of projection matrices. We call this matrix as the complete subset averaging $P$ (CSA-$P$) matrix. Note that the CSA-$P$ matrix is symmetric but not idempotent in general.
	
	Therefore, we can estimate the model by the CSA-2SLS estimator for a given $k$. We will deliberate on the choice of $k$ in the next section and close this section by characterizing the CSA-2SLS estimator and the CSA-$P$ matrix in a broader context.
	
	The CSA-2SLS estimator can be interpreted as the minimizer of the average of 2SLS criterion functions. For each subset of instruments $Z_{m,i}^{k}$, the corresponding moment condition is
	\begin{equation}
	E[Z_{m,i}^{k}\eps_{i}] = E[Z_{m,i}^{k}(y_{i}-X_{i}'\beta)]=0.\label{eq:moment-cond}
	\end{equation}
	The standard 2SLS estimator given the moment condition \eqref{eq:moment-cond} minimizes
	\begin{equation}
	\label{critm}
	(y-X\beta)'Z_{m}^{k}\left(Z_{m}^{k'}Z_{m}^{k}\right)^{-1}Z_{m}^{k'}(y-X\beta).
	\end{equation}
	This equation is the GMM criterion with the weight matrix $(Z_{m}^{k'}Z_{m}^{k})^{-1}$. Conventional model average estimators are based on a weighted average of $\widehat{\beta}^{k}_{m}$, the minimizer of \eqref{critm}, over different models. For this type of model averaging estimators, see \citet{hansen2007least} for the OLS estimator, \citet{lee2015averaged} for the 2SLS estimator, and \citet*{chen2016averaging} for the general approach based on moment conditions.
	
	In contrast, the CSA-2SLS estimator minimizes the average of \eqref{critm} over different models directly:
	\begin{equation}
	\label{crita}
	\widehat{\beta} = \argmin_{\beta}~ \sum_{m=1}^{M}  \left((y-X\beta)'Z_{m}^{k}\left(Z_{m}^{k'}Z_{m}^{k}\right)^{-1}Z_{m}^{k'}(y-X\beta)\right).
	\end{equation}
	The optimal model averaging 2SLS estimator of \citet{kuersteiner2010constructing} can be interpreted similarly as the minimizer of the average of 2SLS criterion functions with data-dependent weights.

	\section{Subset Size Choice and Optimality}\label{sc:mse}
	In this section, we discuss how to choose the subset size $k$. First, we derive the approximate MSE of the CSA-2SLS estimator by the \citet{nagar1959bias} expansion. This high-order expansion shows the bias-variance trade-off that helps improve the finite sample performance of the estimator (e.g. \citet{donald2001choosing}, \citet{kuersteiner2010constructing}, and \citet{carrasco2012regularization}). The subset size $k$ is chosen to minimize the sample counterpart of the approximate MSE whose formula is provided. We prove the optimality of the chosen subset size in the sense of \citet{li1987asymptotic}. 

	\subsection{Approximate MSE}
	In our analysis, the CSA-$P$ matrix $P^{k}$ plays an important role. Since $P^{k}$ is not idempotent unless $k=K(N)$, we cannot directly apply the existing technique in \citet{donald2001choosing} or \citet{kuersteiner2010constructing} to our analysis. 
	We first list regularity conditions. Let $\|A\|=\sqrt{\text{tr}(A'A)}$ denote the Frobenius norm for a matrix $A$.
	\begin{assumption}\
		\begin{enumerate}
			\item[(i)] $\{y_{i},X_{i},z_{i}\}$ are i.i.d. with finite fourth moment and $E[\eps_{i}^{2}]=\sigma_{\varepsilon}^{2}>0$.
			
			\item[(ii)] $E[\eps_{i}|z_{i}]=0$ and $E[u_{i}|z_{i}]=0$.
			
			\item[(iii)] Let $u_{ia}$ be the $a$th element of $u_{i}$. Then $E[\eps_{i}^{r}u_{ia}^{s}|z_{i}]$ are constant and bounded for all $a$ and all $r,s\geq0$ and $r+s\leq4$.
			
			\item[(iv)] $f_{i}$ is bounded.
			
			\item[(v)] $\overline{H} = Ef_{i}f_{i}'$ exists and is nonsingular.
			
			
			\item[(vi)] For each $k\geq d$ and all $m=1,\ldots,M$, $Z_{m}^{k'}Z_{m}^{k}$ is nonsingular with probability approaching one.
			
			\item[(vii)] Let $P_{ii}^{k}$ denote the $i$th diagonal element of $P^{k}$. Then $\max_{i\leq N}P_{ii}^{k}\xrightarrow{p}0$ as $N\rightarrow\infty$.
			
		\end{enumerate}
		\label{A1}
	\end{assumption}
	
	\begin{assumption}
		Let $c>0$ be given. There exists a sequence $\{\underline{k}(N)\}$ such that for all $k(N)\geq \underline{k}(N)$ there exists $\Pi_m^{k(N)}$ that satisfies
		\[\frac{1}{M}\sum_{m=1}^{M}E\|f(z_{i})-\Pi_{m}^{k(N)'}Z_{m,i}^{k(N)}\|^{2} < c\]
		for large enough $N$.
		\label{A2}
	\end{assumption}
	Assumption \ref{A1} collects  standard moment and identification conditions similar to those of \citet{donald2001choosing} and \citet{kuersteiner2010constructing}.
	
	Assumption \ref{A2} controls the speed of the lower bound $\underline{k}(N)$ such that $f(z_i)$ is arbitrarily well-approximated by the size-$k(N)$ complete subset projections. This assumption is not particularly stronger than Assumption 2(ii) of \citet{donald2001choosing} since they are equivalent when $\underline{k}(N)=K(N)$. The approximation error can be small even with $\underline{k}(N) < K(N)$ when instruments are correlated. For example, consider that $f(z_{i})=z_i'\pi_0+o_p(1)$ and all elements of $z_i$ are the linear transformation of $z_{1i}$. Then, Assumption 2 holds with $k=1$ and $M=K$. Below we provide an example where $k(N)$ can grow at arbitrarily slow rate when the instruments are equally correlated at the rate of $\rho$. It is also possible that Assumption \ref{A2} holds only if ${k}(N)/K(N)\rightarrow 1$. A leading case is when the instruments are orthogonal. We provide a detailed analysis for orthogonal instruments in Section \ref{sc:AMSE-diff}. In sum, Assumption 2 can hold for a fixed $k$, a growing $k(N)$ slower than $K(N)$, or a $k(N)$ growing as fast as $K(N)$, depending on the correlation structure of the instruments.
	
	
	We require Assumption \ref{A2} only to derive a valid approximate MSE expression, which gives an intuition on how CSA-2SLS can reduce MSE. It also provides a data adaptive method to choose the subset size $k$ balancing between the bias and the variance in a finite sample. When the bias is relatively large, $k<K$ can be chosen as we confirm both in the simulation studies in Section \ref{sc:simulation} and in the empirical application in Section \ref{sc:empirical}. We also provide an alternative method of choosing $k$ other than the approximate MSE in Section \ref{sc:alternatives} and investigate the approximate MSE under different circumstances in Section \ref{sc:AMSE-diff}.
	
	\begin{example}
		[Equicorrelated Instruments] We provide a special example to further discuss Assumption \ref{A2}. Consider the function
		\[f(z_{i}) = \gamma_{K}\sum_{j=1}^{K}z_{ij}\]
		for $\gamma_{K}\neq0$, $K>1$, and $dim(f)=1$, where $z_{ij}$'s have mean zero, the unit variance, and the constant correlation $\rho$. Since
		\begin{equation*}
		\text{Var}(f(z_{i})) = \gamma_{K}^{2}K(1 + (K-1)\rho),
		\end{equation*}
		where $1+(K-1)\rho>0$, which is the smallest eigenvalue of the covariance matrix of $Z$, we set
		\begin{equation*}
		\gamma_{K} = \frac{1}{\sqrt{K(1 + (K-1)\rho)}}
		\end{equation*}
		so that the variance of $f(z_i)$ is fixed at 1.  Setting $\gamma_{K}$ as a decreasing sequence in $K$ does not necessarily imply weak instruments. Since $K^2/N\rightarrow 0$, $\gamma_K$ converges to zero at a slower rate than $1/\sqrt{N}$ so that $\sqrt{N}$-consistency of the CSA-2SLS estimator is still maintained. This sequence of instruments corresponds to ``nearly weak'' instruments of \citet*{hahn2002discontinuities} and \citet*{antoine2009efficient}. Note that the weak instruments sequence of \citet*{staiger1997instrumental} assumes that the first-stage coefficients are $O(1/\sqrt{N})$ and 2SLS estimators are not $\sqrt{N}$-consistent.
		
		The minimized mean squared approximation error of the CSA with $k$ instruments\footnote{Since the minimization problem is symmetric in $\pi$, we set $\pi_{1}=\pi_{2}=\cdots=\pi_{k}=\pi$ and solve the first-order condition to obtain the minimizer and \eqref{mamse2}.} is
		\begin{align}
		&\min_{{\Pi_{m}^{k}}}\frac{1}{M}\sum_{m=1}^{M}E\|f(z_{i})-\Pi_{m}^{k'}Z_{m,i}^{k}\|^{2} \notag  \\
		&=\min_{\pi_{1},\cdots,\pi_{k}}E[(f(z_{i})-\pi_{1} z_{i1} - \pi_{2} z_{i2} - \cdots -\pi_{k} z_{ik})^{2}] \notag  \\
		\label{mamse2}&= \frac{1-\rho}{1+(k-1)\rho}\left(1-\frac{k}{K}\right).
		\end{align}
		Some interesting observations can be made for the expression \eqref{mamse2}.
		
		First, the approximation error \eqref{mamse2} converges to zero so long as $k = O(K^{\alpha})$ for $0<\alpha\leq1$ as $K\rightarrow\infty$. Since $k$ can grow at an arbitrarily slow rate if $\rho\neq0$, at least for this example, Assumption 2 imposes a very mild condition on $k$. In contrast, if $\rho=0$ or $\rho=1/K$, then \eqref{mamse2} converges to zero if and only if $k/K\rightarrow1$, which would be the least favourable scenario for the CSA-2SLS estimator.  
		
		Second, for any given $(k,K)$, the approximation error gets smaller as $\rho$ increases and it is equal to zero when $\rho=1$. Figure \ref{fg:approxerr} shows the approximation errors when $K=20$ and $k=1,3,5$.
		
		Third, the approximation error is decreasing with $k$ uniformly in $\rho$, which is shown in Figure \ref{fg:approxerr}. Evaluating \eqref{mamse2} at $k$ and $k+1$, respectively, the difference is $(1-\rho)(1+(K-1)\rho)/(1+\rho)\geq 0$.
		
		\begin{figure}[t]
			\centering
			\includegraphics[scale=0.2]{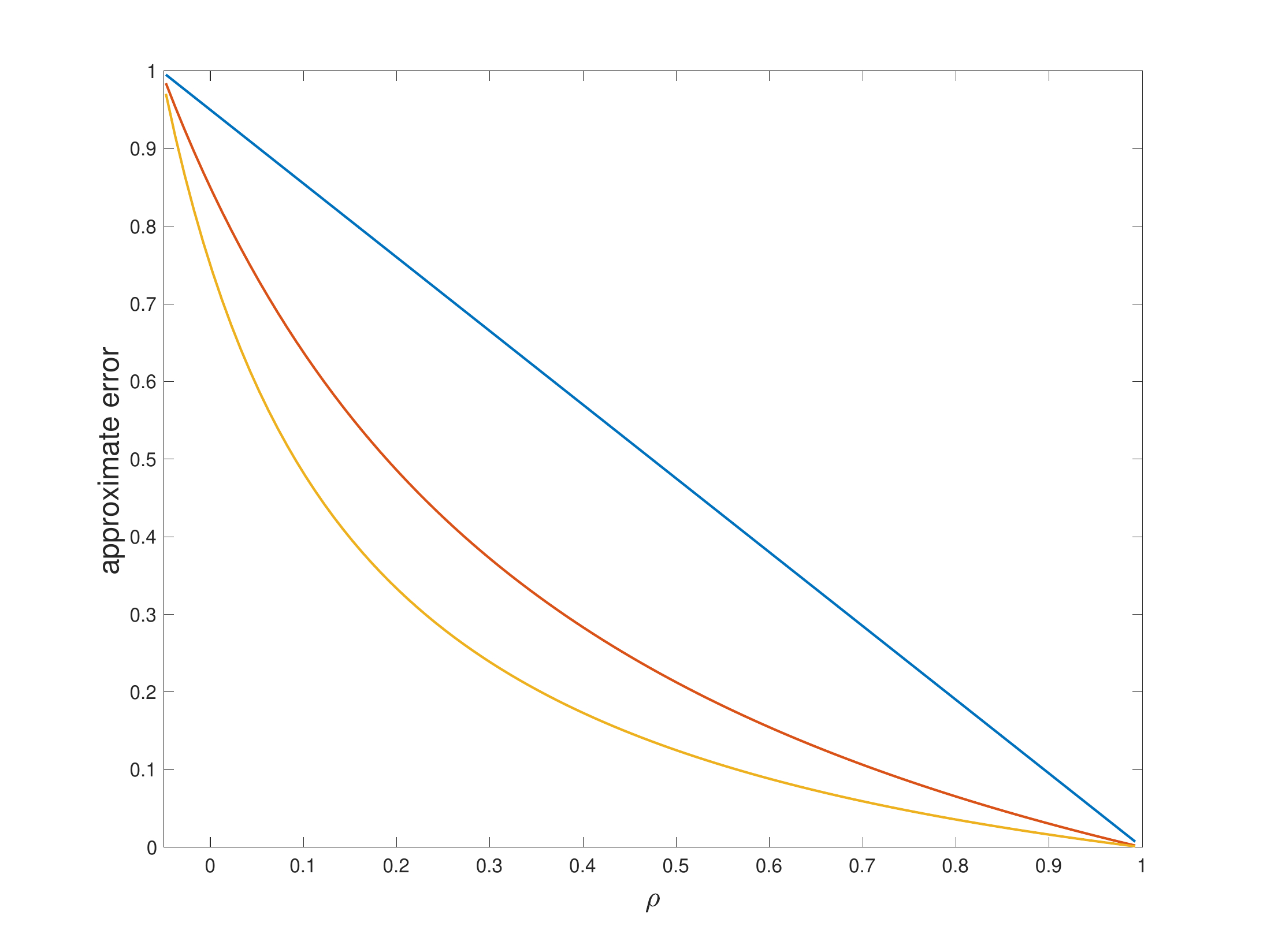}
			\caption{Approximation error when $K=20$: $k=1$ (blue), $k=3$ (red), $k=5$ (yellow)}
			\label{fg:approxerr}
		\end{figure}
		
	\end{example}
	
	%
	
	
	We now provide the approximate MSE formula of CSA-2SLS. Define $H = f'f/N$ and $P_{f} = f(f'f)^{-1}f'$.
	\begin{theorem}
		Suppose that Assumptions \ref{A1}-\ref{A2} are satisfied, $k^{2}/N\rightarrow0$, and $E[u_{i}\eps_{i}|z_{i}] = \sigma_{u\varepsilon}\neq0$. Then, we have the following results for the CSA-2SLS estimator:
		\begin{eqnarray}
		\label{mse1}
		N(\widehat{\beta}-\beta)(\widehat{\beta}-\beta)' &=& \widehat{Q}(k) + \widehat{r}(k),\\
		\label{mse2}
		E\left[\widehat{Q}(k)|Z\right] &=& \sigma_{\eps}^{2}H^{-1} + S(k) + T(k), \\
		\frac{\widehat{r}(k)+T(k)}{\trace(S(k))} & = & o_p(1), \mbox{ as } k \rightarrow \infty, N \rightarrow \infty,
		\end{eqnarray}
		with
		\begin{equation}
		\label{mse}
		S(k) = H^{-1}\left[\sigma_{u\eps}\sigma_{u\eps}'\frac{k^{2}}{N} + \sigma_{\eps}^{2}\frac{f'(I-P^{k})(I-P_{f})(I-P^{k})f}{N}\right]H^{-1}.
		\end{equation}
		\label{Thm1}
	\end{theorem}
	
	In the expansion, $\widehat{r}(k)$ and $T(k)$ are terms of smaller order in probability than those in $S(k)$ and that the term $\sigma_{\eps}^{2}H^{-1}$ is the first-order asymptotic variance under homoskedasticity. The first term of $S(k)$ (ignoring pre- and post-multiplied $H^{-1}$), $\sigma_{u\eps}\sigma_{u\eps}'k^{2}/N$, corresponds to the bias and is similar to that of \citet{donald2001choosing}. The second term is different from the usual higher-order variance in similar expansions in the literature. Let $V(k) = f'(I-P^{k})(I-P_{f})(I-P^{k})f/N$. It can be decomposed as
	\begin{equation}
	\label{highvar}
	V(k) = \frac{f'(I-P^{k})^{2}f}{N} -  \frac{f'(I-P^{k})f}{N}\left(\frac{f'f}{N}\right)^{-1}  \frac{f'(I-P^{k})f}{N}.
	\end{equation}
	In \citet{donald2001choosing}, the higher-order variance term, which takes the form of $f'(I-P^{K})f/N$ where $P^{K}$ is the projection matrix consisting of $K$ instruments, decreases with $K$. This gives the bias-variance trade-off in their expression. In contrast, $V(k)$ is the sum of two components and the monotone decrease with respect to $k$ is not always guaranteed.

	We further elaborate on the behavior of the higher-order variance term $V(k)$ using shrinkage. Since $I-P_{f}$ is idempotent, we can write
	\begin{equation}
	V(k) = \frac{u^{k'}(I-P_f)(I-P_f)u^{k}}{N}
	\end{equation}
	where $u^k\equiv(I-P^k)f$.
	There exist two factors that force to move $V(k)$ into opposite directions as $k$ varies. First, the norm of $u^k$ decreases as $k$ increases since $P^{k^{*}}$ for $k^{*}>k$ is the average of the projection matrices on the space spanned by a larger set of instruments. Second, $P^k$ can cause less shrinkage for $P^{k}f$ for a larger $k$, which makes the norm of $(I-P_f)u^k$ larger.  Note that
	\begin{equation}
	\label{rel}
	(I-P_f)u^k = (I-P_f)(f-P^kf)=f - P_{f}f - P^{k}f + P_{f}P^{k}f = -(I-P_{f})P^{k}f
	\end{equation}
	and that $P^k f$ with less shrinkage lies farther away from the space spanned by $f$. Therefore, if the second factor dominates the first then $V(k)$ increases along with $k$. 

	If there are more than one endogenous variable, then we choose $k$ that minimizes a linear combination of the MSE, $S_{\ld}(k)\equiv  \lambda' S(k)\lambda$ for a user-specified $\lambda$ similarly in \cite{donald2001choosing}. 

	\subsection{Implementation and Optimality of Approximate MSE}
	The approximate MSE $S(k)$ in \eqref{mse} gives useful guidance on the data-adaptive choice of $k$. Both terms in $S(k)$ converge to zero asymptotically, but there is a trade-off between the higher-order bias and variance terms in finite samples. Similarly, in \citet{donald2001choosing} and \citet{kuersteiner2010constructing}, the approximate MSE sheds light on the choice of $k$ only when there exists a non-negligible size of bias for a given sample size $N$ with many instruments.
	
	We first introduce notation to construct the sample counterpart of the approximate MSE. Let $\widetilde{\beta}$ be a preliminary estimator that is fixed across different values of $k$ and $\widetilde{\varepsilon}=y-X\widetilde{\beta}$. Let $\wtd{f}$ be an estimate of $f$. The residual matrix is denoted by $\widetilde{u}=X-\wtd{f}$ and $\wtd{u} = (u_1,u_2,\ldots,u_N)'$ where $\widetilde{u}_{i}$ is a $d\times1$ vector. Define $\widetilde{H} = \widetilde{f}'\widetilde{f}/N$, $\widetilde{\sigma}_{\eps}^{2} = \widetilde{\eps}'\widetilde{\eps}/N$, $\widetilde{\sigma}_{u\eps} = \widetilde{u}'\widetilde{\eps}/N$, $\wtd{\sigma}_{\ld\eps} = \wtd{\ld}'\wtd{H}^{-1}\widetilde{\sigma}_{u\eps}$, and $\wtd{\Sigma}_u=\wtd{u}'\wtd{u}/N$. The feasible criterion function for $S_{\ld}(k)$ is defined as below:
	\begin{align*}
	\widehat{S}_{\ld}(k)
	& = \wtd{\sigma}_{\ld \eps}^2\frac{k^{2}}{N} + \wtd{\sigma}_{\eps}^{2}  \left[ \wtd{\ld}' \wtd{H}^{-1} \wtd{e}_f^{k} \wtd{H}^{-1}\wtd{\ld} - \wtd{\ld}' \wtd{H}^{-1} \wtd{\xi}_f^{k} \wtd{H}^{-1} \wtd{\xi}_f^{k} \wtd{H}^{-1}\wtd{\ld} \right],
	\end{align*}
	where
	\begin{align*}
	\wtd{e}_f^k    & = \frac{X'(I-P^k)^2 X}{N}+\wtd{\Sigma}_u\lt(\frac{2k-\trace((P^k)^2)}{N}\rt), \\
	\wtd{\xi}_f^k & =  \frac{X'(I-P^k) X}{N} + \wtd{\Sigma}_u \frac{k}{N} - \wtd{\Sigma}_u,\\
	\widetilde{\sigma}^{2}_{\ld\eps} &= (\widetilde{\ld}'\widetilde{H}^{-1}\widetilde{\sg}_{u\eps})^{2},\\
	\wtd{\Sigma}_u &=\frac{\wtd{u}'\wtd{u}}{N}.
	\end{align*}
	Then, we choose $\what{k}$ as a minimizer of the feasible approximate MSE function, $\what{S}_{\ld}(k)$. Given $\what{k}$, we can compute the CSA-2SLS estimator defined in \eqref{aiv}.
	
	We remark on two practical issues in implementation. First, our theory requires the preliminary estimator $\wtd{\bt}$ be consistent, and it is simply achieved by 2SLS estimation with any valid IVs. However, the performance of the estimator in a finite sample might rely on the choice. Following \citet{donald2001choosing} and \citet{kuersteiner2010constructing}, we propose choosing the set of IVs based on the first stage Mallows criterion, which brings satisfactory performance results in the extensive simulations experiments reported in Section \ref{sc:simulation}. Second, it could be infeasible to estimate $P^k$ when $K$ is too large. Note that the number of complete subsets (all the subsets with different $k$'s) grows exponentially, $2^K-1$. To deal with this computational issue, we propose to draw a smaller number of submodels randomly as follows. Let $\mathcal{R}^k$ be a class of $R$ subsets with $k$ elements that are randomly selected from $\{1,\ldots,M\}$. Then, the CSA projection matrix with randomly drawn submodels is defined as $\breve{P} = R^{-1} \sum_{r\in\mathcal{R}^k} P^k_r$, where $P^k_r$ is a projection matrix using the set of IVs indexed by $r$. Table \ref{tb:m19-small-R} in Section \ref{sc:all-tables-figures} of the Appendix shows simulation results such that randomly drawn submodels work well with a feasible size of $R$.
	
	We finalize this subsection by proving that the CSA-2SLS estimator achieves the asymptotic optimality in the sense that it minimizes the MSE among the class of CSA-2SLS estimators with different complete subset sizes. We collect additional regularity conditions below.
	

	\begin{assumption}\
		\label{A:consistency-and-rate}
		\begin{enumerate}
			\item[(i)] $\wtd{\sig}^2_{\eps}\xrightarrow{p} \sig^2_{\eps}$, $\wtd{\sig}^2_{u\eps}\xrightarrow{p} \sig^2_{u\eps}$, $\wtd{\ld} \xrightarrow{p} \ld$, $\wtd{H} \xrightarrow{p} \overline{H}$, and  $\ld'\overline{H}^{-1}\sig_{u\eps} \neq0$.
			\item[(ii)] $\lim_{N\rightarrow \infty} \sum_{k=1}^{K(N)} \lt(N S_{\ld}(k)\rt)^{-1} =0$ almost surely in $Z$.
			\item[(iii)] There exist a constant $\phi \in (0,1/2)$ such that $\|\wtd{\Sg}_u- \Sg_u\| = O_p(N^{-1/2+\phi}S_{\ld}(k)^{\phi})$ where $\Sigma_{u} = E[u_{i}u_{i}'|z_{i}]$.
		\end{enumerate}
	\end{assumption}
	Assumptions \ref{A:consistency-and-rate}(i) and (ii) are similar to Assumptions 4--5 in \citet{donald2001choosing}. Assumption \ref{A:consistency-and-rate}(i) is a high-level assumption on the consistency of the preliminary estimators.
	Assumption \ref{A:consistency-and-rate}(ii) is a standard assumption in the model selection or model averaging literature (see, e.g.\ Assumption (A.3) in \citet{li1987asymptotic}). This condition excludes the case that $f$ is perfectly explained by a finite number of instruments.
	Assumption \ref{A:consistency-and-rate}(iii) is a mild requirement on the convergence rate of $\wtd{f}$ that the standard series estimator easily satisfies. For example, consider a series estimator with $\wtd{k}$ b-spline bases as in \citet{newey1997convergence}, where $\sup_{z} \Vert f(z) - \Pi^{\wtd{k} \prime}_0 Z_{\wtd{k}}(z)  \Vert =O_p(\wtd{k}^{-\ap})$ where $\ap=1$ and $Z_{\wtd{k}}(z)\equiv(\psi_{1}(z),\ldots,\psi_{K}(z),x_{1})'$. By choosing the optimal rate for $\wtd{k}$, we get the rate for the error variance, $\Vert \wtd{\Sg}_u - \Sg_u \Vert =O_p(N^{-1/3})$. Since $S_{\ld}(k)=o_p(1)$, we can write $S_{\ld}(k)=O_p(N^{-c})$ for some $c>0$. Then, Assumption \ref{A:consistency-and-rate}(iii) requires $\Vert \wtd{\Sg}_u - \Sg_u \Vert =O_p(N^{-1/2+\phi-c})$. Comparing this rate with $O_p(N^{-1/3})$ above, we can find $\phi >1/6+c$.
	
	The following result shows that the proposed estimator is optimal among the class of the CSA-2SLS estimators.
	\begin{theorem}
		\label{Thm-optimality}
		Under Assumptions \ref{A1}, \ref{A2}, and \ref{A:consistency-and-rate}, as $N\rightarrow\infty$,
		\eq{
			\frac{S_{\ld}(\widehat{k})}{\min_{k } S_{\ld}(k)} \xrightarrow{p} 1.
		}
	\end{theorem}

	\subsection{Cross-validation Method} \label{sc:alternatives}
	
	In this subsection we consider an alternative method of choosing the subset size $k$. The approximate MSE provides useful guidance in choosing $k$. More importantly, it gives an intuition on how CSA-2SLS could reduce MSE further by providing the expression for the higher-order bias and variance. However, the theoretical illustration in the previous subsections depends on a few simplifying assumptions: homoskedasticity and the asymptotic approximation of $f(z_i)$ by the complete subset model averaging (Assumption \ref{A2}). When they are violated, the subset size choice based on the approximate MSE improves the finite sample property of CSA-2SLS less than expected. Therefore, it is worthwhile to propose some alternative methods for the subset size choice.
	
	We propose to choose $k$ by the cross-validation (CV) method that minimizes the first-stage mean squared prediction error. We explain it by the leave-one-out cross-validation. This can be immediately extended to any $b$-fold cross-validation, where $b$ is the number of subsamples. For $k=1,\ldots, K$, we define the following cross-validation objective function:
	\eqs{
		CV_n(k) = \frac{1}{N} \sum_{i=1}^N \lt\Vert X_i - \frac{1}{M} \sum_{m=1}^M \what{\Pi}_{m,-i}^{k \prime} Z_{m,i}^k  \rt\Vert^2
	}where $\what{\Pi}_{m,-i}^{k \prime}$ is the jackknife estimator for $\what{\Pi}_{m}^{k \prime}$, which is estimated by \eqref{eq:pi.hat} without using the $i$-th observation $(X_i, Z_i)$. Then, we choose $k$ that minimizes $CV_n(k)$. \citet{lee2020complete} apply a similar cross-validation method in the context of quantile prediction and show that $\what{k}$ is asymptotically equivalent to the infeasible best subset size. However, the optimality result is not for the second-stage MSE but for the prediction error of the first-stage regression. Therefore, the CV method is robust to the requirement of homoskedasticity and Assumption \ref{A2} but it might perform worse than the approximate MSE when those conditions are satisfied.

	\section{Approximate MSE under Different Circumstances}\label{sc:AMSE-diff}
	\subsection{Irrelevant Instruments}\label{sc:irrelevant}
	
	The higher-order MSE expansion in the previous section assumes an increasing sequence of instruments where the instruments are strong enough for $\sqrt{N}$-consistency and the existence of higher-order terms. This implies that the concentration parameter, which measures the strength of the instruments, increases at the rate of $N$. An important feature of the CSA-2SLS is that no instrument is excluded either in finite samples or asymptotically due to equal-weighted averaging. Assume that the econometrician does not know whether irrelevant instruments are included or not. 
	Although irrelevant instruments can be excluded when a pre-screening method is available, it is important to investigate whether the CSA-2SLS estimator and the MSE expansion are still valid if irrelevant instruments happen to be included in the averaging. 
	In contrast, irrelevant instruments can be excluded by zero or negative weights in \citet{kuersteiner2010constructing}.
	
	In this section, we derive the MSE under a growing set of irrelevant instruments as $N$ and $K$ increase. This makes the concentration parameter smaller in level for given $N$ and $K$ than the case where all the instruments are relevant. By comparing the MSEs with and without irrelevant instruments we can analyze the effect of irrelevant instruments on the MSE. Under the many weak instruments asymptotics of \citet{chao2005consistent}, the growth rate of the concentration parameter can be slower than $N$.
	In this case, the 2SLS estimator is not $\sqrt{N}$-consistent anymore and it would be difficult to compare the higher-order expansion with the other case. Thus, we leave the analysis with many weak instruments for future research.
	
	For the sake of simplicity, we assume that there is no exogenous variables, $x_{1i}$, so the model simplifies to
	\begin{eqnarray}
	y_{i} &=& X_{i}'\beta + \eps_{i},\\
	X_{i} &=& E[X_{i}|z_{i}] + u_{i},~~i=1,\ldots,N,
	\end{eqnarray}
	where $X_{i}$ is a $d\times 1$ vector of endogenous variables and $z_{i}$ is a vector of exogenous variables. We set $f(z_{i}) = E[X_{i}|z_{i}]$. We divide $z_{i}$ into the relevant and irrelevant ones, $z_{1i}$ and $z_{2i}$ and make the following definition:
	\begin{definition}[Definition 4.1]\label{definition:irrelevantIV}
		A vector of instruments $z_{2i}$ is irrelevant if $E[X_{i}|z_{1i},z_{2i}] = E[X_{i}|z_{1i}]$.
	\end{definition}
	Under Definition \ref{definition:irrelevantIV}, the set of instruments $Z_{K,i}$ can be divided into two sets, relevant and irrelevant instruments. Let $K_{1}$ and $K_{2}$ be the number of relevant and irrelevant instruments, respectively. 
	
	We introduce additional definitions. Let $\mathcal{M}(K,k) =\left\{1,2,\ldots,M(K,k)\right\}$ be the index set for all subsets with $k$ instruments and let $\mathcal{M}_{1}(K,K_{1},k)$ and $\mathcal{M}_{2}(K,K_{2},k)$ be the index sets for subsets with at least $d$ relevant instruments and those with less than $d$, respectively. By construction, $\mathcal{M}_{1}(K,K_{1},k)\cup\mathcal{M}_{2}(K,K_{2},k) =\mathcal{M}(K,k)$. Let $M_{1}(K,K_{1},k)$ and $M_{2}(K,K_{2},k)$ be the cardinality of $\mathcal{M}_{1}(K,K_{1},k)$ and $\mathcal{M}_{2}(K,K_{2},k)$, respectively. For brevity, we suppress the dependence on $K$, $K_{1}$, $K_{2}$, and $k$. 
	
	Each element of the index sets $\mathcal{M}_{1}$ and $\mathcal{M}_{2}$ corresponds to a relevant and an irrelevant first stage model (a set of $k$ instruments). Since there are $d$ endogenous variables, those models in $\mathcal{M}_{1}$ satisfy the order condition for identification and (completely) approximate $f(z_{i})$ as $k$ increases. In contrast, those irrelevant models in $\mathcal{M}_{2}$ do not approximate $f(z_{i})$ as $k$ increases. 
	
	We make assumptions that will replace Assumption \ref{A2}. 
	
	\begin{assumption}\
		\begin{enumerate}
			
			\item[(i)]         Let $c>0$ be given. There exists a sequence $\{\underline{k}(N)\}$ such that for all $k(N)\geq \underline{k}(N)$ there exists $\Pi_m^{k(N)}$ that satisfies
			\[\frac{1}{M_1}\sum_{m \in \mathcal{M}_1}E\|f(z_{i})-\Pi_{m}^{k(N)'}Z_{m,i}^{k(N)}\|^{2} < c\]
			for large enough $N$.
			\item[(ii)] For each $k\geq d$ and some positive constant $C$, $0 <C^{-1}\leq M_{1}/M$.
			
			\item[(iii)] $M_{2}/M=o(1/\sqrt{k})$ as $N\rightarrow\infty$.

		\end{enumerate}
		\label{A3}
	\end{assumption}
	Assumption \ref{A3}(i) is a version of Assumption \ref{A2} with relevant instruments. Assumptions \ref{A3}(ii)-(iii) are restrictions on the proportion of relevant and irrelevant first-stage models. Assumption \ref{A3}(ii) sets a non-zero lower bound for the proportion of relevant first-stage models but still allows an increasing number of irrelevant instruments. Assumption \ref{A3}(iii) states that the proportion of irrelevant models decreases to zero at the $1/\sqrt{k}$ rate. Since $M_{2}(K,K_{2},k)=0$ for $k\geq K_{2}+d-1$ by construction (e.g. if $d=1$, $K=10$, and $K_{2}=5$, one cannot choose a subset of $6$ or more irrelevant instruments), Assumption \ref{A3}(iii) controls the growth rate of the sequence $k(N)$ as $N$, $K$, and $K_{2}$ grow. 
	
	To fix ideas, suppose that $d=1$ (one endogenous variable). $M_{1}$ is the number of subsets with $k$ instruments including at least 1 relevant instrument and $M_{2}$ is the number of subsets with $k$ irrelevant instruments. Since $M=M_{1}+M_{2}$ by definition and
	\begin{equation*}
	M_{2} = \left(\begin{array}{c}
	K_{2} \\ k
	\end{array}\right)
	\end{equation*}
	for $k=1,...,K_{2}$ and $M_{2}=0$ for $k=K_{2}+1,...,K$, we have
	\begin{equation*}
	\frac{M_{1}}{M} =\left\{\begin{array}{ll}
	1-\frac{(K-k)(K-k-1)\cdots(K_{2}-k+1)}{K(K-1)\cdots (K_{2}+1)} & \text{ if } 1\leq k \leq K_{2},\\
	1 & \text{ if }K_{2}+1\leq k \leq K.
	\end{array}\right.
	\end{equation*}
	Thus, $M_{1}/M$ is monotonically increasing in $k$ for given $K$ and $K_{1}$ with the minimum $K_{1}/K$ at $k=1$. Thus, Assumption \ref{A3}(ii) holds if there is a positive constant $C$ such that $0< C^{-1} \leq K_{1}/K$.
	
\begin{figure}[t]
		\centering
		\includegraphics[width=0.7\textwidth]{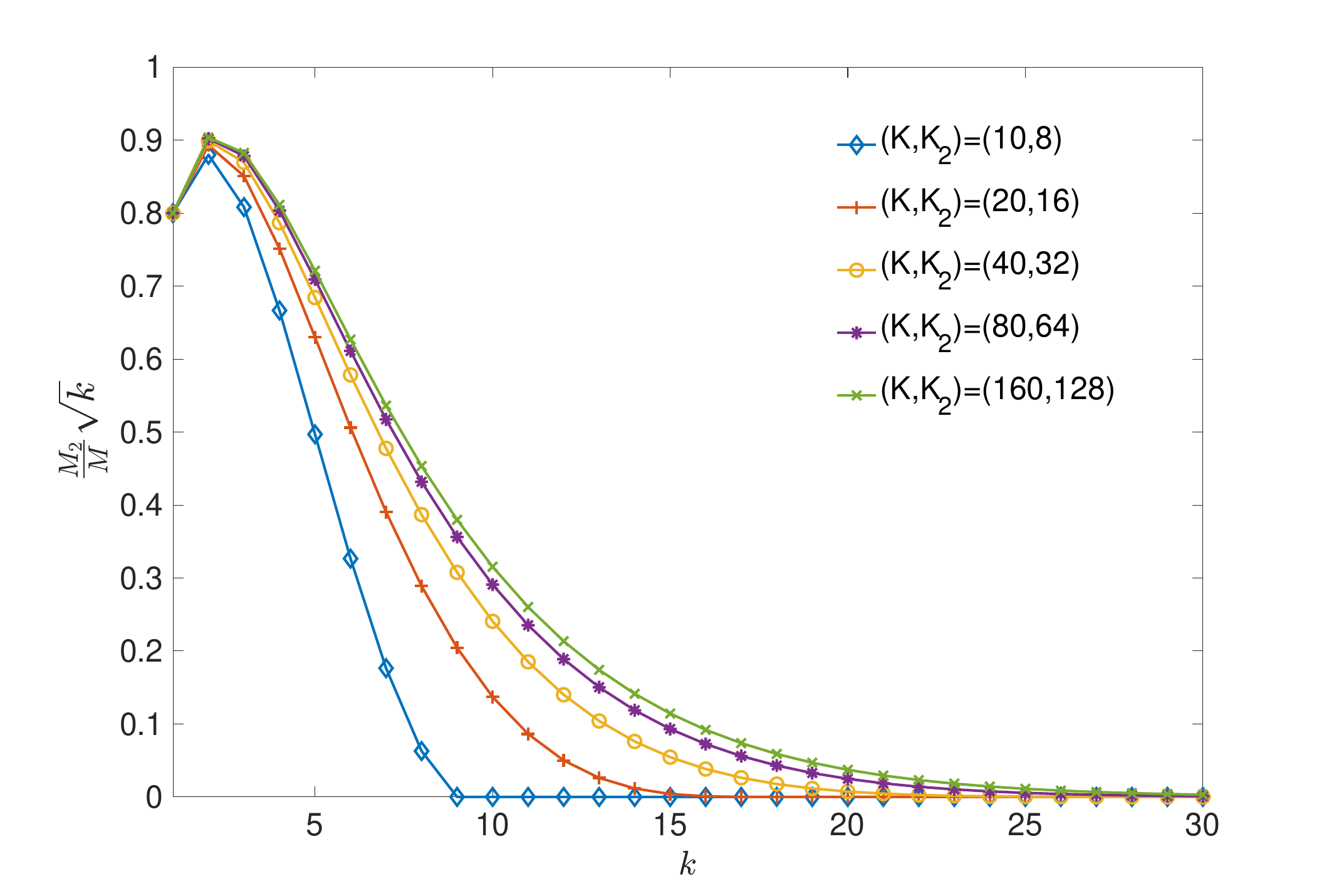}
		\caption{$(M_{2}/M)\sqrt{k}$ for different sets of $(K,K_{2})$ with $K_{2}/K=0.8$}
		\label{fig:M2M}
	\end{figure}

	We can also find a sufficient condition for Assumption \ref{A3}(iii) when $d=1$. By using the bounds for binomial coefficients and Stirling's formula for a large $k$, 
	\begin{equation}
	\frac{M_{2}}{M} = \frac{\left(\begin{array}{c}
		K_{2} \\ k
		\end{array}\right)}{\left(\begin{array}{c}
		K \\ k
		\end{array}\right)} \leq \left(\frac{K_{2}}{K}\right)^{k}\frac{k^{k}}{k!}\sim \left(e\frac{K_{2}}{K}\right)^{k}\frac{1}{\sqrt{k}}.
	\end{equation}
	Thus, Assumption \ref{A3}(iii) holds if $K_{2}/K<e^{-1}$. Since this is a sufficient condition, Assumption \ref{A3}(iii) can hold even when the proportion of irrelevant instruments is much higher than $e^{-1}$. Figure \ref{fig:M2M} shows that $(M_{2}/M)\sqrt{k}\rightarrow0$ as long as $k$ grows sufficiently fast as $(K,K_{2})$ increases when $K_{2}/K=0.8$.
	
	Now we can generalize the main theorem allowing for an increasing number of irrelevant instruments. 
	
	\begin{theorem}
		If Assumptions \ref{A1} and \ref{A3} are satisfied, $k^{2}/N\rightarrow0$, and $\sigma_{u\varepsilon}\neq0$, then for the CSA-2SLS estimator the equations \eqref{mse1}-\eqref{mse2} are satisfied with
		\begin{equation}
		\label{mse_ir}
		S(k) = \left(\frac{M}{M_{1}}\right)^{2}H^{-1}\left[\sigma_{u\eps}\sigma_{u\eps}'\frac{k^{2}}{N} + \sigma_{\eps}^{2}\frac{f'(I-P^{k})(I-P_{f})(I-P^{k})f}{N}\right]H^{-1}.
		\end{equation}
		\label{Thm-MSE2}
	\end{theorem}
	
	Theorem \ref{Thm-MSE2} shows that the approximate MSE formula in the presence of irrelevant instruments is the same as the formula in Theorem \ref{Thm1} except that $(M/M_{1})^{2}$ is multiplied. Since $(M/M_{1})^{2}$ depends on $k$, the optimal $k$ that minimizes \eqref{mse_ir} may differ from the minimizer of \eqref{mse}. 
	
	Specifically, by observing that $(M/M_{1})^{2}$ places larger penalties for smaller $k$, we can conclude that the optimal $k$ will be (weakly) larger than those without the term $(M/M_{1})^{2}$. A practical recommendation is to try larger values of $k$ than the minimizer of the sample approximate MSE if the presence of irrelevant (or very weak) instruments is suspected.

	\subsection{Orthogonal Instruments} \label{sc:orthogonal-IV}
	
	There are special cases when the CSA-2SLS estimator coincides exactly with the 2SLS estimator using all the instruments. The two estimators are the same if (a) $k=K$, which is trivial, or (b) the instruments are mutually orthogonal. We elaborate on the second case.
	
	The projection matrix of the orthogonal instruments is equal to the sum of the projection matrices of each of the instruments.  Thus, the projection matrix of the 2SLS estimator, $P$, becomes
	\eqs{
		P=\widetilde{P}^1_1 + \widetilde{P}^1_2 + \cdots + \widetilde{P}^1_K,
	}
	where $\widetilde{P}^{1}_{j}$ for $j=1,...,K$ is the projection matrix based on the orthogonal instrument $j$ with the subset size 1.  Now consider the CSA-$P$ matrix with subset size $k$ constructed from the  orthogonal instruments:
	\eqs{
		P^k \equiv \frac{1}{M(K,k)}\lt(P^k_1+\cdots P^k_M\rt) = \frac{\binom{K-1}{k-1} }{M(K,k)}\lt(\widetilde{P}^1_1+\cdots + \widetilde{P}^1_K\rt)= \frac{k }{K}P.
	}
	Since the term $k/K$ is cancelled out when we plug-in $(k/K)P$ into Equation \eqref{aiv}, the CSA-2SLS estimator becomes identical to the 2SLS estimator for any $k$.
	
	The identity result does not hold in general when instruments are correlated although they can be always orthogonalized without affecting the column space they span. We illustrate this point by a simple example with $K=2$ and $k=1$. Let $(Z_1,Z_2)$ be the vector of instruments and $(\wtd{Z}_1,\wtd{Z}_2)$ be the corresponding vector of orthogonalized instruments. Then,
	\eqs{
		P^1 (\text{CSA-$P$ matrix with $k=1$}) = \frac{1}{2} \lt(P^1_1 + P^1_2\rt) \neq \frac{1}{2} \lt(\wtd{P}^1_1 + \wtd{P}^1_2\rt) \equiv \frac{1}{2} \wtd{P}^2 = \frac{1}{2} P^2,
	}where $\wtd{P}$ denotes a generic projection matrix of the orthogonalized instruments. Therefore, the CSA-2SLS estimator using $P^1$ does not give the same estimate as that using $\frac{1}{2} P^2$, the 2SLS estimator.
	
	The approximate MSE in \eqref{mse} is simplified further when instruments are orthogonal. Let $\Delta_{K} = \text{tr}\left(\frac{f'(I-P)f}{N}\right)$. Because of $P^k=(k/K)P$, $(I-P_{f})f=0$, and the idempotent property of $I-P$, we have
	\begin{align}
	\notag \frac{f'(I-P^{k})(I-P_{f})(I-P^{k})f}{N} &= \left(\frac{k}{K}\right)^{2}\frac{f'(I-P)(I-P_{f})(I-P)f}{N}\\
	\notag &= \left(\frac{k}{K}\right)^{2}\left(\frac{f'(I-P)f}{N} - \frac{f'(I-P)f}{N}\left(\frac{f'f}{N}\right)^{-1}\frac{f'(I-P)f}{N}\right)\\
	\label{ortho2} &= \left(\frac{k}{K}\right)^{2}\frac{f'(I-P)f}{N} +O_{p}(\Delta_{K}^{2}).
	\end{align}
	Thus, \eqref{mse} can be written as
	\begin{equation}
	\label{ort:mse}
	S(k) = \left(\frac{k}{K}\right)^{2}\widetilde{S}(K) + O_{p}(\Delta_{K}^{2})
	\end{equation}
	where
	\begin{equation*}
	\widetilde{S}(K) = H^{-1}\left[\sigma_{u\eps}\sigma_{u\eps}'\frac{K^{2}}{N} + \sigma_{\eps}^{2}\frac{f'(I-P)f}{N}\right]H^{-1}.
	\end{equation*}
	Note that $\widetilde{S}(K)$ is the approximate MSE for 2SLS using $K$ instruments as shown in \citet*{donald2001choosing} and is independent of $k$. We show in the following Corollary that the approximate MSE in \eqref{mse} is asymptotically equivalent to $\widetilde{S}(K)$ when instruments are orthogonal.
	\begin{corollary}\label{col:orthogonal}
		Suppose that Assumptions \ref{A1}--\ref{A2} are satisfied with orthogonal instruments. Then, for any non-zero $d\times 1$ vector $\lambda$, as $N \rightarrow \infty$,
		\eq{
			\frac{S_{\ld}(k)}{\widetilde{S}_{\ld}(K)} \parrow 1. \label{eq:S-tilde}
		}
	\end{corollary}
	Corollary \ref{col:orthogonal} holds because Assumption \ref{A1}--\ref{A2} imply $k/K\rightarrow1$ and $\Delta_{K}\xrightarrow{p}0$ in \eqref{ort:mse} when instruments are orthogonal. The full proof is given in Appendix A.
	
	Recall that CSA-2SLS does not depend on $k$ when instruments are orthogonal. Corollary \ref{col:orthogonal} establishes the logical consistency by showing that the approximate MSE becomes independent of $k$ as well when instruments are orthogonal.
	

	\section{Simulation} \label{sc:simulation}
	
	We investigate the finite sample properties of the CSA-2SLS estimator by conducting Monte Carlo simulation studies\footnote{The replication R codes for both the Monte Carlo experiments and empirical applications are available at \url{https://github.com/yshin12/ls-csa}.}. We consider the following simulation design:
	\begin{align}
	y_{i} &= \bt_0 + \beta_1 Y_{i} + \eps_{i}\\
	Y_{i} &= \pi'Z_{i} + u_{i},~~i=1,\ldots,N,
	\end{align}
	where $Y_{i}$ is a scalar, $(\bt_0,\bt_1)$ is set to be (0, 0.1), $\beta_1$ is the parameter of interest, and $Z_{i}\sim \text{i.i.d. } N(0,\Sigma_{z})$. The diagonal terms of $\Sigma_{z}$ are ones and off-diagonal terms are $\rho_z$'s. Thus, $\rho_{z}$ denotes the correlation between instruments and is set to $\rho_z=0$ or $0.5$. The error terms $(\eps_{i},u_{i})$ are i.i.d.\ over $i$, bivariate normal with variances 1 and covariance $\sig_{u \eps}$. Note that $\sig_{u\eps}$ denotes the severity  of endogeneity and is set to $\sig_{u\eps}=0.1$ or $0.9$. We consider three designs for $\pi$: flat, decreasing, half-zero signals depending on the element values of the vector $\pi$. In each design, the signal-noise ratio is controlled by $R_f^2$, which is either 0.01 or 0.1. The exact formula of $\pi$ and $R_f^2$ are provided in the online supplement. 
	The number of observations and the number of instruments are set to $(N,K)=(100,20)$ and $(1000,30)$. 
	Note that the flat and decreasing designs of $\pi$ with $\rho_z=0$ coincide with those used in \citet{donald2001choosing} and \citet{kuersteiner2010constructing}.
	The simulation results are based on 400 replications.

	In addition to the CSA-2SLS (denoted by CSA hereafter) estimator based on the approximate MSE, we also estimate the model by OLS, 2SLS with full instruments, the optimal 2SLS by \citet{donald2001choosing} denoted by DN, and the model average estimator by \citet{kuersteiner2010constructing} denoted by KO and compare their performance.\footnote{Since we build up the idea of CSA based on 2SLS, we focus on the comparison with similar 2SLS type estimators in various situations. We leave it for future research to develop a CSA estimator in different classes (e.g.\ LIML or JIVE) and compare the performance with other types of estimators.}

	Tables \ref{tb:m19-24-small}--\ref{tb:m7-12-large} report the simulation results for designs with $\sig_{u \eps}=0.9$. The complete simulation results including $\sig_{u \eps}=0.1$ are collected in the Appendix. In the tables, we report the mean squared errors (MSE), bias (Bias), median absolute deviation (MAD), median bias (Median Bias), interdecile range (Range), and the coverage probability of the 95\% confidence interval (Coverage). The mean and median of $\what{k}$ are also reported for DN and CSA.

	As expected from the theory, the simulation results are quite promising for CSA when the instruments are correlated ($\rho_z=0.5$). Table \ref{tb:m19-24-small} reports the results of $(N,K)=(100,20)$. It confirms that CSA chooses smaller $k$'s and shows smaller variances. DN also shows a similar bias level but MSE is much larger. KO performs worse than CSA in terms of both bias and MSE in all designs in the table. However, MSE of KO is still comparable to CSA and is much smaller than that of DN. This indicates that the estimators based on model averaging mitigate the well-known `moments problem' caused by the large outliers with a small number of instruments, e.g. \citet{mariano1972existence}.\footnote{We thank the referee for pointing this out.}
	
	Next, we compare some robust statistics such as MAD and Median Bias. CSA shows better performance in both measures than DN except two designs (weak IV signal/decreasing $\pi_0$ and strong IV signal/flat $\pi_0$), where the losses are much smaller than the gains in other designs. Overall, KO shows smaller MAD but larger Median Bias than CSA.   
	The coverage of CSA is close to the nominal level across different specifications except weak IV signal/half-zero $\pi_0$ but that of KO is far from the nominal level especially when the instrument signal is weak.\footnote{We construct the confidence intervals by assuming the instrument/weight/subset selection in each method is correct. See Section \ref{sc:se} of the Appendix for details.} 
	
	Finally, the performance of CSA.CV, which chooses $\hat{k}$ by the 10-fold cross-validation, is slightly worse than CSA in these simulation designs. 
	In Table \ref{tb:m19-24-large}, we observe similar patterns when $(N,K)=(1000,30)$ and $R_f^2=0.01$. However, the performance of DN, KO, and CSA is quite similar in the lower panel when the signal from the instruments is strong.

	Tables \ref{tb:m7-12-small}--\ref{tb:m7-12-large} report the simulation results when the instruments are uncorrelated. As expected from the theory, the performance of CSA is quite similar to 2SLS. We partially confirm the results in \citet{donald2001choosing} with the same designs. The coverage of DN is better than other estimators that show quite similar performance but MAD of DN is much larger.\footnote{We could duplicate the MAD statistics in \citet{donald2001choosing} only by the formula $med\lt\vert \what{\bt}_{1} -\bt_{1,0} \rt\vert$, i.e. re-centered  at the true parameter value. However, we use the formula $med\lt\vert \what{\bt}_{1} -med (\what{\bt}_{1}) \rt\vert$ for MAD in this paper. See, for example, \citet{rousseeuw1993alternatives} for the definition.}

	\begin{table}
		\caption{Simulation Results\\($N=100$, $K=20$, $\sig_{u\eps}=0.9$, $\rho_z=0.5$ )}\label{tb:m19-24-small}
		\begin{center}
			\resizebox{\textwidth}{!}{
				\begin{tabular}{lcccccccc}
					\hline
					& MSE & Bias & MAD & Median Bias & Range & Coverage & Mean($\what{k}$) & Med($\what{k}$) \\
					\hline
					\multicolumn{9}{c}{\underline{$R_f^2=0.01$ (weak IV signal)}} \\
					\multicolumn{3}{l}{\underline{$\pi_0:$ flat}} \\
					OLS & 0.671 & 0.817 & 0.036 & 0.816 & 0.130 & 0.000 & NA & NA \\
					TSLS & 0.359 & 0.589 & 0.069 & 0.589 & 0.283 & 0.005 & NA & NA \\
					DN & 5.492 & -0.011 & 0.226 & 0.134 & 1.058 & 0.825 & 1.430 & 1.000 \\
					KO & 0.232 & 0.460 & 0.088 & 0.468 & 0.366 & 0.172 & NA & NA \\
					CSA.AMSE & 0.090 & 0.029 & 0.178 & 0.071 & 0.648 & 0.890 & 1.058 & 1.000 \\
					CSA.CV & 0.120 & 0.179 & 0.164 & 0.244 & 0.725 & 0.630 & 3.680 & 3.000 \\
					\multicolumn{3}{l}{\underline{$\pi_0:$ decreasing}} \\
					OLS & 0.798 & 0.892 & 0.030 & 0.893 & 0.117 & 0.000 & NA & NA \\
					TSLS & 0.744 & 0.856 & 0.071 & 0.858 & 0.263 & 0.000 & NA & NA \\
					DN & 13938.210 & -4.575 & 0.373 & 0.596 & 2.900 & 0.635 & 1.840 & 1.000 \\
					KO & 0.707 & 0.831 & 0.080 & 0.832 & 0.316 & 0.003 & NA & NA \\
					CSA.AMSE & 0.552 & 0.665 & 0.199 & 0.650 & 0.761 & 0.287 & 3.433 & 1.000 \\
					CSA.CV & 0.592 & 0.697 & 0.177 & 0.715 & 0.734 & 0.247 & 2.803 & 1.000 \\
					\multicolumn{3}{l}{\underline{$\pi_0:$ half-zero}} \\
					OLS & 0.798 & 0.892 & 0.031 & 0.893 & 0.119 & 0.000 & NA & NA \\
					TSLS & 0.746 & 0.857 & 0.072 & 0.860 & 0.270 & 0.000 & NA & NA \\
					DN & 94.560 & 0.668 & 0.341 & 0.722 & 2.424 & 0.568 & 2.292 & 1.000 \\
					KO & 0.726 & 0.843 & 0.078 & 0.848 & 0.303 & 0.003 & NA & NA \\
					CSA.AMSE & 0.591 & 0.696 & 0.196 & 0.711 & 0.779 & 0.253 & 4.107 & 1.000 \\
					CSA.CV & 0.620 & 0.718 & 0.176 & 0.719 & 0.827 & 0.207 & 2.743 & 1.000 \\
					\multicolumn{9}{c}{\underline{$R_f^2=0.1$ (strong IV signal)}} \\
					\multicolumn{3}{l}{\underline{$\pi_0:$ flat}} \\
					OLS & 0.179 & 0.420 & 0.037 & 0.417 & 0.142 & 0.000 & NA & NA \\
					TSLS & 0.021 & 0.124 & 0.052 & 0.130 & 0.195 & 0.585 & NA & NA \\
					DN & 0.011 & 0.030 & 0.062 & 0.041 & 0.259 & 0.905 & 3.752 & 4.000 \\
					KO & 0.013 & 0.075 & 0.055 & 0.082 & 0.213 & 0.772 & NA & NA \\
					CSA.AMSE & 0.010 & -0.005 & 0.063 & 0.002 & 0.240 & 0.948 & 1.038 & 1.000 \\
					CSA.CV & 0.009 & 0.036 & 0.065 & 0.037 & 0.226 & 0.892 & 6.103 & 6.000 \\
					\multicolumn{3}{l}{\underline{$\pi_0:$ decreasing}} \\
					OLS & 0.663 & 0.813 & 0.037 & 0.813 & 0.134 & 0.000 & NA & NA \\
					TSLS & 0.345 & 0.577 & 0.071 & 0.579 & 0.278 & 0.010 & NA & NA \\
					DN & 4.320 & 0.034 & 0.203 & 0.105 & 0.903 & 0.843 & 1.435 & 1.000 \\
					KO & 0.195 & 0.417 & 0.087 & 0.426 & 0.363 & 0.220 & NA & NA \\
					CSA.AMSE & 0.087 & 0.034 & 0.179 & 0.072 & 0.652 & 0.890 & 1.050 & 1.000 \\
					CSA.CV & 0.125 & 0.187 & 0.163 & 0.256 & 0.723 & 0.605 & 3.962 & 3.000 \\
					\multicolumn{3}{l}{\underline{$\pi_0:$ half-zero}} \\
					OLS & 0.666 & 0.814 & 0.037 & 0.813 & 0.138 & 0.000 & NA & NA \\
					TSLS & 0.352 & 0.582 & 0.070 & 0.588 & 0.288 & 0.007 & NA & NA \\
					DN & 3.987 & 0.028 & 0.250 & 0.179 & 1.195 & 0.785 & 1.567 & 1.000 \\
					KO & 0.258 & 0.487 & 0.087 & 0.495 & 0.356 & 0.110 & NA & NA \\
					CSA.AMSE & 0.094 & 0.048 & 0.177 & 0.085 & 0.633 & 0.873 & 1.160 & 1.000 \\
					CSA.CV & 0.131 & 0.201 & 0.171 & 0.258 & 0.709 & 0.593 & 4.018 & 3.000 \\
					\hline
				\end{tabular}
			}
		\end{center}
		\footnotesize
		\renewcommand{\baselineskip}{11pt}
		\textbf{Note:} We report mean squared errors (MSE), mean biases (Bias), median absolute deviations (MAD), median biases (Median Bias), 10-90\% ranges of the estimator (Range), coverages for the 95\% confidence interval (Coverage), means of $\what{k}$ and medians of $\what{k}$. For estimators DN, KO, and CSA.AMSE, we apply the Mallows criterion for the preliminary estimator. We set the 10-fold cross-validation method for CSA.CV.
	\end{table}

	\begin{table}[ht]
		\caption{Simulation Results\\($N=1000$, $K=30$, $\sig_{u\eps}=0.9$, $\rho_z=0.5$)} \label{tb:m19-24-large}
		\begin{center}
			\resizebox{\textwidth}{!}{
				\begin{tabular}{lcccccccc}
					\hline
					& MSE & Bias & MAD & Median Bias & Range & Coverage & Mean($\what{k}$) & Med($\what{k}$) \\
					\hline
					\multicolumn{9}{c}{\underline{$R_f^2=0.01$ (weak IV signal)}} \\
					\multicolumn{3}{l}{\underline{$\pi_0:$ flat}} \\
					OLS & 0.608 & 0.779 & 0.011 & 0.778 & 0.042 & 0.000 & NA & NA \\
					TSLS & 0.027 & 0.144 & 0.039 & 0.146 & 0.160 & 0.417 & NA & NA \\
					DN & 0.010 & 0.024 & 0.055 & 0.028 & 0.211 & 0.907 & 3.885 & 4.000 \\
					KO & 0.011 & 0.059 & 0.047 & 0.061 & 0.181 & 0.835 & NA & NA \\
					CSA.AMSE & 0.006 & -0.002 & 0.052 & 0.001 & 0.197 & 0.955 & 1.012 & 1.000 \\
					CSA.CV & 0.008 & 0.037 & 0.050 & 0.041 & 0.191 & 0.887 & 7.575 & 7.000 \\
					\multicolumn{3}{l}{\underline{$\pi_0:$ decreasing}} \\
					OLS & 0.794 & 0.891 & 0.009 & 0.891 & 0.037 & 0.000 & NA & NA \\
					TSLS & 0.468 & 0.678 & 0.060 & 0.672 & 0.236 & 0.000 & NA & NA \\
					DN & 3.709 & 0.051 & 0.207 & 0.116 & 0.970 & 0.843 & 1.415 & 1.000 \\
					KO & 0.243 & 0.473 & 0.091 & 0.471 & 0.323 & 0.113 & NA & NA \\
					CSA.AMSE & 0.074 & 0.045 & 0.162 & 0.076 & 0.603 & 0.892 & 1.157 & 1.000 \\
					CSA.CV & 0.143 & 0.229 & 0.176 & 0.297 & 0.688 & 0.540 & 4.468 & 3.000 \\
					\multicolumn{3}{l}{\underline{$\pi_0:$ half-zero}} \\
					OLS & 0.794 & 0.891 & 0.009 & 0.891 & 0.037 & 0.000 & NA & NA \\
					TSLS & 0.467 & 0.677 & 0.063 & 0.672 & 0.238 & 0.000 & NA & NA \\
					DN & 182.108 & 0.541 & 0.246 & 0.135 & 1.297 & 0.810 & 1.387 & 1.000 \\
					KO & 0.299 & 0.530 & 0.088 & 0.525 & 0.331 & 0.037 & NA & NA \\
					CSA.AMSE & 0.081 & 0.065 & 0.159 & 0.093 & 0.611 & 0.877 & 1.275 & 1.000 \\
					CSA.CV & 0.155 & 0.246 & 0.173 & 0.312 & 0.693 & 0.515 & 4.607 & 3.000 \\
					\multicolumn{9}{c}{\underline{$R_f^2=0.1$ (strong IV signal)}} \\
					\multicolumn{3}{l}{\underline{$\pi_0:$ flat}} \\
					OLS & 0.115 & 0.335 & 0.011 & 0.330 & 0.043 & 0.000 & NA & NA \\
					TSLS & 0.001 & 0.016 & 0.015 & 0.014 & 0.058 & 0.907 & NA & NA \\
					DN & 0.001 & 0.006 & 0.016 & 0.007 & 0.062 & 0.935 & 9.617 & 10.000 \\
					KO & 0.001 & 0.009 & 0.015 & 0.009 & 0.059 & 0.920 & NA & NA \\
					CSA.AMSE & 0.001 & -0.001 & 0.016 & -0.001 & 0.059 & 0.948 & 1.000 & 1.000 \\
					CSA.CV & 0.001 & 0.006 & 0.015 & 0.005 & 0.059 & 0.935 & 12.825 & 13.000 \\
					\multicolumn{3}{l}{\underline{$\pi_0:$ decreasing}} \\
					OLS & 0.657 & 0.810 & 0.010 & 0.809 & 0.040 & 0.000 & NA & NA \\
					TSLS & 0.039 & 0.184 & 0.044 & 0.191 & 0.179 & 0.290 & NA & NA \\
					DN & 0.010 & 0.022 & 0.061 & 0.028 & 0.238 & 0.917 & 3.225 & 3.000 \\
					KO & 0.011 & 0.063 & 0.054 & 0.068 & 0.216 & 0.850 & NA & NA \\
					CSA.AMSE & 0.009 & -0.004 & 0.063 & 0.003 & 0.232 & 0.960 & 1.000 & 1.000 \\
					CSA.CV & 0.012 & 0.058 & 0.061 & 0.070 & 0.218 & 0.830 & 8.398 & 8.000 \\
					\multicolumn{3}{l}{\underline{$\pi_0:$ half-zero}} \\
					OLS & 0.657 & 0.810 & 0.011 & 0.809 & 0.041 & 0.000 & NA & NA \\
					TSLS & 0.038 & 0.184 & 0.045 & 0.191 & 0.176 & 0.285 & NA & NA \\
					DN & 0.014 & 0.030 & 0.073 & 0.039 & 0.282 & 0.907 & 3.038 & 3.000 \\
					KO & 0.016 & 0.092 & 0.052 & 0.098 & 0.207 & 0.770 & NA & NA \\
					CSA.AMSE & 0.010 & -0.003 & 0.064 & 0.003 & 0.244 & 0.960 & 1.010 & 1.000 \\
					CSA.CV & 0.014 & 0.076 & 0.056 & 0.089 & 0.214 & 0.782 & 10.252 & 10.000 \\
					\hline
				\end{tabular}
			}
		\end{center}
		\footnotesize
		\renewcommand{\baselineskip}{11pt}
		\textbf{Note:} See the Note below Table \ref{tb:m19-24-small} for details.
	\end{table}

	\begin{table}[ht]
		\caption{Simulation Results \\($N=100$, $K=20$, $\sig_{u\eps}=0.9$, $\rho_z=0$)}\label{tb:m7-12-small}
		\begin{center}
			\resizebox{\textwidth}{!}{
				\begin{tabular}{lcccccccc}
					\hline
					& MSE & Bias & MAD & Median Bias & Range & Coverage & Mean($\what{k}$) & Med($\what{k}$) \\
					\hline
					\multicolumn{9}{c}{\underline{$R_f^2=0.01$ (weak IV signal)}} \\
					\multicolumn{3}{l}{\underline{$\pi_0:$ flat}} \\
					OLS & 0.799 & 0.892 & 0.032 & 0.895 & 0.119 & 0.000 & NA & NA \\
					TSLS & 0.747 & 0.857 & 0.071 & 0.854 & 0.265 & 0.000 & NA & NA \\
					DN.M & 32.443 & 1.214 & 0.260 & 0.881 & 2.010 & 0.540 & 2.312 & 1.000 \\
					KO.M & 0.747 & 0.857 & 0.076 & 0.858 & 0.289 & 0.000 & NA & NA \\
					CSA.AMSE & 0.737 & 0.850 & 0.076 & 0.849 & 0.282 & 0.000 & 5.145 & 1.000 \\
					CSA.CV & 0.736 & 0.850 & 0.077 & 0.846 & 0.299 & 0.000 & 2.895 & 1.000 \\
					\multicolumn{3}{l}{\underline{$\pi_0:$ decreasing}} \\
					OLS & 0.798 & 0.892 & 0.030 & 0.893 & 0.116 & 0.000 & NA & NA \\
					TSLS & 0.744 & 0.856 & 0.072 & 0.859 & 0.254 & 0.000 & NA & NA \\
					DN.M & 16.810 & 0.426 & 0.322 & 0.726 & 2.348 & 0.565 & 2.013 & 1.000 \\
					KO.M & 0.718 & 0.839 & 0.077 & 0.842 & 0.308 & 0.003 & NA & NA \\
					CSA.AMSE & 0.732 & 0.848 & 0.069 & 0.855 & 0.270 & 0.003 & 3.730 & 1.000 \\
					CSA.CV & 0.729 & 0.846 & 0.068 & 0.850 & 0.286 & 0.003 & 2.902 & 1.000 \\
					\multicolumn{3}{l}{\underline{$\pi_0:$ half-zero}} \\
					OLS & 0.797 & 0.892 & 0.033 & 0.892 & 0.125 & 0.000 & NA & NA \\
					TSLS & 0.743 & 0.855 & 0.074 & 0.855 & 0.281 & 0.000 & NA & NA \\
					DN.M & 3.517 & 0.887 & 0.275 & 0.899 & 1.796 & 0.542 & 2.380 & 1.000 \\
					KO.M & 0.747 & 0.856 & 0.076 & 0.857 & 0.303 & 0.000 & NA & NA \\
					CSA.AMSE & 0.735 & 0.849 & 0.086 & 0.844 & 0.300 & 0.000 & 5.065 & 1.000 \\
					CSA.CV & 0.733 & 0.847 & 0.083 & 0.844 & 0.297 & 0.000 & 2.910 & 1.000 \\
					\multicolumn{9}{c}{\underline{$R_f^2=0.1$ (strong IV signal)}} \\
					\multicolumn{3}{l}{\underline{$\pi_0:$ flat}} \\
					OLS & 0.665 & 0.814 & 0.033 & 0.812 & 0.137 & 0.000 & NA & NA \\
					TSLS & 0.349 & 0.580 & 0.075 & 0.582 & 0.283 & 0.007 & NA & NA \\
					DN.M & 44.540 & 0.361 & 0.269 & 0.571 & 2.280 & 0.642 & 2.455 & 1.000 \\
					KO.M & 0.344 & 0.573 & 0.075 & 0.577 & 0.312 & 0.020 & NA & NA \\
					CSA.AMSE & 0.316 & 0.546 & 0.081 & 0.550 & 0.354 & 0.030 & 3.155 & 1.000 \\
					CSA.CV & 0.319 & 0.549 & 0.082 & 0.555 & 0.345 & 0.030 & 5.938 & 6.000 \\
					\multicolumn{3}{l}{\underline{$\pi_0:$ decreasing}} \\
					OLS & 0.664 & 0.813 & 0.035 & 0.813 & 0.137 & 0.000 & NA & NA \\
					TSLS & 0.347 & 0.579 & 0.071 & 0.576 & 0.269 & 0.005 & NA & NA \\
					DN.M & 3.424 & 0.009 & 0.196 & 0.185 & 1.198 & 0.787 & 1.590 & 1.000 \\
					KO.M & 0.231 & 0.460 & 0.084 & 0.461 & 0.333 & 0.175 & NA & NA \\
					CSA.AMSE & 0.310 & 0.541 & 0.085 & 0.542 & 0.319 & 0.045 & 1.127 & 1.000 \\
					CSA.CV & 0.318 & 0.548 & 0.079 & 0.554 & 0.312 & 0.043 & 5.978 & 6.000 \\
					\multicolumn{3}{l}{\underline{$\pi_0:$ half-zero}} \\
					OLS & 0.663 & 0.812 & 0.037 & 0.812 & 0.135 & 0.000 & NA & NA \\
					TSLS & 0.349 & 0.578 & 0.081 & 0.571 & 0.302 & 0.018 & NA & NA \\
					DN.M & 5932.040 & 4.732 & 0.309 & 0.733 & 1.790 & 0.605 & 2.623 & 1.000 \\
					KO.M & 0.337 & 0.566 & 0.083 & 0.570 & 0.330 & 0.028 & NA & NA \\
					CSA.AMSE & 0.315 & 0.545 & 0.092 & 0.541 & 0.337 & 0.040 & 3.623 & 1.000 \\
					CSA.CV & 0.318 & 0.547 & 0.093 & 0.549 & 0.346 & 0.045 & 6.110 & 6.000 \\
					\hline
				\end{tabular}
			}
		\end{center}
		\footnotesize
		\renewcommand{\baselineskip}{11pt}
		\textbf{Note:} See the Note below Table \ref{tb:m19-24-small} for details.
	\end{table}

	\begin{table}[ht]
		\caption{Simulation Results \\($N=1000$, $K=30$, $\sig_{u\eps}=0.9$, $\rho_z=0$)}\label{tb:m7-12-large}
		
		\begin{center}
			\resizebox{\textwidth}{!}{
				
				\begin{tabular}{lcccccccc}
					\hline
					& MSE & Bias & MAD & Median Bias & Range & Coverage & Mean($\what{k}$) & Med($\what{k}$) \\
					\hline
					\multicolumn{9}{c}{\underline{$R_f^2=0.01$ (weak IV signal)}} \\
					\multicolumn{3}{l}{\underline{$\pi_0:$ flat}} \\
					OLS & 0.793 & 0.891 & 0.009 & 0.891 & 0.037 & 0.000 & NA & NA \\
					TSLS & 0.462 & 0.673 & 0.062 & 0.678 & 0.235 & 0.003 & NA & NA \\
					DN.M & 496.283 & -0.472 & 0.348 & 0.643 & 2.155 & 0.620 & 2.303 & 1.000 \\
					KO.M & 0.455 & 0.667 & 0.067 & 0.671 & 0.249 & 0.005 & NA & NA \\
					CSA.AMSE & 0.457 & 0.669 & 0.065 & 0.670 & 0.242 & 0.003 & 5.878 & 1.000 \\
					CSA.CV & 0.455 & 0.667 & 0.066 & 0.670 & 0.251 & 0.003 & 7.050 & 7.000 \\
					\multicolumn{3}{l}{\underline{$\pi_0:$ decreasing}} \\
					OLS & 0.794 & 0.891 & 0.009 & 0.891 & 0.037 & 0.000 & NA & NA \\
					TSLS & 0.467 & 0.677 & 0.064 & 0.678 & 0.236 & 0.000 & NA & NA \\
					DN.M & 42.391 & 0.203 & 0.245 & 0.210 & 1.461 & 0.780 & 1.528 & 1.000 \\
					KO.M & 0.305 & 0.536 & 0.088 & 0.529 & 0.331 & 0.040 & NA & NA \\
					CSA.AMSE & 0.460 & 0.671 & 0.065 & 0.673 & 0.246 & 0.000 & 1.292 & 1.000 \\
					CSA.CV & 0.462 & 0.673 & 0.067 & 0.674 & 0.248 & 0.000 & 7.032 & 7.000 \\
					\multicolumn{3}{l}{\underline{$\pi_0:$ half-zero}} \\
					OLS & 0.794 & 0.891 & 0.010 & 0.891 & 0.037 & 0.000 & NA & NA \\
					TSLS & 0.464 & 0.675 & 0.062 & 0.673 & 0.222 & 0.000 & NA & NA \\
					DN.M & 5.060 & 0.883 & 0.297 & 0.843 & 1.897 & 0.575 & 2.965 & 1.000 \\
					KO.M & 0.456 & 0.667 & 0.069 & 0.667 & 0.257 & 0.003 & NA & NA \\
					CSA.AMSE & 0.459 & 0.671 & 0.063 & 0.670 & 0.231 & 0.000 & 8.432 & 1.000 \\
					CSA.CV & 0.461 & 0.672 & 0.066 & 0.672 & 0.246 & 0.000 & 7.000 & 7.000 \\
					\multicolumn{9}{c}{\underline{$R_f^2=0.1$ (strong IV signal)}} \\
					\multicolumn{3}{l}{\underline{$\pi_0:$ flat}} \\
					OLS & 0.655 & 0.809 & 0.011 & 0.809 & 0.041 & 0.000 & NA & NA \\
					TSLS & 0.039 & 0.188 & 0.047 & 0.186 & 0.166 & 0.268 & NA & NA \\
					DN.M & 4.709 & -0.107 & 0.143 & 0.218 & 0.972 & 0.762 & 2.185 & 2.000 \\
					KO.M & 0.048 & 0.208 & 0.049 & 0.210 & 0.181 & 0.237 & NA & NA \\
					CSA.AMSE & 0.038 & 0.183 & 0.048 & 0.180 & 0.167 & 0.315 & 1.000 & 1.000 \\
					CSA.CV & 0.039 & 0.187 & 0.048 & 0.184 & 0.167 & 0.282 & 22.767 & 23.000 \\
					\multicolumn{3}{l}{\underline{$\pi_0:$ decreasing}} \\
					OLS & 0.656 & 0.810 & 0.011 & 0.809 & 0.040 & 0.000 & NA & NA \\
					TSLS & 0.039 & 0.184 & 0.046 & 0.193 & 0.176 & 0.295 & NA & NA \\
					DN.M & 0.012 & 0.044 & 0.064 & 0.054 & 0.237 & 0.877 & 4.730 & 5.000 \\
					KO.M & 0.013 & 0.075 & 0.056 & 0.084 & 0.207 & 0.807 & NA & NA \\
					CSA.AMSE & 0.037 & 0.179 & 0.047 & 0.186 & 0.178 & 0.338 & 1.000 & 1.000 \\
					CSA.CV & 0.038 & 0.182 & 0.047 & 0.192 & 0.179 & 0.307 & 22.650 & 23.000 \\
					\multicolumn{3}{l}{\underline{$\pi_0:$ half-zero}} \\
					OLS & 0.656 & 0.810 & 0.011 & 0.810 & 0.042 & 0.000 & NA & NA \\
					TSLS & 0.038 & 0.184 & 0.050 & 0.184 & 0.179 & 0.295 & NA & NA \\
					DN.M & 15609.030 & 7.069 & 0.409 & 0.766 & 2.577 & 0.782 & 1.073 & 1.000 \\
					KO.M & 0.026 & 0.145 & 0.049 & 0.148 & 0.190 & 0.522 & NA & NA \\
					CSA.AMSE & 0.037 & 0.180 & 0.051 & 0.182 & 0.185 & 0.318 & 1.000 & 1.000 \\
					CSA.CV & 0.038 & 0.183 & 0.049 & 0.182 & 0.180 & 0.297 & 22.688 & 23.000 \\
					
					\hline
				\end{tabular}
			}
		\end{center}
		\footnotesize
		\renewcommand{\baselineskip}{11pt}
		\textbf{Note:} See the Note below Table \ref{tb:m19-24-small} for details.
	\end{table}

	

	We finish this section by a remark on the computation. When we compute the CSA-P matrix, we use 100 random draws from complete subsets when the complete subset size is bigger than 100. Table \ref{tb:m19-small-R} in the Appendix reports simulation results with different sizes of random draws in one design, and it shows that CSA performs satisfactorily with a relatively small number of draws.

	\section{Empirical Illustration}\label{sc:empirical}
	In this section we illustrate the usefulness of the CSA-2SLS estimator by estimating a logistic demand function for automobiles in \citet*{berry1995automobile}. The model specification is
	\eqs{
		\log(S_{it})-\log(S_{0t}) &= \ap_0 P_{it} + X_{it}'\bt_0 + \eps_{it},\\
		P_{it} & = Z_{it}'\dt_0 +X_{it}'\gm_0 + u_{it},
	}where $S_{it}$ is the market share of the product $i$ in market $t$ with product 0 denoting the outside
	option, $P_{it}$ is the endogenous price variable, $X_{it}$ is a vector of included exogenous variables, and $Z_{it}$ is a vector of instruments. The parameter of interest is $\ap_0$ from which we can calculate the price elasticity of demand.
	
	\begin{table}[ht]
		\begin{center}
			\caption{Logistic Demand Function for Automobiles}\label{tb:empirical}
			\begin{tabular}{p{2cm}ccccccc}
				\hline
				& \multirow{2}{*}{$\hat{k}$}& \multirow{2}{*}{$\what{\ap}$} &\# of Inelastic & & \multirow{2}{*}{$\hat{k}$}  & \multirow{2}{*}{$\what{\ap}$} & \# of Inelastic   \\
				&                              &                               & Demand        & &                                 &                              & Demand \\
				\hline
				\noalign{\vskip 2mm}
				&  \multicolumn{3}{l}{\underline{Original Design}} & & \multicolumn{3}{l}{\underline{Extended Design}}\\
				\noalign{\vskip 2mm}
				OLS & -- & -0.0886 & 1502 &  & -- & -0.0991 & 1405 \\
				&    & (0.0114) &  &    &    & (0.0124) & \\
				2SLS &  -- & -0.1357 & 746 & & -- & -0.1273 & 874 \\
				&     & (0.0464) & &  &    & (0.0246) &  \\
				DN & 10 & -0.1357 & 746 & & 47 & -0.1271 & 876 \\
				&   &(0.0464) & & &    &(0.0245) & \\
				KO & -- &-0.1357 &  746 & & -- &-0.1273 &  874 \\
				&    & (0.0464) & &      &    & (0.0246) & \\
				CSA.AMSE & 9 & -0.1426 & 659 & & 1 & -0.2515  & 7 \\
				&   & (0.0491)  & &        &   & (0.0871) & \\
				CSA.CV & 5 & -0.1918 & 107 & & 2 & -0.2526  & 7 \\
				&   & (0.0634)  & &        &   & (0.0818) & \\
				\noalign{\vskip 2mm}
				\hline
			\end{tabular}
		\end{center}
		\footnotesize
		\renewcommand{\baselineskip}{11pt}
		\textbf{Note:} The original design uses 5 regressors and 10 instruments and the extended design does 24 regressors and 48 instruments. The sample size is 2,217. The heteroskedasticity and cluster robust standard errors of $\what{\ap}$ are provided inside parentheses.
	\end{table}
	
	We first estimate the model by using the same set of regressors and instruments used by \citet*{berry1995automobile}. The vector of covariates $X_{it}$ includes 5 variables: a constant, an air conditioning dummy, horsepower divided by weight, miles per dollar, and vehicle size. The vector of original instruments $Z_{it}$ includes 10 variables and is constructed by the characteristics of other car models. We also consider an extended design by adopting 48 instruments and 24 regressors constructed by \citet*{chernozhukov2015post}. We presume that all instruments are valid and relevant. Based on the previous simulation results, we adopt two methods for choosing $k$ for CSA: (a) the approximate MSE (AMSE) method and (b) the cross-validation(CV) method.
	We use the Mallows criterion for selecting IVs in AMSE and apply the 10-fold cross-validation for CV.\
	For both estimators, we set $R=100$ when calculating the CSA-P matrix.
	
	Table \ref{tb:empirical} summarizes the estimation results. In addition to two CSA estimates, we also estimate the model using OLS, 2SLS, DN, and KO for comparison.
	For each design we report the choice of $k$ in DN and CSAs, the estimate of $\ap$, and the heteroskedasticity and cluster robust standard errors of $\ap$ given that we have chosen the correct model for $k$ or the optimal weight of KO. Finally, we report the number of products whose price elasticity of demand is inelastic, which is computed by the following formula:
	\eqs{
		\sum_{i,t} 1\{\vert \what{\ap} \times  P_{it} \times (1-S_{it}) \vert < 1\},
	}where $1\{A\}$ is an indicator function equal to 1 if $A$ holds.
	
	It is interesting to note that the estimation results of CSA contrast sharply with those of other estimation methods in the extended design.
	Recall that the economic theory predicts elastic demand in this market.
	Similar to the result in \citet*{berry1995automobile}, the OLS estimate is biased towards zero and makes 1,405 out of 2,217 products (63\%)  have inelastic own price demand. The 2SLS estimator mitigates the bias but not enough. Still, 874 (41\%) products show inelastic own price elasticity. It is interesting to note that DN and KO are not particularly better than 2SLS in this empirical example although they are supposed to correct the bias caused by many instruments. The estimation result of DN comes close to 2SLS by choosing most instruments, 47 out of 48, and that of KO coincides exactly with 2SLS by putting the whole weight to the largest set of instruments. On the contrary, only 7 products (0.3\%) have inelastic demand according to both estimation results of CSA. The $\ap$ estimates by CSA is about twice as large as those by other estimators in absolute term. Since the bias caused by many instruments is towards the OLS estimates, this result can be viewed as a correction for the many instrument bias. However, the standard error of CSA is larger than others and there is a potential trade-off between the bias and the variance. Finally, the original design has fewer instruments and the bias correction by CSAs is not as large as in the extended design. However, CSA.CV shows a larger bias correction and its robustness to heteroskedasticity could be a potential reason.
	
	In sum, 2SLS with all the available instruments suffers from many instruments bias in this application. DN and KO do not correct the bias enough to make the estimation results consistent with the prediction by the economic theory. In contrast, the CSA point estimates reduce the bias substantially in the extended design. Therefore, it is worthwhile to estimate a model with CSA and compare it with other existing methods when the model contains many instruments.


	
 \appendix
\section*{{\LARGE{Appendices}}}

\section{Proofs and Lemmas}

Let $e_{f}^{k} = f'(I-P^{k})^{2}f/N$, $\xi_{f}^{k} = f'(I-P^{k})f/N$, and $\Delta_{k} = \text{tr}(e_{f}^{k})$. In addition, for $\widehat{P}^{k}$ and $\widetilde{P}^{k}$ defined in Section \ref{sc:AMSE-diff} of the main text, let $\widehat{e}_{f}^{k} = f'(I-\widehat{P}^{k})^{2}f/N$, $\widehat{\xi}_{f}^{k} = f'(I-\widehat{P}^{k})f/N$, $\widehat{\Delta}_{k} = \text{tr}(\widehat{e}_{f}^{k})$, $\widetilde{\zeta}_{f}^{k} =f'\widetilde{P}^{k}f/N$, and $\widetilde{\Delta}_{k} = \text{tr}(\widetilde{\zeta}_{f}^{k})$.

\subsection{Proof of Theorem \ref{Thm1}}
The complete subset averaging 2SLS estimator is
\begin{equation}
\notag \sqrt{N}\left(\widehat{\beta}-\beta\right) = \left( \frac{X'P^{k}X}{N}\right)^{-1}\frac{X'P^{k}\eps}{\sqrt{N}}.
\end{equation}
By expanding
\begin{eqnarray}
\label{expan1} \frac{X'P^{k}X}{N} &=& \frac{f'f}{N} -\frac{f'(I-P^{k})f}{N} + \frac{u'f + f'u}{N} + \frac{u'P^{k}u }{N} -\frac{u'(I-P^{k})f + f'(I-P^{k})u}{N} \\
\label{expan2} \frac{X'P^{k}\eps}{\sqrt{N}} &=& \frac{f'\eps}{\sqrt{N}} -\frac{f'(I-P^{k})\eps}{\sqrt{N}} + \frac{u'P^{k}\eps}{\sqrt{N}}
\end{eqnarray}
we can write
\begin{equation}
\sqrt{N}\left(\widehat{\beta}-\beta\right) = \widehat{H}^{-1}\widehat{h},
\end{equation}
where
\begin{eqnarray}
\nonumber \widehat{H} &=& H  +T_{1}^{H} + T_{2}^{H} + Z^{H},\\
\nonumber H &=& \frac{f'f}{N},\\
\nonumber T_{1}^{H} &=& -\frac{f'(I-{P}^{k})f}{N},\\
\nonumber T_{2}^{H} &=&  \frac{u'f + f'u}{N},\\
\nonumber Z^{H} &=& -\frac{u'(I-{P}^{k})f + f'(I-{P}^{k})u}{N}+\frac{u'P^{k}u }{N},
\end{eqnarray}
and
\begin{eqnarray}
\nonumber \widehat{h} &=& h + T_{1}^{h} + T_{2}^{h}\\
\nonumber h &=& \frac{f'\eps}{\sqrt{N}},\\
\nonumber T_{1}^{h} &=&  -\frac{f'(I-{P}^{k})\eps}{\sqrt{N}},\\
\nonumber T_{2}^{h} &=& \frac{u'P^{k}\eps}{\sqrt{N}}.
\end{eqnarray}

Based on Lemmas \ref{L_P}-\ref{L2}, we specify the convergence rate of each term. By Lemma \ref{L2}(ii), $T_{1}^{H} = O_{p}(\Delta_{k}^{1/2})$. By CLT, $T_{2}^{H} = O_{p}(1/\sqrt{N})$. By Lemmas \ref{L2}(iii)-(v),
\begin{equation}
Z^{H} =  O_{p}\left(\frac{\Delta_{k}^{1/2}}{\sqrt{N}}\right) + O_{p}\left(\frac{k}{N}\right)=o_p\left(\frac{k^2}{N}+\Delta_k\right).
\end{equation}
By Lemma \ref{L2}(ii), $T_{1}^{h}=O_{p}(\Delta_{k}^{1/2})$. By Lemma \ref{L2}(iv), $T_{2}^{h} = O_{p}(k/\sqrt{N})$. By WLLN, $H\xrightarrow{p}Ef_{i}f_{i}'$ and thus $H=O_{p}(1)$. By CLT, $h=O_{p}(1)$. By Assumption \ref{A1}(v) and the continuous mapping theorem, $\widehat{H}^{-1} = O_{p}(1)$. Let $T^{h} = T_{1}^{h} + T_{2}^{h}$ and $T^{H} = T_{1}^{H} + T_{2}^{H}$.

We show that
\begin{equation}
S(k) = H^{-1}\left[\frac{k^{2}}{N}\sigma_{u\eps}\sigma_{u\eps}' + \sigma_{\eps}^{2}e_{f}^{k} - \sigma_{\eps}^{2}\xi_{f}^{k}H^{-1}\xi_{f}^{k}\right]H^{-1}.
\end{equation}
The leading orders are $k^{2}/N$ and $\Delta_{k}$ by Lemma \ref{L2}.
The proof proceed by showing that $\widehat{r}(k) +T(k) = o_{p}(\rho_{k,N})$ as $k,N\rightarrow\infty$ where $\rho_{k,N}$ is the \textit{lower} order (the slower) between $k^{2}/N$ and $\Delta_{k}$.

Our expansion is a non-trivial extension of \citet{donald2001choosing} and \citet{kuersteiner2010constructing}
because we need to specify additional terms that are supposed to be small in those papers. This is due to the fact that our $P^{k}$ matrix not being idempotent. In particular, Lemma 1 of \citet{donald2001choosing} cannot be applied because $\|T^{H}\|\cdot \|T^{h}\|$ is not $\opr$. In addition, Lemma A.1 of \citet{kuersteiner2010constructing} cannot be applied because $\|T^{H}\|^{2}$ is not $\opr$.

We use the following expansion. Let $\widehat{H} = \widetilde{H} + Z^{H}$ and $\widetilde{H}  = H + T^{H}$. Using
\begin{eqnarray}
\label{Hhat1}
\widehat{H}^{-1} &=& \widetilde{H}^{-1} - \widehat{H}^{-1}(\widehat{H}-\widetilde{H})\widetilde{H}^{-1}
\end{eqnarray}
we can write
\begin{equation}
\label{eq1}
\widehat{H}^{-1}\widehat{h} = \widetilde{H}^{-1}\widehat{h} + o_{p}(\rho_{k,N})
\end{equation}
because $Z^{H} = o_{p}(\rho_{k,N})$, $\widehat{H}^{-1}=O_{p}(1)$, and $\widetilde{H}^{-1}=O_{p}(1)$. Furthermore, using
\begin{eqnarray}
\widetilde{H}^{-1} &=& H^{-1} - \widetilde{H}^{-1}(\widetilde{H}-H)H^{-1},
\end{eqnarray}
and that $\|T_{2}^{H}\|^{2}= O_{p}(1/N) = o_{p}(\rho_{k,N})$, $\|T_{1}^{H}\|\cdot \|T_{2}^{H}\| = O_p ( (\Delta_k/N)^{1/2} ) = o_{p}(\rho_{k,N})$ by Lemma \ref{L2}(v), and $\|T_{1}^{H}\|^{3} = O_{p}(\Delta_{k}^{3/2}) = o_{p}(\Delta_{k})$ by Lemma \ref{L2}(i), \eqref{eq1} can be further expanded as
\begin{equation}
\label{eq2}
\widehat{H}^{-1}\widehat{h} = H^{-1}\widehat{h} - H^{-1}T^{H}H^{-1}\widehat{h} + H^{-1}T^{H}_{1}H^{-1}T^{H}_{1}H^{-1}\widehat{h}+o_{p}(\rho_{k,N}).
\end{equation}
Thus,
\begin{eqnarray}
\label{exp1} \widehat{H}^{-1}\widehat{h}\widehat{h}'\widehat{H}^{-1} &=& H^{-1}\widehat{h}\widehat{h}'H^{-1}\\
\label{exp2} &&- H^{-1}\what{h}\what{h}'H^{-1}{T^{H}}'H^{-1} - H^{-1}T^H H^{-1} \what{h}\what{h}' H^{-1} \label{eq:expHhhH}\\ 
\label{exp3} && + H^{-1}\what{h}\what{h}'H^{-1}{T_1^{H}}'H^{-1}{T_1^H}'H^{-1}+ H^{-1}T_1^H H^{-1} T_1^H H^{-1} \what{h} \what{h}'H^{-1} \\
\label{exp4} &&  +  H^{-1}T^H_{1} H^{-1} \what{h} \what{h}' H^{-1}{T^H_{1}}' H^{-1} \\ 
\notag &&  + \opr.
\end{eqnarray}
The higher-order terms in the MSE are obtained by taking the conditional expectation on both sides of the above expansion.

First, take \eqref{exp1}. We derive the conditional expectation of $\what{h}\what{h}' = hh' + hT^{h'} + T_{h}h' + T^{h}T^{h'}$. By Lemma \ref{L2},
\begin{eqnarray}
\notag E[hh'|z] &=& E\left[\frac{f'\eps\eps'f}{N}|z\right] = \sigma_{\eps}^{2}H,\\
\notag E[T_{1}^{h}T_{1}^{h'}|z] &=& \sigma_{\eps}^{2}\frac{f'(I-{P}^{k})^{2}f}{N} = \sigma_{\eps}^{2}e_{f}^{k},\\
\notag E[T_{2}^{h}T_{2}^{h'}|z] &=& \sigma_{u\eps}\sigma_{u\eps}' \frac{k^{2}}{N} + O_{p}\left(\frac{k}{N}\right)= \sigma_{u\eps}\sigma_{u\eps}' \frac{k^{2}}{N} + o_{p}\left(\frac{k^2}{N}\right) = \sigma_{u\eps}\sigma_{u\eps}' \frac{k^{2}}{N} + \opr,\\
\notag E[hT_{1}^{h'}|z] &=& -\sigma_{\eps}^{2}\frac{f'(I-{P}^{k})f}{N} = -\sigma_{\eps}^{2}\xi_{f}^{k},\\
\notag E[hT_{2}^{h'}|z] &=& E\left[\frac{f'\eps\eps'P^{k}u}{N}|z\right] = O_{p}\left(\frac{k}{N}\right)=o_p\lt(\frac{k^2}{N}\rt)=\opr,\\
\notag \label{h1h2} E[T_{1}^{h}T_{2}^{h'}|z] &=& -E\left[\frac{f'(I-P^{k})\eps\eps'P^{k}u}{N}|z\right] = o_{p}\left(\sqrt{k}\frac{ \Delta_{k}^{1/2}}{\sqrt{N}}\right) = o_{p}\left(\frac{k}{N} + \Delta_{k}\right)=\opr.
\end{eqnarray}
In the last equation of the above display, the third equality holds by the inequality $\sqrt{xy}\le 2^{-1}(x+y)$ with $x=k/N$ and $y=\Delta_{k}$. Thus, we have
\begin{equation}
\label{Ehh}
E[\what{h}\what{h}'|z] = \sigma_{\eps}^{2}H + \sigma_{\eps}^{2}e_{f}^{k} +  \sigma_{u\eps}\sigma_{u\eps}' \frac{k^{2}}{N} - 2\sigma_{\eps}^{2}\xi_{f}^{k} + o_{p}(\rho_{k,N}).
\end{equation}

Next, take \eqref{exp2}. Since $\|T_{1}^{h}\|\cdot \|T_{2}^{H}\| = \opr$, $\|T_{2}^{h}\|\cdot \|T_{2}^{H}\| = \opr$, $\|T^{h}\|^{2}\cdot \|T^{H}\| = \opr$,
\begin{eqnarray}
\label{hhHT}
\notag \what{h}\what{h}'H^{-1}T^{H'} &=& hh'H^{-1}T_{1}^{H'} + hh'H^{-1}T_{2}^{H'} + hT_{1}^{h'}H^{-1}T_{1}^{H'} + hT_{2}^{h'}H^{-1}T_{1}^{H'}\\
&& + T_{1}^{h}h'H^{-1}T_{1}^{H'} + T_{2}^{h}h'H^{-1}T_{1}^{H'} + \opr.
\end{eqnarray}
By Lemma \ref{L2},
\begin{eqnarray}
\notag E[hh'H^{-1}T_{1}^{H'}|z] &=& -E\left[\frac{f'\eps\eps'f}{N}H^{-1}\xi_{f}^{k}|z\right] = -\sigma_{\eps}^{2}\xi_{f}^{k},\\
\notag E[hh'H^{-1}T_{2}^{H'}|z] &=&E[hh'H^{-1}u'f/N|z] + E[hh'H^{-1}f'u/N|z] = O_{p}\left(1/N\right)= o_{p}(\rho_{k,N}),\\
\notag E[hT_{1}^{h'}H^{-1}T_{1}^{H'}|z] &=& E\left[\frac{f'\eps\eps'(I-{P}^{k})f}{N}H^{-1}\xi_{f}^{k}|z\right] = \sigma_{\eps}^{2}\xi_{f}^{k}H^{-1}\xi_{f}^{k},\\
\notag E[T_{1}^{h}h'H^{-1}T_{1}^{H'}|z] &=&E\left[\frac{f'(I-{P}^{k})\eps\eps'f}{N}H^{-1}\xi_{f}^{k}|z\right] = \sigma_{\eps}^{2}\xi_{f}^{k}H^{-1}\xi_{f}^{k},\\
\notag E[hT_{2}^{h'}H^{-1}T_{1}^{H'}|z] &=& E\left[\frac{f'\eps\eps'P^{k}u}{N}H^{-1}\xi_{f}^{k}|z\right] = O_{p}(\Delta_{k}^{1/2}k/N) = \opr,\\
\notag E[T_{2}^{h}h'H^{-1}T_{1}^{H'}|z] &=& E\left[\frac{u'P^{k}\eps f'\eps}{N}H^{-1}\xi_{f}^{k}|z\right] = \opr.
\end{eqnarray}
Thus,
\begin{equation}
E[\what{h}\what{h}'H^{-1}T^{H'}|z] = -\sigma_{\eps}^{2}\xi_{f}^{k} + 2\sigma_{\eps}^{2}\xi_{f}^{k}H^{-1}\xi_{f}^{k} +\opr.
\end{equation}
By symmetry,
\begin{equation}
E[T^{H}H^{-1}\what{h}\what{h}'|z] = -\sigma_{\eps}^{2}\xi_{f}^{k} + 2\sigma_{\eps}^{2}\xi_{f}^{k}H^{-1}\xi_{f}^{k} +\opr.
\end{equation}

Finally, take \eqref{exp3} and \eqref{exp4}. Since $\|T_{1}^{H}\|^{2} = O_{p}(\Delta_{k})$,
\begin{eqnarray}
\what{h}\what{h}'H^{-1}{T_1^{H}}'H^{-1}{T_1^H}' &=& hh'H^{-1}{T_1^{H}}'H^{-1}{T_1^H}' + \opr,\\
T^H_{1} H^{-1} \what{h} \what{h}' H^{-1}{T^H_{1}}' &=& T^H_{1} H^{-1} hh' H^{-1}{T^H_{1}}' + \opr.
\end{eqnarray}
Thus,
\begin{eqnarray}
E[hh'H^{-1}{T_1^{H}}'H^{-1}{T_1^H}' |z] &=& E\left[\frac{f'\eps\eps'f}{N}H^{-1}\xi_{f}^{k}H^{-1}\xi_{f}^{k}|z\right] = \sigma_{\eps}^{2}\xi_{f}^{k}H^{-1}\xi_{f}^{k},\\
E[T_1^H H^{-1} T_1^H H^{-1} hh' |z] &=& \sigma_{\eps}^{2}\xi_{f}^{k}H^{-1}\xi_{f}^{k},\\
E[T^H_{1} H^{-1} hh' H^{-1}{T^H_{1}}'|z] &=&\sigma_{\eps}^{2}\xi_{f}^{k}H^{-1}\xi_{f}^{k}.
\end{eqnarray}

Combining the results together, we get
\begin{eqnarray}
\widehat{H}^{-1}\widehat{h}\widehat{h}'\widehat{H}^{-1} &=& H^{-1}\left(\sigma_{\eps}^{2}H + \sigma_{\eps}^{2}e_{f}^{k} +  \sigma_{u\eps}\sigma_{u\eps}'\frac{k^{2}}{N} - 2\sigma_{\eps}^{2}\xi_{f}^{k}+2\sigma_{\eps}^{2}\xi_{f}^{k}   \right.\\
\notag && \left. - 4\sigma_{\eps}^{2}\xi_{f}^{k}H^{-1}\xi_{f}^{k} + 3\sigma_{\eps}^{2}\xi_{f}^{k}H^{-1}\xi_{f}^{k} + o_{p}(\rho_{k,N})\right)H^{-1}\\
\notag &=&  H^{-1}\left(\sigma_{\eps}^{2}H +  \sigma_{u\eps}\sigma_{u\eps}'\frac{k^{2}}{N} +\sigma_{\eps}^{2}e_{f}^{k}- \sigma_{\eps}^{2}\xi_{f}^{k}H^{-1}\xi_{f}^{k}\right)H^{-1}+ o_{p}(\rho_{k,N}).
\end{eqnarray}
The desired result is established by noting that
\begin{eqnarray}
e_{f}^{k} - \xi_{f}^{k}H^{-1}\xi_{f}^{k} &=& \frac{f'(I-P^{k})^{2}f}{N} - \frac{f'(I-P^{k})f}{N}\left(\frac{f'f}{N}\right)^{-1}\frac{f'(I-P^{k})f}{N} \\
&=& \frac{f'(I-P^{k})(I-P_{f})(I-P^{k})f}{N}
\end{eqnarray}
where $P_{f} = f(f'f)^{-1}f'$. \qed

\subsection{Proof of Theorem \ref{Thm-optimality}}
In the following proof, let $0<C<\infty$ be a generic constant.

Recall that
\begin{align*}
S_{\ld}(k) &= \ld'H^{-1}\left[\sigma_{u\eps}\sigma_{u\eps}'\frac{k^{2}}{N} + \sigma_{\eps}^{2}\frac{f'(I-P^{k})(I-P_{f})(I-P^{k})f}{N}\right]H^{-1}\ld \\
& = \sigma_{\ld \eps}^2\frac{k^{2}}{N} + \sigma_{\eps}^{2}  \left[ \ld' H^{-1} e_f^{k} H^{-1}\ld - \ld' H^{-1} \xi_f^{k} H^{-1} \xi_f^{k} H^{-1}\ld \right],
\end{align*}
where 
$\sigma_{\ld\eps} = \ld'H^{-1}\sg_{u\eps}$.
The feasible criterion is obtained by replacing the unknown population quantities $\sg_{\ld \eps}, \sg_{\eps}$, $\ld$, and $f$ with their sample counterparts, which come from the preliminary estimates. Note that these preliminary estimators do not depend on $k$. Then, we can rewrite $\widehat{S}_{\ld}(k)$ as
\begin{align*}
\widehat{S}_{\ld}(k)
& = \wtd{\sigma}_{\ld \eps}^2\frac{k^{2}}{N} + \wtd{\sigma}_{\eps}^{2}  \left[ \wtd{\ld}' \wtd{H}^{-1} \wtd{e}_f^{k} \wtd{H}^{-1}\wtd{\ld} - \wtd{\ld}' \wtd{H}^{-1} \wtd{\xi}_f^{k} \wtd{H}^{-1} \wtd{\xi}_f^{k} \wtd{H}^{-1}\wtd{\ld} \right]
\end{align*}
where
\begin{align*}
\wtd{e}_f^k    & = \frac{X'(I-P^k)^2 X}{N}+\wtd{\Sigma}_u\lt(\frac{2k-\trace((P^k)^2)}{N}\rt), \\
\wtd{\xi}_f^k & =  \frac{X'(I-P^k) X}{N} + \wtd{\Sigma}_u \frac{k}{N} - \wtd{\Sigma}_u,\\
\widetilde{\sigma}^{2}_{\ld\eps} &= (\widetilde{\ld}'\widetilde{H}^{-1}\widetilde{\sg}_{u\eps})^{2},\\
\wtd{\Sigma}_u &=\frac{\wtd{u}'\wtd{u}}{N}.
\end{align*}
Note that if we use the above expression, an additional term $\tlh \wtd{\Sg}_u \wtd{H}^{-1} \wtd{\Sg}_u \thl$ will be added to $\what{S}_{\ld}(k)$ defined in the main text. However, the additional term does not depend on $k$ and is irrelevant to get $\what{k}$. By Lemma A.9 in \citet{donald2001choosing}, it suffices to show
\begin{equation}
\sup_{k\in\mathcal{K}_N} \frac{\lt\vert \what{S}_{\ld}(k) - S_{\ld}(k) \rt\vert}{S_{\ld}(k)} \parrow 0. \label{eq:s.hat-s}
\end{equation}

We first define
\begin{align*}
V_f^k & =\ld' H^{-1} e_f^{k} H^{-1}\ld - \ld' H^{-1} \xi_f^{k} H^{-1} \xi_f^{k} H^{-1}\ld, \\
\wtd{V}_f^k & = \wtd{\ld}' \wtd{H}^{-1} \wtd{e}_f^{k} \wtd{H}^{-1}\wtd{\ld} - \wtd{\ld}' \wtd{H}^{-1} \wtd{\xi}_f^{k} \wtd{H}^{-1} \wtd{\xi}_f^{k} \wtd{H}^{-1}\wtd{\ld},
\end{align*}
and rewrite the LHS of \eqref{eq:s.hat-s} as
\begin{align}
\frac{\lt\vert \what{S}_{\ld}(k) - S_{\ld}(k) \rt\vert}{S_{\ld}(k)}
&=\frac{\lt\vert (\wtd{\sg}^2_{\ld \eps}- {\sg}^2_{\ld \eps})k^2/N + (\wtd{\sg}^2_{\eps}-\sg^2_{\eps})V_f^k + \wtd{\sg}^2_{\eps}(\wtd{V}_f^k - V_f^k) \rt\vert }{S_{\ld}(k)} \nonumber \\
& \leq
\lt\vert \wtd{\sg}^2_{\ld \eps}- {\sg}^2_{\ld \eps} \rt\vert \frac{k^2/N}{S_{\ld}(k)}
+ \lt\vert \wtd{\sg}_{\eps}^2 - \sg_{\eps}^2 \rt\vert \frac{\vert V_f(k) \vert }{S_{\ld}(k)}
+ \wtd{\sg}_{\eps}^2\frac{\lt\vert \wtd{V}_f^k - V_f^k\rt\vert}{S_{\ld}(k)}.
\label{eq:S.exp}
\end{align}
Since $\sg_{\ld\eps}\neq 0 $ and $\sg^2_{\eps}\neq 0$, it holds that
\begin{equation*}
\frac{k^2/N}{S_\ld(k)} =O_p(1)~~\text{and}~~\frac{\vert V_f(k)\vert }{S_{\ld}(k)}=O_p(1)
\end{equation*}
uniformly over all $k$. Thus, consistency of $\wtd{\sg}_{\ld\eps}^{2}$ and $\wtd{\sg}^2_{\eps}$ implies that the first two terms in \eqref{eq:S.exp} are $o_p(1)$ uniformly over $k$.

Let $\mathcal{K}_N=\{1,\ldots,K(N)\}$. Since $\wtd{\sg}^2_{\eps} = O_{p}(1)$ by Assumption \ref{A:consistency-and-rate}, it remains to show that
\begin{equation}
\label{uniV}
\sup_{k\in\mathcal{K}_N} \frac{\lt\vert \wtd{V}_f^k - V_f^k\rt\vert}{S_{\ld}(k)} \parrow 0.
\end{equation}
By the triangle inequality
\begin{align}
\label{eq:V.exp1} \frac{\lt\vert \wtd{V}_f^k - V_f^k\rt\vert}{S_{\ld}(k)}  \leq& \frac{\lt\vert \wtd{\ld}' \wtd{H}^{-1} \wtd{e}_f^{k} \wtd{H}^{-1}\wtd{\ld} -\ld' H^{-1} e_f^{k} H^{-1}\ld \rt\vert}{S_{\ld}(k)} \\
\label{eq:V.exp2}&+ \frac{\lt\vert \wtd{\ld}' \wtd{H}^{-1} \wtd{\xi}_f^{k} \wtd{H}^{-1} \wtd{\xi}_f^{k} \wtd{H}^{-1}\wtd{\ld}  - \ld' H^{-1} \xi_f^{k} H^{-1} \xi_f^{k} H^{-1}\ld\rt\vert}{S_{\ld}(k)}.
\end{align}
We first show the uniform convergence of the RHS of \eqref{eq:V.exp1}. Expanding $\tlh\te\thl$ and applying the triangle inequality, we get
\begin{align}
\notag &\lt\vert \wtd{\ld}' \wtd{H}^{-1} \wtd{e}_f^{k} \wtd{H}^{-1}\wtd{\ld} -\ld' H^{-1} e_f^{k} H^{-1}\ld \rt\vert\\
\notag & \leq \lt\vert \lh \e (\thl-\hl) \rt\vert + \lt\vert \lh(\te-\e)\hl \rt\vert + \lt\vert \lh(\te-\e)(\thl-\hl) \rt\vert \\
\notag & \hskip10pt + \lt\vert (\tlh-\lh)\e \hl \rt\vert + \lt\vert (\tlh-\lh)\e (\thl - \hl)  \rt\vert \\
\notag & \hskip10pt +\lt\vert (\tlh-\lh)(\te-\e) \hl \rt\vert +\lt\vert (\tlh-\lh)(\te-\e) (\thl-\hl) \rt\vert.
\end{align}
Since $\Vert\lh\Vert=O_p(1)$, $\Vert \tlh -\lh\Vert=o_p(1)$, and $\Vert\e\Vert/S_{\ld}(k)=O_p(1)$ uniformly over $k$, it is enough to show that
\eqs{
	\sup_{k\in\mathcal{K}_N} \frac{\lt\Vert \te - \e \rt\Vert}{S_{\ld}(k)} =o_p(1)
}for the uniform convergence of the RHS of \eqref{eq:V.exp1}. Since the dimension of $\e$ is fixed, we abuse notation and use $\te-\e$ for the maximum element of the $d\times d$ matrix. Let $\breve{e}_f^k:=\tilde{e}_f^k-N^{-1}u'u$. Since $N^{-1}u'u$ does not depend on $k$, we can prove the uniform convergence for $\breve{e}_f^k$ instead of $\wtd{e}_f^k$. Recentering each term and applying the triangle inequality, we have
\begin{align*}
\frac{\lt\vert  \breve{e}_f^k - e_f^k \rt\vert }{S_{\ld}(k)}
&\le \frac{2\lt\vert f'(I-P^k)^2 u  \rt\vert }{NS_{\ld}(k)} + \frac{2\lt\vert u'P^ku-\Sg_u k\rt\vert}{NS_{\ld}(k)} + \frac{\lt\vert u'(P^k)^2u-\Sg_{u}\trace((P^k)^2)\rt\vert}{NS_{\ld}(k)}  \\
& \hskip10pt  + (\wtd{\Sg}_{u}-\Sg_{u})  \lt(\frac{-2N^{-1}k+N^{-1}\trace((P^k)^2)}{S_{\ld}(k)}\rt)\\
& \equiv \text{I.1} + \text{I.2} + \text{I.3} +  \text{I.4}.
\end{align*}
We show that terms I.1--I.4 converge to zero in probability uniformly over $k$.

We first look at I.1. Given any $\dt>0$, it holds almost surely for $z$ that
\begin{align*}
\Pr\lt(\sup_{k\in\mathcal{K}_N} \frac{\lt\vert f'(I-P^k)^2 u \rt\vert}{NS_{\ld}(k)} > \dt | z \rt)
& \le \sum_{k=1}^{K(N)} \frac{E\lt[\lt\vert f'(I-P^k)^2  u \rt\vert ^2|z\rt]}{\dt^{2} \lt(NS_{\ld}(k)\rt)^2} \\
& \le  \frac{C}{\dt^2} \sum_{k=1}^{K(N)} \frac{ f' (I-P^k)^4 f }{  \lt(NS_{\ld}(k)\rt)^2}\\
& \le \frac{C}{\dt^2} \sum_{k=1}^{K(N)} \frac{(N S_{\ld}(k))}{\lt(NS_{\ld}(k)\rt)^2} \\
& = \frac{C}{\dt^2} \sum_{k=1}^{K(N)} {(N S_{\ld}(k))^{-1}} \xrightarrow{p} 0.
\end{align*}
The first inequality holds by the Markov inequality and the second inequality holds by Theorem 2 (7) in \citet{whittle1960bounds}. The third inequality holds by Lemma \ref{L4} and from the definition of $S_{\ld}(k)$. The final main convergence result comes from Assumption \ref{A:consistency-and-rate}(iii).

We next take our attention to I.2. It holds almost surely for $z$ that
\begin{align*}
\Pr\left(\sum_{k \in \mathcal{K}_N}\frac{ \vert u' P^k u - \Sg_u k \vert  }{NS_{\ld}(k)}>\dt |z\right)
&\leq \sum_{k=1}^{K(N)} \frac{E\lt[\vert u' P^k u - \Sg_u \trace(P^k) \vert ^2|z\rt]}{\dt^{2} \lt(NS_{\ld}(k)\rt)^2}\\
&\leq \frac{C}{\dt^{2}}\sum_{k=1}^{K(N)} \frac{\text{tr}((P^{k})^2)}{\lt(NS_{\ld}(k)\rt)^2}\\
&\leq \frac{C}{\dt^{2}}\sum_{k=1}^{K(N)} \frac{NS_{\ld}(k)}{\lt(NS_{\ld}(k)\rt)^2}  \\
&= \frac{C}{\dt^2} \sum_{k=1}^{K(N)} {(N S_{\ld}(k))^{-1}} \xrightarrow{p} 0.
\end{align*}
The first and the second inequalities hold by the Markov inequality and by Theorem 2 (8) in \citet{whittle1960bounds}, respectively. Third third  holds by Lemma \ref{L_P}(ii) and the definition of $S_{\ld}(k)$.

Similarly, we can show the uniform convergence of I.3 as follows:
\begin{align*}
\Pr\left(\sup_{k \in \mathcal{K}_N}\frac{ \vert u' (P^k)^2 u - \Sg_u \trace((P^k)^2) \vert  }{NS_{\ld}(k)}>\dt |z\right)
&\leq \sum_{k=1}^{K(N)} \frac{E\lt[\vert u' (P^k)^2 u - \Sg_u \trace((P^k)^2) \vert ^2|z\rt]}{\dt^{2} \lt(NS_{\ld}(k)\rt)^2}\\
&\leq \frac{C}{\dt^{2}}\sum_{k=1}^{K(N)} \frac{\text{tr}((P^{k})^4)}{\lt(NS_{\ld}(k)\rt)^2}\\
&\leq \frac{C}{\dt^{2}}\sum_{k=1}^{K(N)} \frac{NS_{\ld}(k)}{\lt(NS_{\ld}(k)\rt)^2} \\
&= \frac{C}{\dt^2} \sum_{k=1}^{K(N)} {(N S_{\ld}(k))^{-1}} \xrightarrow{p} 0.
\end{align*}The third inequality holds by Lemma \ref{L_P}(ii)--(iii) and the definition of $S_{\ld}(k)$.

The uniform convergence of I.4 immediately follow from $({-2N^{-1}k+N^{-1}\trace((P^k)^2)})/{S_{\ld}(k)}=O_p(1)$ uniformly over $k$ and $\wtd{\Sg}_{u} - \Sg_u=o_p(1)$.

We now show the uniform convergence of \eqref{eq:V.exp2}. For the same arguments above, it is enough to show that
\eq{
	\label{eq:unif-xi} \sup_{k\in\mathcal{K}_N} \frac{\lt\vert \wtd{\xi}_f^k \wtd{H}^{-1} \wtd{\xi}_f^k - \xi_f^k H^{-1} \xi_f^k \rt\vert}{S_{\ld}(k)} = o_p(1).
}We expand $\wtd{\xi}_f^k \wtd{H}^{-1} \wtd{\xi}_f^k$ and apply the triangular inequality to get
\eqs{
	& \lt\vert \wtd{\xi}_f^k \wtd{H}^{-1} \wtd{\xi}_f^k - \xi_f^k H^{-1} \xi_f^k \rt\vert\\
	& \le \lt\vert \xik H^{-1} (\txik - \xik) \rt\vert + \lt\vert \xik (\wtd{H}^{-1} -H^{-1}) \xik\rt\vert  + \lt\vert \xik (\wtd{H}^{-1} -H^{-1}) (\txik-\xik) \rt\vert \\
	& \hskip10pt + \lt\vert (\txik-\xik) H^{-1} \xik \rt\vert + \lt\vert (\txik-\xik) H^{-1} (\txik - \xik) \rt\vert \\
	& \hskip10pt + \lt\vert (\txik-\xik) (\wtd{H}^{-1}- H^{-1}) \xik \rt\vert + \lt\vert (\txik-\xik) (\wtd{H}^{-1}- H^{-1}) (\txik - \xik) \rt\vert.
}Note again that $\Vert H^{-1}\Vert=O_p(1)$, $\Vert \wtd{H}^{-1} - H^{-1}\Vert =o_p(1)$, and $\xik/\sqrt{S_{\ld}(k)}=O_p(1)$ uniformly over $k$. Therefore, the uniform convergence in \eqref{eq:unif-xi} is implied by
\eq{
	\sup_{k\in\mathcal{K}_N} \frac{\vert \txik - \xik \vert}{\sqrt{S_{\ld}(k)}} =o_p(1).
}Recentering each term and applying the triangle inequality, we have
\eqs{
	\frac{\vert \txik - \xik \vert}{\sqrt{S_{\ld}(k)}} & = \frac{2\vert f'(I-P^k)u\vert}{N\sqrt{S_{\ld}(k)}} + \frac{\vert N^{-1}u'u-\Sg_u\vert}{\sqrt{S_{\ld}(k)}} + \frac{\vert u'P^ku - \Sg_u k \vert}{N \sqrt{S_{\ld}(k)}} + \frac{\vert \wtd{\Sg}_u-\Sg_u \vert}{\sqrt{S_{\ld}(k)}} + \vert \wtd{\Sg}_u-\Sg_u \vert \frac{k/N}{\sqrt{S_{\ld}(k)}}\\
	&\equiv \mbox{II.1}+\mbox{II.2}+\mbox{II.3}+\mbox{II.4}+\mbox{II.5}
}

We first show the uniform convergence of II.1:
\eqs{
	\Pr \lt(\sup_{k\in\mathcal{K}_N} \frac{\vert f'(I-P^k)u\vert}{N\sqrt{S_{\ld}(k)}} > \dt |z\rt)
	& \le \sum_{k=1}^{K(N)} \frac{E\lt[\lt\vert f'(I-P^k)  u \rt\vert ^2|z\rt]}{\dt^2 N^2 S_{\ld}(k) } \\
	& \le  \frac{C}{\dt^2} \sum_{k=1}^{K(N)} \frac{ f' (I-P^k)^2 f }{ N^2S_{\ld}(k)}\\
	& \le  \frac{C}{\dt^2} \sum_{k=1}^{K(N)} \frac{ N S_{\ld}(k) }{ N^2S_{\ld}(k)}\\
	& \le  \frac{C}{\dt^2} \sum_{k=1}^{K(N)} \frac{ 1 }{ N } \rightarrow 0.
}The same arguments above apply to the first four inequalities. The final convergence holds from $K(N)^2/N\rightarrow 0$.

We next show the uniform convergence of II.2:
\eqs{
	\sup_{k\in\mathcal{K}_N} \frac{\vert N^{-1}u'u-\Sg_u\vert}{\sqrt{S_{\ld}(k)}} &\le \sup_{k\in\mathcal{K}_N} \frac{1}{\sqrt{S_{\ld}(k)}} O_p\lt(\frac{1}{\sqrt{N}}\rt) \\
	& \le \sup_{k\in\mathcal{K}_N} \frac{1}{\sqrt{N S_{\ld}(k)}}O_p\lt(1\rt) \xrightarrow{p} 0,
}where the first inequality holds from the central limit theorem and the second inequality holds from Assumption \ref{A:consistency-and-rate}(ii), $\sum_{k=1}^{K(N)} (NS_{\ld}(k))^{-1} \rightarrow 0$.

We next show the uniform convergence of II.3, which holds by the Markov inequality and the Whittle inequality:
\begin{align*}
\Pr\left(\sup_{k \in \mathcal{K}_N}\frac{ \vert u' P^k u - \Sg_u k \vert  }{N\sqrt{S_{\ld}(k)}}>\dt |z\right)
&\leq \sum_{k=1}^{K(N)} \frac{E\lt[\vert u' P^k u - \Sg_u \trace(P^k) \vert ^2|z\rt]}{\dt^{2} N^2 S_{\ld}(k)}\\
&\leq \frac{C}{\dt^{2}}\sum_{k=1}^{K(N)} \frac{\text{tr}((P^{k})^2)}{ N^2S_{\ld}(k)}\\
&\leq \frac{C}{\dt^{2}}\sum_{k=1}^{K(N)} \frac{NS_{\ld}(k)}{N^2S_{\ld}(k)}  \\
&= \frac{C}{\dt^2} \sum_{k=1}^{K(N)} \frac{1}{N} \rightarrow 0.
\end{align*}

We next show the uniform convergence of II.4:
\eqs{
	\sup_{k\in\mathcal{K}_N} \frac{\vert \wtd{\Sg}_u-\Sg_u \vert}{\sqrt{S_{\ld}(k)}}
	& =   \sup_{k\in\mathcal{K}_N} \frac{O_p(N^{-1/2+\phi}S_{\ld}(k)^{\phi})}{\sqrt{S_{\ld}(k)}} \\
	& =   \sup_{k\in\mathcal{K}_N} \frac{(N S_{\ld}(k))^{\phi}}{\sqrt{N S_{\ld}(k)}} \cdot O_p(1)\\
	& = \sup_{k\in\mathcal{K}_N} (N S_{\ld}(k))^{\phi-\frac{1}{2}} \cdot O_p(1) \xrightarrow{p} 0.
}
The first equality holds from Assumption \ref{A:consistency-and-rate}(iii). The final convergence is implied from Assumption \ref{A:consistency-and-rate}(ii)  and from $\phi \in (0, 1/2)$.

The uniform convergence of II.5 follows from $\vert \wtd{\Sg}_u - \Sg_u\vert=o_p(1)$ and $(k/N)/\sqrt{S_{\ld}(k)}=O_p(1)$ uniformly over $k$. \qed

\subsection{Proof of Theorem \ref{Thm-MSE2}}
Define
\begin{equation*}
\widehat{P}^{k} = \frac{1}{M_{1}}\sum_{m\in \mathcal{M}_{1}}P^{k}_{m}~~\text{and}~~\widetilde{P}^{k} = \frac{1}{M_{2}}\sum_{m\in\mathcal{M}_{2}}P^{k}_{m}
\end{equation*}
so that
\begin{equation}
\notag P^{k} = \frac{M_{1}}{M}\widehat{P}^{k} + \frac{M_{2}}{M}\widetilde{P}^{k}.
\end{equation}

Decompose
\begin{eqnarray}
\nonumber \frac{f'(I-P^{k})f}{N} &=& \frac{M_{1}}{M}\frac{f'(I-\widehat{P}^{k})f}{N} +  \frac{M_{2}}{M}\frac{f'(I-\widetilde{P}^{k})f}{N}\\
\label{fIPkf} &=& \frac{M_{1}}{M}\frac{f'(I-\widehat{P}^{k})f}{N} + \frac{M_{2}}{M}\frac{f'f}{N} - \frac{M_{2}}{M}\frac{f'\widetilde{P}^{k}f}{N}
\end{eqnarray}
and similarly
\begin{eqnarray}
\nonumber \frac{f'(I-P^{k})u}{N} &=& \frac{M_{1}}{M}\frac{f'(I-\widehat{P}^{k})u}{N} + \frac{M_{2}}{M}\frac{f'u}{N} - \frac{M_{2}}{M}\frac{f'\widetilde{P}^{k}u}{N},\\
\nonumber \frac{f'(I-P^{k})\eps}{\sqrt{N}} &=& \frac{M_{1}}{M}\frac{f'(I-\widehat{P}^{k})\eps}{\sqrt{N}} + \frac{M_{2}}{M}\frac{f'\eps}{\sqrt{N}} - \frac{M_{2}}{M}\frac{f'\widetilde{P}^{k}\eps}{N}.
\end{eqnarray}
Now the expansion \eqref{expan1} and \eqref{expan2} can be further expanded as
\begin{eqnarray}
\nonumber \frac{X'P^{k}X}{N} &=& \frac{f'f}{N} -\frac{f'(I-P^{k})f}{N} + \frac{u'f + f'u}{N} + \frac{u'P^{k}u }{N} -\frac{u'(I-P^{k})f + f'(I-P^{k})u}{N} \\
\nonumber &=& \frac{M_{1}}{M}\left( \frac{f'f}{N} -\frac{f'(I-\widehat{P}^{k})f}{N} + \frac{u'f + f'u}{N} + \frac{M}{M_{1}}\frac{u'P^{k}u }{N}\right.\\
\nonumber && \left. -\frac{u'(I-\widehat{P}^{k})f + f'(I-\widehat{P}^{k})u}{N} + \frac{M_{2}}{M_{1}}\frac{f'\widetilde{P}^{k}f + f'\widetilde{P}^{k}u + u'\widetilde{P}^{k}f }{N}\right)
\end{eqnarray}
and
\begin{eqnarray}
\nonumber \frac{X'P^{k}\eps}{\sqrt{N}} &=& \frac{f'\eps}{\sqrt{N}} -\frac{f'(I-P^{k})\eps}{\sqrt{N}} + \frac{u'P^{k}\eps}{\sqrt{N}}\\
\nonumber &=& \frac{M_{1}}{M}\left(\frac{f'\eps}{\sqrt{N}} -\frac{f'(I-\widehat{P}^{k})\eps}{\sqrt{N}} + \frac{M}{M_{1}}\frac{u'P^{k}\eps }{\sqrt{N}} + \frac{M_{2}}{M_{1}}\frac{f'\widetilde{P}^{k}\eps}{\sqrt{N}}\right).
\end{eqnarray}
Therefore,
\[\sqrt{N}(\hat{\beta}-\beta_{0}) = \left(\frac{X'P^{k}X}{N}\right)^{-1}\frac{X'P^{k}\varepsilon}{\sqrt{N}} = \widehat{H}^{-1}\widehat{h},\]
where
\begin{eqnarray}
\nonumber \widehat{H} &=& H  +T_{1}^{H} + T_{2}^{H} + Z^{H},\\
\nonumber H &=& \frac{f'f}{N},\\
\nonumber T_{1}^{H} &=& -\frac{f'(I-\widehat{P}^{k})f}{N},\\
\nonumber T_{2}^{H} &=&  \frac{u'f + f'u}{N},\\
\nonumber T_{3}^{H} &=& \frac{M_{2}}{M_{1}}\frac{f'\widetilde{P}^{k}f}{N},\\
\nonumber Z^{H} &=& -\frac{u'(I-\widehat{P}^{k})f + f'(I-\widehat{P}^{k})u}{N}+\frac{M}{M_{1}}\frac{u'P^{k}u }{N}+\frac{M_{2}}{M_{1}}\frac{f'\widetilde{P}^{k}u+ u'\widetilde{P}^{k}f}{N},
\end{eqnarray}
and
\begin{eqnarray}
\nonumber \widehat{h} &=& h + T_{1}^{h} + T_{2}^{h} + T_{3}^{h},\\
\nonumber h &=& \frac{f'\eps}{\sqrt{N}},\\
\nonumber T_{1}^{h} &=&  -\frac{f'(I-\widehat{P}^{k})\varepsilon}{\sqrt{N}},\\
\nonumber T_{2}^{h} &=& \frac{M}{M_{1}}\frac{u'P^{k}\varepsilon}{\sqrt{N}},\\
\nonumber T_{3}^{h} &=& \frac{M_{2}}{M_{1}}\frac{f'\widetilde{P}^{k}\eps}{\sqrt{N}} .
\end{eqnarray}
These expansions simplify to the expansions in the proof of Theorem \ref{Thm1} when $M_{2}=0$ and $M=M_{1}$.\footnote{If $M_{2}=0$, $\widetilde{P}^{k}$ is not defined. We may define $\widetilde{P}^{k}$ as the null matrix in this case.} This happens if (i) all instruments are relevant or (ii) $k\geq K-K_{1}+1$. Since these cases are trivial, we assume $M_{2}>0$ in the proof. Also, note that by Assumption \ref{A3}(ii)-(iii), 
\begin{equation*}
\frac{M_{2}}{M_{1}}=\frac{M}{M_{1}}\frac{M_{2}}{M} =o\left(\frac{1}{\sqrt{k}}\right).
\end{equation*}


The proof proceeds similar to that of Theorem \ref{Thm1}. In particular, the current expansion has additional terms, $T_{3}^{H}$ and $T_{3}^{h}$, that do not appear in the proof of Theorem \ref{Thm1}. Characterising the leading (lower-order) terms including $T_{3}^{H}$ and $T_{3}^{h}$ is the main task of the proof. The remainder term $Z^{H}$ also includes additional terms, but $Z^{H}$ will be shown to be higher-order than the leading terms so it will not affect the AMSE formula.

Recall that $\widehat{e}_{f}^{k} = f'(I-\widehat{P}^{k})^{2}f/N$, $\widehat{\xi}_{f}^{k} = f'(I-\widehat{P}^{k})f/N$, $\widehat{\Delta}_{k} = \text{tr}(\widehat{e}_{f}^{k})$, $\widetilde{\zeta}_{f}^{k}=f'\widetilde{P}^{k}f/N$, and $\widetilde{\Delta}_{k} = \text{tr}(\widetilde{\zeta}_{f}^{k})$.

We first specify the convergence rate of those terms in $T^{H}$ and $T_{h}$. First, Lemma \ref{L_P} holds with $\widehat{P}^{k}$ and $\widetilde{P}^{k}$ because we do not use the instrument strength in the proof at all. Second, Lemma \ref{L2} holds with $\widehat{P}^{k}$, $\widehat{\Delta}_{k}$, and $\widehat{\xi}_{f}^{k}$ under Assumptions \ref{A1} and \ref{A3}.  Thus, $T_{1}^{H}=O_{p}(\widehat{\Delta}_{k}^{1/2})$, $T_{2}^{H}=O_{p}(1/\sqrt{N})$, $T_{1}^{h}=O_{p}(\widehat{\Delta}_{k}^{1/2})$, and $T_{2}^{h} = ((M/M_{1})k/\sqrt{N})$ which are the same with those in the proof of Theorem \ref{Thm1} except that  $\widehat{\Delta}_{k}$ is used in place of $\Delta_{k}$. Note that the order of convergence of the term $\widehat{\Delta}_{k}$ may be slower than that of $\Delta_{k}$ in the proof of Theorem \ref{Thm1} because $\widehat{P}^{k}$ is the average projection matrix based on the instrument matrix including some irrelevant instruments. Finally, we use Lemmas \ref{L_P}-\ref{L3} for $T_{3}^{H}=O_{p}((M_{2}/M_{1})\widetilde{\Delta}_{k})$ and $T_{3}^{h} = O_{p}((M_{2}/M_{1})\widetilde{\Delta}_{k}^{1/2})$.  

In the remainder of the proof, we show that the leading orders are  $k^{2}/N$, $\widehat{\Delta}_{k}$, and $(M_{2}/M_{1})\widetilde{\Delta}_{k}$, which are $o_{p}(1)$ under Assumption \ref{A3}. Similar to the proof of Theorem \ref{Thm1}, let $\rho_{k,N} = k^{2}/N + \widehat{\Delta}_{k} + (M_{2}/M_{1})\widetilde{\Delta}_{k}$ so that the higher order terms can be conveniently written as $\opr$. Note that $\rho_{k,N}$ may have a different convergence rate from the one used in the proof of Theorem \ref{Thm1}. We use the same notation for simplicity.

Using Lemma \ref{L2} and \ref{L3}, and Assumption \ref{A3},
\begin{align*}
Z^{H} =& O_{p}\left(\frac{\widehat{\Delta}_{k}^{1/2}}{\sqrt{N}}\right) + O_{p}\left(\frac{M}{M_{1}}\frac{k}{N}\right) + O_{p}\left(\left(\frac{M_{2}}{M_{1}}\right)^{1/2}\frac{\widetilde{\Delta}_{k}^{1/2}}{\sqrt{N}}\right)\\
=& o_{p}\left(\frac{k}{N}+\Delta_{k}\right) + o_{p}\left(\frac{k^{2}}{N}\right) + o_{p}\left(\frac{M_{2}}{M_{1}}\widetilde{\Delta}_{k}\right)= o_{p}(\rho_{k,N}).
\end{align*}
Thus, the expansion arguments \eqref{Hhat1}-\eqref{eq1} hold. The next expansion \eqref{eq2} holds if $\|T_{3}^{H}\|^{2}=o_{p}(\rho_{k,N})$, $\|T_{1}^{H}\|\cdot \|T_{3}^{H}\|=o_{p}(\rho_{k,N})$, and $\|T_{2}^{H}\|\cdot \|T_{3}^{H}\|=o_{p}(\rho_{k,N})$. These hold because $\|T_{1}^{H}\| = O_{p}(\widehat{\Delta}_{k}^{1/2})$, $\|T_{2}^{H}\| = O_{p}(1/\sqrt{N})$, and $\|T_{3}^{H}\|=O_{p}((M_{2}/M_{1})\widetilde{\Delta}_{k})$.

Now we derive the conditional expectations of \eqref{exp1}-\eqref{exp4} with $\widehat{h} = h+T_{1}^{h} + T_{2}^{h}+T_{3}^{h}$ and $T^{H} = T_{1}^{H}+T_{2}^{H}+T_{3}^{H}$. In particular, it is sufficient to derive those terms including $T_{3}^{h}$ and $T_{3}^{H}$ because the conditional expectation of the other terms can be found in the proof of Theorem \ref{Thm1}. Note the expansion
\begin{equation*}
\widehat{h}\widehat{h}' = hh' + hT_{1}^{h'} +hT_{2}^{h'}+hT_{3}^{h'} + T_{1}^{h}h'+T_{2}^{h}h'+T_{3}^{h}h' + T^{h}T^{h'}.
\end{equation*}

First take \eqref{exp1}. By Lemma \ref{L3}, the terms that include $T_{3}^{h}$ are
\begin{align}
E[hT_{3}^{h'}|z] &= \frac{M_{2}}{M_{1}}E\left[\frac{f'\eps\eps'\widetilde{P}^{k}f}{N}|z\right] = \sigma_{\eps}^{2}\frac{M_{2}}{M_{1}}\frac{f'\widetilde{P}^{k}f}{N},\\
\label{Dkn1} E[T_{1}^{h}T_{3}^{h'}|z] &= -\frac{M_{2}}{M_{1}}E\left[\frac{f'(I-\widehat{P}^{k})\eps\eps'\widetilde{P}^{k}f}{N}|z\right] = -\sigma_{\eps}^{2}\frac{M_{2}}{M_{1}}\frac{f'(I-\widehat{P}^{k})\widetilde{P}^{k}f}{N}=\opr ,\\
\label{Dkn2} E[T_{2}^{h}T_{3}^{h'}|z] &= \frac{MM_{2}}{M_{1}^{2}}E\left[\frac{u'P^{k}\eps\eps'\widetilde{P}^{k}f}{N}|z\right] = \frac{MM_{2}}{M_{1}^{2}}o_{p}\left(\frac{\widetilde{\Delta}_{k}^{1/2}\sqrt{k}}{\sqrt{N}}\right) = \opr,\\
E[T_{3}^{h}T_{3}^{h'}|z] &= \left(\frac{M_{2}}{M_{1}}\right)^{2}E\left[\frac{f'\widetilde{P}^{k}\eps\eps'\widetilde{P}^{k}f}{N}|z\right] = \sigma_{\eps}^{2} \left(\frac{M_{2}}{M_{1}}\right)^{2}\frac{f'\widetilde{P}^{k}\widetilde{P}^{k}f}{N} = \opr,
\end{align}
where \eqref{Dkn1}-\eqref{Dkn2} hold by the inequality $\sqrt{xy}\le 2^{-1}(x+y)$ and
\begin{align*}
\frac{M_{2}}{M_{1}}O_{p}\left(\widehat{\Delta}_{k}^{1/2}\widetilde{\Delta}_{k}^{1/2}\right) =& \left(\frac{M_{2}}{M_{1}}\right)^{1/2}O_{p}\left(\widehat{\Delta}_{k}^{1/2}\left(\frac{M_{2}}{M_{1}}\widetilde{\Delta}_{k}\right)^{1/2}\right)\\
\leq& o(1)O_{p}\left(\widehat{\Delta}_{k} + \frac{M_{2}}{M_{1}}\widetilde{\Delta}_{k}\right) = \opr,\\
\frac{MM_{2}}{M_{1}^{2}}o_{p}\left(\frac{\widetilde{\Delta}_{k}^{1/2}\sqrt{k}}{\sqrt{N}}\right) =& \frac{M}{M_{1}}\left(\frac{M_{2}}{M_{1}}\right)^{1/2}o_{p}\left(\left(\frac{M_{2}}{M_{1}}\widetilde{\Delta}_{k}\right)^{1/2}\sqrt{\frac{k}{N}}\right)\\
\leq& o(1)o_{p}\left(\frac{M_{2}}{M_{1}}\widetilde{\Delta}_{k}+\frac{k}{N}\right)=\opr.
\end{align*}
In addition, we note that
\begin{equation}
E[T_{2}^{h}T_{2}^{h'}|z] = \left(\frac{M}{M_{1}}\right)^{2}\sigma_{u \eps}\sigma_{u \eps}'\frac{k^{2}}{N} + \opr.
\end{equation}
Thus, \eqref{Ehh} can be written as
%
\begin{equation}
E[\what{h}\what{h}'|z] = \sigma_{\eps}^{2}H + \sigma_{\eps}^{2}\widehat{e}_{f}^{k} +  \left(\frac{M}{M_{1}}\right)^{2}\sigma_{u\eps}\sigma_{u\eps}' \frac{k^{2}}{N} - 2\sigma_{\eps}^{2}\widehat{\xi}_{f}^{k}+ 2\sigma_{\eps}^{2}\frac{M_{2}}{M_{1}}\frac{f'\widetilde{P}^{k}f}{N}+ o_{p}(\rho_{k,N}).
\end{equation}

Next take \eqref{exp2}. Since we can verify that $\|T_{3}^{h}\|\cdot\|T^{H}\| = \opr$, $\|T^{h}\|\cdot\|T_{3}^{H}\| = \opr$ and $\|T^{h}\|^{2}\cdot \|T^{H}\| = \opr$, only $hh'H^{-1}T_{3}^{H'}$ is added to \eqref{hhHT}. Since
\begin{equation*}
E[hh'H^{-1}T_{3}^{H'}|z]=E\left[\frac{f'\eps\eps'f}{N}H^{-1}\frac{M_{2}}{M_{1}}\frac{f'\widetilde{P}^{k}f}{N}|z\right] = \sigma_{\eps}^{2}\frac{M_{2}}{M_{1}}\frac{f'\widetilde{P}^{k}f}{N}.
\end{equation*}
Thus, by symmetry
\begin{align}
E[\what{h}\what{h}'H^{-1}T^{H'}|z] &= -\sigma_{\eps}^{2}\widehat{\xi}_{f}^{k} + 2\sigma_{\eps}^{2}\widehat{\xi}_{f}^{k}H^{-1}\widehat{\xi}_{f}^{k} +\sigma_{\eps}^{2}\frac{M_{2}}{M_{1}}\frac{f'\widetilde{P}^{k}f}{N}+\opr\\
& = E[T^{H}H^{-1}\what{h}\what{h}'|z].
\end{align}

Finally, consider the remaining terms in \eqref{exp3} and \eqref{exp4}. Since $\|T_{3}^{h}\|\cdot\|T_{1}^{H}\| = \opr$, the terms we consider in \eqref{exp3} and \eqref{exp4} are identical to those in the proof of Theorem \ref{Thm1}.

Combining the results together, we get
\begin{align*}
\widehat{H}^{-1}\widehat{h}\widehat{h}'\widehat{H}^{-1} =& H^{-1}\left(\sigma_{\eps}^{2}H + \sigma_{\eps}^{2}\widehat{e}_{f}^{k} +  \left(\frac{M}{M_{1}}\right)^{2}\sigma_{u\eps}\sigma_{u\eps}'\frac{k^{2}}{N} - 2\sigma_{\eps}^{2}\widehat{\xi}_{f}^{k}+2\sigma_{\eps}^{2}\frac{M_{2}}{M_{1}}\frac{f'\widetilde{P}^{k}f}{N}\right.\\
& \left. +2\sigma_{\eps}^{2}\widehat{\xi}_{f}^{k}  - 4\sigma_{\eps}^{2}\widehat{\xi}_{f}^{k}H^{-1}\widehat{\xi}_{f}^{k}-2\sigma_{\eps}^{2}\frac{M_{2}}{M_{1}}\frac{f'\widetilde{P}^{k}f}{N} + 3\sigma_{\eps}^{2}\widehat{\xi}_{f}^{k}H^{-1}\widehat{\xi}_{f}^{k} + o_{p}(\rho_{k,N})\right)H^{-1}\\
=&  H^{-1}\left(\sigma_{\eps}^{2}H +  \left(\frac{M}{M_{1}}\right)^{2}\sigma_{u\eps}\sigma_{u\eps}'\frac{k^{2}}{N} +\sigma_{\eps}^{2}\widehat{e}_{f}^{k}- \sigma_{\eps}^{2}\widehat{\xi}_{f}^{k}H^{-1}\widehat{\xi}_{f}^{k}\right)H^{-1}+ o_{p}(\rho_{k,N}).
\end{align*}
Note that the leading order term due to irrelevant instruments, $\sigma_{\eps}^{2}(M_{2}/M_{1})f'\widetilde{P}^{k}f/N$, cancel out and does not appear in the approximate MSE expression. 

Now we rewrite the higher-order variance term 
\begin{equation}
\label{ir_var}\sigma_{\eps}^{2}\widehat{e}_{f}^{k}- \sigma_{\eps}^{2}\widehat{\xi}_{f}^{k}H^{-1}\widehat{\xi}_{f}^{k} = \frac{f'(I-\widehat{P}^{k})(I-P_{f})(I-\widehat{P}^{k})f}{N}
\end{equation}
in terms of $P^{k}$. Observe that
\begin{align*}
\frac{f'(I-P^{k})^{2}f}{N} =& \left(\frac{M_{1}}{M}\right)^{2}\frac{f'(I-\widehat{P}^{k})^{2}f}{N} + 2\frac{M_{1}}{M}\frac{M_{2}}{M}\frac{f'(I-\widehat{P}^{k})f}{N}\\
& + \left(\frac{M_{2}}{M}\right)^{2}\frac{f'f}{N} + \left(\frac{M_{2}}{M}\right)^{2}\frac{f'\widetilde{P}^{k}\widetilde{P}^{k}f}{N}\\
&-\frac{M_{1}}{M}\frac{M_{2}}{M}\left(\frac{f'(I-\widehat{P}^{k})\widetilde{P}^{k}f}{N}+\frac{f'\widetilde{P}^{k}(I-\widehat{P}^{k})f}{N}\right)-2\left(\frac{M_{2}}{M}\right)^{2}\frac{f'\widetilde{P}^{k}f}{N}.
\end{align*}
Using \eqref{fIPkf} and the above expansion,
\begin{align*}
\frac{f'(I-P^{k})(I-P_{f})(I-P^{k})f}{N}=& \frac{f'(I-P^{k})^{2}f}{N} - \frac{f'(I-P^{k})f}{N}\left(\frac{f'f}{N}\right)^{-1}\frac{f'(I-P^{k})f}{N}\\
=& \left(\frac{M_{1}}{M}\right)^{2}\frac{f'(I-\widehat{P}^{k})(I-P_{f})(I-\widehat{P}^{k})f}{N} + \widetilde{Z}
\end{align*}
where
\begin{align*}
\widetilde{Z} =& \left(\frac{M_{2}}{M}\right)^{2}\frac{f'\widetilde{P}^{k}\widetilde{P}^{k}f}{N}-\frac{M_{1}}{M}\frac{M_{2}}{M}\left(\frac{f'(I-\widehat{P}^{k})\widetilde{P}^{k}f}{N}+\frac{f'\widetilde{P}^{k}(I-\widehat{P}^{k})f}{N}\right)\\
& + \frac{M_{1}}{M}\frac{M_{2}}{M}\left(\frac{f'(I-\widehat{P}^{k})f}{N}\left(\frac{f'f}{N}\right)^{-1}\frac{f'\widetilde{P}^{k}f}{N} + \frac{f'\widetilde{P}^{k}f}{N}\left(\frac{f'f}{N}\right)^{-1}\frac{f'(I-\widehat{P}^{k})f}{N}\right)\\
& - \left(\frac{M_{2}}{M}\right)^{2}\frac{f'\widetilde{P}^{k}f}{N}\left(\frac{f'f}{N}\right)^{-1}\frac{f'\widetilde{P}^{k}f}{N}.
\end{align*}
By Lemma \ref{L3},
\begin{equation*}
\widetilde{Z}= O_{p}\left(\left(\frac{M_{2}}{M}\right)^{2}\widetilde{\Delta}_{k}\right) +  O_{p}\left(\frac{M_{1}}{M}\frac{M_{2}}{M}\widehat{\Delta}_{k}^{1/2}\widetilde{\Delta}_{k}\right) + O_{p}\left( \left(\frac{M_{2}}{M}\right)^{2}\widetilde{\Delta}_{k}^{2}\right)=\opr.
\end{equation*}
Thus, we can write
\begin{equation}
\label{fff}
\frac{f'(I-P^{k})(I-P_{f})(I-P^{k})f}{N} = \left(\frac{M_{1}}{M}\right)^{2}\frac{f'(I-\widehat{P}^{k})(I-P_{f})(I-\widehat{P}^{k})f}{N} + o_{p}(\rho_{k,N}).
\end{equation}

Replacing \eqref{fff} into \eqref{ir_var}, the approximate MSE can be written as
\begin{equation}
\label{amset3}
S(k) = \left(\frac{M}{M_{1}}\right)^{2}H^{-1}\left[\sigma_{u\eps}\sigma_{u\eps}'\frac{k^{2}}{N} + \sigma_{\eps}^{2}\frac{f'(I-P^{k})(I-P_{f})(I-P^{k})f}{N}\right]H^{-1}.
\end{equation}
This proves Theorem \ref{Thm-MSE2}.
\qed

\subsection{Proof of Corollary \ref{col:orthogonal}}
From \eqref{ort:mse} and $\widetilde{S}(K)=O_p\left(K^2/N+\Delta_K\right)$, it is enough to show that (i) $\Delta_K \xrightarrow{p} 0$ and (ii) $k/K\rightarrow 1$ as $N \rightarrow \infty$. First, note that $\Delta_k \xrightarrow{p} 0$ for any $k\ge d$ by Lemma \ref{L2}(i), where $\Delta_k = \trace(f'(I-P^k)f/N)$. Setting $k=K$, we have
\eqs{
	\Delta_k := \trace \left(  \frac{f'(I-P^k)^2f}{N} \right)
	=  \trace \left(  \frac{f'(I-P^K)^2f}{N} \right)
	=  \trace \left(  \frac{f'(I-P)f}{N} \right)
	\equiv \Delta_K,
}since $P^K=P$ and $I-P$ is idempotent. Therefore, (i) is shown. We next turn our attention to (ii). Plugging-in $P^k = (k/K)P$, we have
\begin{equation}
\label{ort:Del}
\frac{f'(I-P^{k})^{2}f}{N}= \frac{k}{K}\left(2-\frac{k}{K}\right)\frac{f'(I-P)f}{N} +\left(1-\frac{k}{K}\right)^{2}\frac{f'f}{N}.
\end{equation}
By taking the trace of both sides,
\begin{equation}
\label{ort:deltak}
\Delta_{k}= \frac{k}{K}\left(2-\frac{k}{K}\right)\Delta_{K} + \left(1-\frac{k}{K}\right)^{2}\text{tr}(H).
\end{equation}
Under the maintained assumptions, $\Delta_k=o_p(1)$, $\Delta_K=o_p(1)$, and $H\rightarrow \overline{H}$, where $\overline{H}$ is nonsingular. Therefore, (ii) follows by taking $N\rightarrow\infty$ on both side of \eqref{ort:deltak}.
\qed

\subsection{Lemmas}

\begin{lemma}\
	Under Assumption \ref{A1}, the followings hold for all $k\geq d$.
	\begin{enumerate}[label=(\roman*)]
		\item $\text{tr}(P^{k})=k$,
		\item $\text{tr}((P^{k})^{2})\leq k$,
		\item $\text{tr}((P^{k})^{s+1})\leq \text{tr}((P^{k})^{s})$ for all positive integer $s$,
		\item $\sum_{i}(P^{k}_{ii})^{2} = o_{p}(k)$,
		\item $\sum_{i\neq j}P^{k}_{ii}P^{k}_{jj} = k^{2} + o_{p}(k)$,
		\item $\sum_{i\neq j} P^{k}_{ij}P^{k}_{ij} = \text{tr}((P^{k})^{2}) + o_{p}(k)=O_p(k)$,
	\end{enumerate}
	\label{L_P}
\end{lemma}

\subsubsection*{Proof of Lemma \ref{L_P}: }
(i)
\begin{equation}
\nonumber \text{tr}(P^{k}) = \text{tr}\left(\frac{1}{M}\sum_{m=1}^{M}P_{m}^{k}\right) = \frac{1}{M}\sum_{m=1}^{M}\text{tr}(P_{m}^{k}) = k.
\end{equation}

(ii)
Since $P_{m}^{k}$ is positive semidefinite, so are $P^{k}$ and $I-P^{k}$. From
\begin{equation}
\notag \text{tr}(P^{k}) - \text{tr}((P^{k})^{2}) = \text{tr}(P^{k}(I-P^{k}))\geq0,
\end{equation}
it follows that $\text{tr}((P^{k})^{2})\leq k$.

(iii)
Since $(P^{k})^{s}$ is symmetric, $(P^{k})^{s}$ is positive semidefinite. Thus,
\begin{equation}
\notag \text{tr}((P^{k})^{s}) - \text{tr}((P^{k})^{s+1}) = \text{tr}((P^{k})^{s}(I-P^{k}))\geq0.
\end{equation}

(iv)
\begin{eqnarray}
\nonumber \sum_{i}\left(P^{k}_{ii}\right)^{2} &\leq& \max_{i}P^{k}_{ii}\cdot \sum_{i}P^{k}_{ii} = \max_{i}P^{k}_{ii}\cdot \text{tr}(P^{k}) = k\max_{i}P_{ii}^{K}=o_{p}(k)
\end{eqnarray}
by Lemma \ref{L_P}(i) and Assumption \ref{A1}(viii).\\

(v) By Lemma \ref{L_P}(i) and Lemma \ref{L_P}(iv),
\begin{equation}
\nonumber \sum_{i\neq j}P^{k}_{ii}P^{k}_{jj} = \sum_{i}P^{k}_{ii}\sum_{j}P^{k}_{jj} - \sum_{i}\left(P^{k}_{ii}\right)^{2} = k^{2}+o_{p}(k).
\end{equation}

(vi)
By Lemma \ref{L_P}(iii) and symmetry of $P^{k}$,
\begin{eqnarray}
\nonumber \sum_{i\neq j} P^{k}_{ij}P^{k}_{ij} &=& \sum_{i}\sum_{j}P^{k}_{ij}P^{k}_{ij} - \sum_{i}\left(P^{k}_{ii}\right)^{2} =\text{tr}((P^{k})^{2})+ o_{p}(k)=O_p(k).
\end{eqnarray}The last equality comes from Lemma \ref{L_P}(ii).
\qed

\begin{lemma}\
	Under Assumptions \ref{A1} and \ref{A2}, the followings hold for all $k\geq d$.
	\begin{enumerate}[label=(\roman*)]
		\item $\Delta_{k}=o_{p}(1)$,
		\item $\xi_{f}^{k}=O_{p}(\Delta_{k}^{1/2})$,
		\item $f'(I-P^{k})\eps/\sqrt{N} = O_{p}(\Delta_{k}^{1/2})$ and $f'(I-P^{k})u/N = O_{p}(\Delta_{k}^{1/2}/\sqrt{N})$,
		\item $u'P^k\eps = O_{p}(k)$ and $u'P^k u = O_{p}(k)$,
		\item $\Delta_{k}^{1/2}/\sqrt{N} = o_{p}\left(k/N+\Delta_{k}\right)$,
		\item $E[u'P^{k}\eps\eps'P^{k}u|z] = \sigma_{u\eps}\sigma_{u\eps}'k^{2} + O_{p}(k)$,
		\item $E[f'\eps\eps'P^{k}u|z] = \sum_{i}^{N}f_{i}P_{ii}^{k}E[\eps_{i}^{2}u_{i}'|z_{i}] = O_{p}(k)$,
		\item $E[f'(I-P^{k})\eps\eps'P^{k}u/N|z] = o_{p}\left(\Delta_{k}^{1/2}\sqrt{k}/\sqrt{N}\right)$,
		\item $E[hh'H^{-1}u'f|z] = \sum_{i=1}^{N}f_{i}f_{i}'H^{-1}E[\eps_{i}^{2}u_{i}|z_{i}]f_{i}'/N^{2} = O_{p}(1/N)$.
	\end{enumerate}
	\label{L2}
\end{lemma}

\subsubsection*{Proof of Lemma \ref{L2}: }
(i) By the matrix version of the Cauchy-Schwarz inequality (Corollary 9.3.9. of Bernstein (2009)) and Jensen's inequality,
\begin{eqnarray}
\nonumber \text{tr}\left(\frac{f'(I-P^{k})^{2}f}{N}\right) &=& \frac{1}{M^{2}}\sum_{m=1}^{M}\sum_{l=1}^{M}\text{tr}\left(\frac{f'(I-P^{k}_{m})(I-P^{k}_{l})f}{N}\right)\\
\nonumber &\leq& \frac{1}{M}\sum_{m=1}^{M}\sqrt{\text{tr}\left(\frac{f'(I-P^{k}_{m})f}{N}\right)}\frac{1}{M}\sum_{l=1}^{M}\sqrt{\text{tr}\left(\frac{f'(I-P^{k}_{l})f}{N}\right)}\\
\nonumber &\leq& \text{tr}\left(\frac{f'(I-P^{k})f}{N}\right).
\end{eqnarray}

Since $\Delta_{k}\geq0$, it suffices to show
\begin{equation}
\label{Xi}
\text{tr}\left(\frac{f'(I-P^{k})f}{N}\right) = o_{p}(1).
\end{equation}
Using $(I-P_{m}^{k})Z_{m}^{k}\Pi_{m}^{k} = 0$ and the fact that $P^{k}_{m}$ and $I-P_{m}^{k}$ are positive semi-definite,
\begin{eqnarray}
\nonumber E\left[\text{tr}\left(\frac{f'(I-P^{k})f}{N}\right)\right] &=& \frac{1}{MN}\sum_{m=1}^{M}E\left[\text{tr}\left(f'(I-P_{m}^{k})f\right)\right] \\
\nonumber &=&\frac{1}{MN}\sum_{m=1}^{M}E\left[\text{tr}\left((f-Z_{m}^{k}\Pi_{m}^{k})'(I-P_{m}^{k})(f-Z_{m}^{k}\Pi_{m}^{k})\right)\right]\\
\nonumber &\leq& \frac{1}{MN}\sum_{m=1}^{M}E\left[\text{tr}\left((f-Z_{m}^{k}\Pi_{m}^{k})'(f-Z_{m}^{k}\Pi_{m}^{k})\right)\right]\\
\nonumber &=& \frac{1}{M}\sum_{m=1}^{M}E\|f(z_{i})-\Pi_{m}^{k'}Z_{m,i}^{k}\|^{2}\rightarrow 0
\end{eqnarray}
as $k\rightarrow\infty$ by Assumptions \ref{A1}(i) and \ref{A2}. By Markov inequality, \eqref{Xi} is shown.\\

(ii)  By the trace inequality, the norm equivalence, the fact that $\text{tr}(A^{2})\leq \left(\text{tr}(A)\right)^{2}$ for a positive semi-definite matrix $A$, and $\text{tr}(H) = O_{p}(1)$,
\begin{eqnarray}
\left\Vert \xi_f^k \right\Vert^{2} &=&\text{tr}\left(\frac{(I-P^{k})ff'(I-P^{k})}{N}\frac{ff'}{N}\right)\\
&\leq& \left\Vert \frac{(I-P^{k})ff'(I-P^{k})}{N}\right\Vert \text{tr}(H)\\
&\leq& \sqrt{\text{tr}\left(e_{f}^{k}e_{f}^{k}\right)} O_{p}(1)\\
&\leq& \Delta_{k}\cdot O_{p}(1).
\end{eqnarray}
Note that $\text{tr}(H) = O_{p}(1)$ by the Markov inequality under Assumptions \ref{A1}(i) and \ref{A1}(v).\\

(iii) To show $f'(I-P^{k})u/\sqrt{N} = O_{p}(\Delta_{k}^{1/2})$, it suffices to show it holds for each column vector of $u$, which corresponds to each element of $u_{i}$. We use the fact that the expectation and trace are linear operators, $E[\eps\eps'|z]=\sigma_{\eps}^{2}I$, $E[u_{a}u_{a}'|z]=\sigma^{2}_{u,a}I$ where $u_{a}$ is any column vector of $u$, and the trace inequality to get
\begin{eqnarray}
\nonumber E\left\|\Delta_{k}^{-1/2}\frac{f'(I-P^{k})\varepsilon}{\sqrt{N}}\right\|^{2} &=& E\left[\Delta_{k}^{-1}E\left[\left.\text{tr}\left(\frac{f'(I-P^{k})\varepsilon\varepsilon'(I-P^{k})f}{N}\right)\right|z\right]\right]=\sigma_{\eps}^{2},\\
\nonumber E\left\|\Delta_{k}^{-1/2}\frac{f'(I-P^{k})u_{a}}{\sqrt{N}}\right\|^{2} &=& E\left[\Delta_{k}^{-1}E\left[\left.\text{tr}\left(\frac{f'(I-P^{k})u_{a}u_{a}'(I-P^{k})f}{N}\right)\right|z\right]\right] = \sigma_{u,a}^{2}.
\end{eqnarray}
By the Markov inequality, for $a>0$,
\begin{eqnarray}
\nonumber P\left(\left\|\Delta_{k}^{-1/2}\frac{f'(I-P^{k})\varepsilon}{\sqrt{N}}\right\|\geq a\right) &=& P\left(\left\|\Delta_{k}^{-1/2}\frac{f'(I-P^{k})\varepsilon}{\sqrt{N}}\right\|^{2}\geq a^{2}\right) \\
\nonumber &\leq& \frac{E\left\|\Delta_{k}^{-1/2}\frac{f'(I-P^{k})\varepsilon}{\sqrt{N}}\right\|^{2}}{a^{2}} = \frac{\sigma_{\eps}^{2}}{a^{2}}
\end{eqnarray}
and similarly
\begin{eqnarray}
\nonumber P\left(\left\|\Delta_{k}^{-1/2}\frac{f'(I-P^{k})u_{a}}{\sqrt{N}}\right\|\geq a\right) &\leq & \frac{\sigma_{u,a}^{2}}{a^{2}}.
\end{eqnarray}
Since $\sigma_{\eps}^{2}$ and $\Sigma_{u}$ are finite, the desired conclusion follows by taking $a\rightarrow\infty$. \\

(iv) Since
\begin{eqnarray}
E[u'P^{k}\eps|z] &=& \sum_{i=1}^{N}P_{ii}^{k}E[u_{i}\eps_{i}|z_{i}] = \sigma_{u\eps}k,\\
E[u'P^{k}u|z] &=& \sum_{i=1}^{N}P_{ii}^{k}E[u_{i}u_{i}|z_{i}]  + \sum_{i\neq j}P_{ij}^{k}E[u_{i}u_{j}'|z_{i},z_{j}]= \Sigma_{u}k,
\end{eqnarray}
the statement of the Lemma follows by the Markov inequality.\\

(v) This is Lemma A.3(vi) of \citet{donald2001choosing}.\\

(vi) By the same argument of the proof of Lemma A.3(iv) of \citet{donald2001choosing} and using our Lemma \ref{L_P},
\begin{eqnarray}
\notag E[u'P^{k}\eps\eps'P^{k}u|z] &=& \sum_{i}(P_{ii}^{k})^{2}E[\eps_{i}^{2}u_{i}u_{i}'|z_{i}] + \sum_{i\neq j} P_{ii}^{k}P_{jj}^{k}E[u_{i}\eps_{i}|z_{i}]E[\eps_{j}u_{j}'|z_{j}]\\
\notag && + \sum_{i\neq j} (P_{ij}^{k})^{2}E[u_{i}\eps_{i}|z_{i}]E[\eps_{j}u_{j}'|z_{j}] + \sum_{i\neq j}P_{ij}^{k}P_{ji}^{k} E[u_{i}u_{i}'|z_{i}]E[\eps_{j}^{2}|z_{j}]\\
\notag &=& o_{p}(k) + (k^{2}+o_{p}(k))\sigma_{u\eps}\sigma_{u\eps}' + O_{p}(k)\\
\notag &=& \sigma_{u\eps}\sigma_{u\eps}'k^{2} + O_{p}(k).
\end{eqnarray}

(vii) This is Lemma A.3(v) of \citet{donald2001choosing}.\\

(viii) The proof proceeds using a similar argument of the proof of Lemma A.3(viii) of \citet{donald2001choosing}. Let $Q^{k} = I-P^{k}$. Note that $Q^{k}$ is not idempotent. For some $a$ and $b$, let $f_{i,a} = f_{a}(z_{i})$ and $\mu_{i,b}^{k} = E[\varepsilon_{i}^{2}u_{ib}|z_{i}]P^{k}_{ii}$. Let $f_{a}$ and $\mu_{b}^{k}$ be stacked matrices over $i=1,\ldots,N$. Then by the Cauchy-Schwarz inequality the absolute value of the $(a,b)$th element of $E[f'(I-P^{k})\varepsilon\varepsilon'P^{k}u|z]$ satisfies
\begin{eqnarray}
\nonumber \left|E\left[\sum_{i,j,l,m}f_{i,a}Q^{k}_{ij}\varepsilon_{j}\varepsilon_{l}P^{k}_{lm}u_{mb}|z\right]\right| &=& \left|\sum_{i,j}f_{i,a}Q^{k}_{ij}E\left[\varepsilon_{j}^{2}u_{jb}|z_{j}\right]P^{k}_{jj}\right|\\
\nonumber &=& \left|f_{a}'Q^{k}\mu_{b}^{k}\right|\leq \left(f_{a}'Q^{k}Q^{k}f_{a}\right)^{1/2}\cdot \left(\mu_{b}^{k'}\mu_{b}^{k}\right)^{1/2}.
\end{eqnarray}
Now $f_{a}'Q^{k}Q^{k}f_{a}/N = O_{p}(\Delta_{k})$ by the definition of $\Delta_{k}$. In addition, for some constant $0<C<\infty$,
\begin{eqnarray}
\nonumber \mu_{b}^{k'}\mu_{b}^{k} = \sum_{i=1}^{N}E[\varepsilon_{i}^{2}u_{ib}'|z_{i}]\left(P^{k}_{ii}\right)^{2}E[\varepsilon_{i}^{2}u_{ib}|z_{i}]\leq C\sum_{i=1}^{N}\left(P^{k}_{ii}\right)^{2} = o_{p}(k)
\end{eqnarray}
by Assumption \ref{A1}(iii) and Lemma \ref{L2}(iii). Combining these results, the desired conclusion follows.\\

(ix) This is Lemma A.3(vii) of \citet{donald2001choosing}.\qed

\begin{lemma}\
	Assume $M_{2}>0$. Under Assumption \ref{A1}, the followings hold for all $k\geq d$.
	\begin{enumerate}[label=(\roman*)]
		\item $\widetilde{\Delta}_{k}=O_{p}(\sqrt{k})$
		\item $\text{tr}\left(\frac{f'\widetilde{P}^{k}\widetilde{P}^{k}f}{N}\right)= O_{p}(\widetilde{\Delta}_{k})$,
		\item $f'\widetilde{P}^{k}\eps/\sqrt{N} = O_{p}(\widetilde{\Delta}_{k}^{1/2})$ and $f'\widetilde{P}^{k}u/N = O_{p}(\widetilde{\Delta}_{k}^{1/2}/\sqrt{N})$,
		\item $f'(I-\widehat{P}^{k})\widetilde{P}^{k}f/N = O_{p}\left(\widehat{\Delta}_{k}^{1/2}\widetilde{\Delta}_{k}^{1/2}\right)$,
		\item $E[f'\widetilde{P}^{k}\eps\eps'P^{k}u/N|z] = o_{p}(\widetilde{\Delta}_{k}^{1/2}\sqrt{k}/\sqrt{N})$.
	\end{enumerate}
	\label{L3}
\end{lemma}

\subsubsection*{Proof of Lemma \ref{L3}: }
(i) By the trace inequality (i.e. for any square matrices $A$ and $B$ such that $A$ is symmetric and $B\geq0$, $\text{tr}(AB)\leq \|A\|\text{tr}(B)$)
\begin{align*}
\widetilde{\Delta}_{k} =& \frac{1}{M_{2}}\sum_{m\in\mathcal{M}_{2}}\text{tr}\left(\frac{P_{m}^{k}ff'}{N}\right)\\
\leq& \frac{1}{M_{2}}\sum_{m\in\mathcal{M}_{2}}\left\|P_{m}^{k}\right\|\text{tr}\left(\frac{ff'}{N}\right)\\
=& \frac{1}{M_{2}}\sum_{m\in\mathcal{M}_{2}}\sqrt{\text{tr}(P_{m}^{k}P_{m}^{k})}\text{tr}(H)\\
=& \frac{1}{M_{2}}\sum_{m\in\mathcal{M}_{2}}\sqrt{\text{tr}(P_{m}^{k})}\text{tr}(H) = O_{p}(\sqrt{k}).
\end{align*}

(ii) By the matrix version of the Cauchy-Schwarz inequality and Jensen's inequality,
\begin{align*}
\text{tr}\left(\frac{f'\widetilde{P}^{k}\widetilde{P}^{k}f}{N}\right) =& \frac{1}{M_{2}^{2}}\sum_{m\in\mathcal{M}_{2}}\sum_{l\in\mathcal{M}_{2}}\text{tr}\left(\frac{f'P^{k}_{m}P^{k}_{l}f}{N}\right)\\
\leq& \frac{1}{M_{2}}\sum_{m\in\mathcal{M}_{2}}\sqrt{\text{tr}\left(\frac{f'P^{k}_{m}f}{N}\right)}\frac{1}{M_{2}}\sum_{l\in\mathcal{M}_{2}}\sqrt{\text{tr}\left(\frac{f'P^{k}_{l}f}{N}\right)}\\
\leq& \text{tr}\left(\frac{f'\widetilde{P}^{k}f}{N}\right)=\widetilde{\Delta}_{k}.
\end{align*}

(iii) The proof is the same with that of Lemma \ref{L2}(iii) by replacing its $I-P^{k}$ and $\Delta_{k}$ with $\widetilde{P}^{k}$ and $\widetilde{\Delta}_{k}$, respectively. \\

(iv) By the matrix version of the Cauchy-Schwarz inequality and Jensen's inequality,
\begin{align*}
\text{tr}\left(\frac{f'(I-\widehat{P}^{k})\widetilde{P}^{k}f}{N}\right) =& \frac{1}{M_{1}M_{2}}\sum_{m\in\mathcal{M}_{1}}\sum_{l\in\mathcal{M}_{2}}\frac{f'(I-P_{m}^{k})P_{l}^{k}f}{N}\\
\leq& \frac{1}{M_{1}}\sum_{m\in\mathcal{M}_{1}}\sqrt{\text{tr}\left(\frac{f'(I-P^{k}_{m})f}{N}\right)}\frac{1}{M_{2}}\sum_{l\in\mathcal{M}_{2}}\sqrt{\text{tr}\left(\frac{f'P^{k}_{l}f}{N}\right)}\\
\leq& \sqrt{\text{tr}\left(\frac{f'(I-\widehat{P}^{k})f}{N}\right)} \sqrt{\text{tr}\left(\frac{f'\widetilde{P}^{k}f}{N}\right)}=\widehat{\Delta}_{k}^{1/2}\widetilde{\Delta}_{k}^{1/2}.
\end{align*}

(iv)  The proof is the same with that of Lemma \ref{L2}(viii) by replacing $Q^{k}$ and $\Delta_{k}$ with $\widetilde{P}^{k}$ and $\widetilde{\Delta}_{k}$, respectively.   \qed

\begin{lemma}\
	Under Assumption \ref{A1}, for all $k\geq d$,
	\[(I-P^{k})^{4}\leq (I-P^{k})^{3}\leq (I-P^{k})^{2}\leq I-P^{k}.\]
	\label{L4}
\end{lemma}
\subsubsection*{Proof of Lemma \ref{L4}: } Since $P_{m}^{k}$ and $I-P_{m}^{k}$ for $m=1,.,,,M$ are idempotent, they are both positive semi-definite. Thus, $P^{k}$ and $I-P^{k}$ are also positive semi-definite. Since $P^{k}(I-P^{k})$ is symmetric (and thus is normal), $P^{k}(I-P^{k})\geq0$. From this, we deduce that
\begin{equation}
0\leq (I-P^{k})^{2}\leq I-P^{k}.
\end{equation}
By Theorem 1(i) of \citet{furuta1987b} with $A=I-P^{k}$, $B=(I-P^{k})^{2}$, $p=q=4$, and $r=1$, we have
\begin{equation}
(I-P^{k})^{3}\leq (I-P^{k})^{2},
\end{equation}
and with $p=q=2$ and $r=1$, we have
\begin{equation}
(I-P^{k})^{4}\leq (I-P^{k})^{3}.
\end{equation}
Since $(I-P^{k})^{4}\geq0$ this proves the lemma. \qed

\section{Inference with CSA-2SLS}\label{sc:se}

In this section we briefly discuss the inference procedure of the CSA-2SLS estimator.
If the bias is the major concern of an estimation problem, then the bias-minimizing CSA-2SLS can be obtained by setting $k=1$. In other words, we calculate the equal-weighted average of the fitted values of the endogenous variable using only one instrument at a time and use that averaged fitted value as the instrument in the second stage. Since the choice of $k$ is not data-dependent, the researcher can proceed the standard inference procedure.

More generally, we could use sample-splitting method to make inference with the optimal $\widehat{k}$. For example, a half of the sample is used to obtain $\widehat{k}$ and the other half is used to estimate $\widehat{\beta}$ given $\widehat{k}$. This method is popular in machine-learning literature where various machine-learning methods are used for model selection or averaging before inference.  \citet{wager2018estimation} use sample-splitting to allow asymptotic inference with causal forests. \citet*{chernozhukov2018double} use cross-fitting to remove bias arising from the machine-learning estimates of nonparametric functions.

Either with a fixed $k$ or an optimal $\widehat{k}$ by sample-splitting, suppose that the CSA-2SLS point estimate is obtained. The next task is to calculate the standard error, which is robust to heteroskedasticity and clustering. This is important because many empirical studies report heteroskedasticity-and-cluster robust standard errors as a measure of estimation uncertainty.

To present the standard error formula robust to heteroskedasticity and clustering, we first introduce some definitions and notations for clustered data. Let $G$ be the number of clusters. The number of observations in each cluster is $N_{g}$ for $g=1,\ldots,G$. We assume that the clusters are independent but allow for arbitrary dependence within the cluster. Let $y_{g}$, $X_{g}$, and $P^{k}_{g}$ be the $N_{g}\times1$, $N_{g}\times d$, and $N_{g}\times N$ submatrix of $P^{k}$ corresponding the $g$th cluster, respectively. Define $\widehat{\eps}_{g} = y_{g}-X_{g}\widehat{\beta}$. For i.i.d. sampling, set $N_{g}=1$ and $G=N$. Let $\Sigma$ be the covariance matrix of $\sqrt{N}(\widehat{\beta}-\beta)$ under the standard large $N$ and fixed $k$ asymptotics. \citet{hansen2019asymptotic} provide sufficient conditions for consistency, asymptotic normality, and consistency of variance estimators.

A covariance matrix estimator robust to heteroskedasticity and clustering is given by
\begin{equation}
\widehat{\Sigma} =N \left(X'P^{k}X\right)^{-1}\sum_{g=1}^{G}X'P^{k'}_{g}\widehat{\eps}_{g}\widehat{\eps}_{g}'P^{k}_{g}X\left(X'P^{k}X\right)^{-1}.
\end{equation}
The standard error can be obtained by taking the diagonal elements of $\sqrt{\widehat{\Sigma}/N}$.

\section{Simulation}\label{sc:all-tables-figures}
In this section we provide the details of the simulation designs and the addtional simulation results. Recall the baseline simulation design:
\begin{align}
y_{i} &= \bt_0 + \beta_1 Y_{i} + \eps_{i}\\
Y_{i} &= \pi'Z_{i} + u_{i},~~i=1,\ldots,N,
\end{align}
where $Y_{i}$ is a scalar, $(\bt_0,\bt_1)$ is set to be (0, 0.1), $\beta_1$ is the parameter of interest, and $Z_{i}\sim \text{i.i.d. } N(0,\Sigma_{z})$. The diagonal terms of $\Sigma_{z}$ are ones and off-diagonal terms are $\rho_z$'s. The error terms $(\eps_{i},u_{i})$ are i.i.d.\ over $i$, bivariate normal with variances 1 and covariance $\sig_{u \eps}$.

We set the parameter value for $\pi$ while controlling the explanatory power of instruments. As instruments are possibly correlated to each other, the theoretical first stage R-squared now becomes
\begin{equation}\label{eq:Rf2}
R_{f}^{2} = \frac{\pi'E[Z_{i}Z_{i}']\pi}{\pi'E[Z_{i}Z_{i}']\pi + 1} = \frac{\sum_{k=1}^{K}\pi_{k}^{2} + \sum_{k=1}^{K}\sum_{j\neq k}\pi_{k}\pi_{j}\rho_{z}}{\sum_{k=1}^{K}\pi_{k}^{2} + \sum_{k=1}^{K}\sum_{j\neq k}\pi_{k}\pi_{j}\rho_{z}+1},
\end{equation}
where $\pi_{k}$ is the $k$th element of the $k\times1$ vector $\pi$ and $K$ is the total number of instruments. 
Thus, we can set the value of $\pi$ given $R_f^2$ and $\rho_z$ by solving Equation \eqref{eq:Rf2}. Specifically, we consider the following three designs:
\begin{eqnarray}
\nonumber \text{Flat Signal: } && \pi_{k} = \sqrt{\frac{R_{f}^{2}}{(K+K(K-1)\rho_{z})(1-R_{f}^{2})}}\\
\nonumber \text{Decreasing Signal: } && \pi_{k} = C_D\left(1-\frac{k}{K+1}\right)^{4} \\
\nonumber \text{Half-zero Signal: } && \pi_{k} = \left\{
\begin{array}{ll}
0, & \text{ for }k\leq K/2\\
C_H\left(1-\frac{k-K/2}{K/2+1}\right)^{4}, & \text{ for }k>K/2\\
\end{array}\right.
\end{eqnarray}
where $C_D$ and $C_H$ are defined as
\[C_D = \sqrt{\frac{R_{f}^{2}}{1-R_{f}^{2}}\cdot\frac{1}{\sum_{k=1}^{K}\left(1-\frac{k}{K+1}\right)^{8} + \sum_{k=1}^{K}\sum_{j\neq k}\left(1-\frac{k}{K+1}\right)^{4}\left(1-\frac{j}{K+1}\right)^{4}\rho_{z}}}\]
and
\begin{footnotesize}
	\begin{align*}
	C_H = \sqrt{\frac{R_{f}^{2}}{1-R_{f}^{2}}\cdot\frac{1}{\sum_{k=K/2+1}^{K}\left(1-\frac{k-K/2}{K/2+1}\right)^{8} + \sum_{k=K/2+1}^{K}\sum_{j\neq k,j=K/2+1}\left(1-\frac{k-K/2}{K/2+1}\right)^{4}\left(1-\frac{j-K/2}{K/2+1}\right)^{4}\rho_{z}}}.
	\end{align*}
\end{footnotesize}
The number of observations and the number of instruments are set to $(N,K)=(100,20)$ and $(1000,30)$. The first stage R-squared  is set to $R_{f}^{2}= 0.01$ and $0.1$. The IV correlation parameter is set to $\rho_{z}=0$ or $0.5$. The endogeneity parameter is set to $\sigma_{u \eps}=0$ or $0.9$.

We use the Mallows criterion for the preliminary estimates required for DN, KO, and CSA when we use the approximate MSEs. We also consider two different methods in choosing $k$ of CSA: (i) the 10-fold cross-validation (CV) method (CSA.CV) and (ii) $k=1$ (CSA.1). We omit to report the results of CSA.1 because the results are similar to the case of the approximate MSE.

The following tables summarize the complete simulation results.


\begin{table}[ht]
	\caption{Simulation Results \\($N=100$, $K=20$, $\sig_{u\eps}=0.1$, $\rho_z=0$)}\label{tb:m1-6-small}
	
	\begin{center}
		\resizebox{\textwidth}{!}{
			
			\begin{tabular}{lcccccccc}
				\hline
				& MSE & Bias & MAD & Median Bias & Range & Coverage & Mean($\what{k}$) & Med($\what{k}$) \\
				\hline
				\multicolumn{9}{c}{\underline{$R_f^2=0.01$ (weak IV signal)}} \\
				\multicolumn{3}{l}{\underline{$\pi_0:$ flat}} \\
				OLS & 0.020 & 0.099 & 0.064 & 0.102 & 0.258 & 0.830 & NA & NA \\
				TSLS & 0.063 & 0.098 & 0.147 & 0.098 & 0.563 & 0.930 & NA & NA \\
				DN.M & 4.937 & 0.174 & 0.439 & 0.091 & 2.583 & 0.983 & 3.775 & 1.000 \\
				KO.M & 0.069 & 0.101 & 0.153 & 0.098 & 0.608 & 0.927 & NA & NA \\
				CSA.AMSE & 0.071 & 0.101 & 0.162 & 0.115 & 0.579 & 0.935 & 4.853 & 1.000 \\
				CSA.CV & 0.070 & 0.101 & 0.159 & 0.116 & 0.590 & 0.938 & 3.087 & 2.000 \\
				
				\multicolumn{3}{l}{\underline{$\pi_0:$ decreasing}} \\
				OLS & 0.020 & 0.099 & 0.068 & 0.100 & 0.260 & 0.823 & NA & NA \\
				TSLS & 0.061 & 0.098 & 0.140 & 0.102 & 0.567 & 0.930 & NA & NA \\
				DN.M & 2.882 & 0.054 & 0.403 & 0.081 & 2.045 & 0.980 & 3.938 & 2.000 \\
				KO.M & 0.065 & 0.096 & 0.159 & 0.099 & 0.618 & 0.935 & NA & NA \\
				CSA.AMSE & 0.066 & 0.100 & 0.163 & 0.101 & 0.585 & 0.935 & 4.240 & 1.000 \\
				CSA.CV & 0.066 & 0.099 & 0.168 & 0.100 & 0.604 & 0.943 & 3.035 & 2.000 \\
				
				\multicolumn{3}{l}{\underline{$\pi_0:$ half-zero}} \\
				OLS & 0.020 & 0.100 & 0.065 & 0.101 & 0.262 & 0.828 & NA & NA \\
				TSLS & 0.064 & 0.101 & 0.148 & 0.097 & 0.585 & 0.932 & NA & NA \\
				DN.M & 7.701 & 0.331 & 0.439 & 0.087 & 2.722 & 0.985 & 4.207 & 1.000 \\
				KO.M & 0.071 & 0.104 & 0.145 & 0.096 & 0.610 & 0.927 & NA & NA \\
				CSA.AMSE & 0.070 & 0.103 & 0.163 & 0.094 & 0.606 & 0.932 & 5.220 & 1.000 \\
				CSA.CV & 0.072 & 0.100 & 0.160 & 0.092 & 0.636 & 0.932 & 3.050 & 1.000 \\
				
				\multicolumn{9}{c}{\underline{$R_f^2=0.1$ (strong IV signal)}} \\
				\multicolumn{3}{l}{\underline{$\pi_0:$ flat}} \\
				OLS & 0.017 & 0.089 & 0.064 & 0.089 & 0.236 & 0.843 & NA & NA \\
				TSLS & 0.040 & 0.066 & 0.118 & 0.074 & 0.460 & 0.927 & NA & NA \\
				DN.M & 0.796 & 0.059 & 0.264 & 0.066 & 1.469 & 0.973 & 7.143 & 4.000 \\
				KO.M & 0.043 & 0.066 & 0.129 & 0.069 & 0.495 & 0.927 & NA & NA \\
				CSA.AMSE & 0.043 & 0.061 & 0.129 & 0.075 & 0.472 & 0.950 & 4.327 & 1.000 \\
				CSA.CV & 0.042 & 0.062 & 0.134 & 0.065 & 0.472 & 0.935 & 6.152 & 6.000 \\
				\multicolumn{3}{l}{\underline{$\pi_0:$ decreasing}} \\
				OLS & 0.018 & 0.090 & 0.061 & 0.091 & 0.248 & 0.833 & NA & NA \\
				TSLS & 0.037 & 0.065 & 0.128 & 0.072 & 0.450 & 0.938 & NA & NA \\
				DN.M & 0.233 & 0.001 & 0.202 & 0.005 & 0.758 & 0.950 & 6.128 & 5.000 \\
				KO.M & 0.041 & 0.056 & 0.135 & 0.051 & 0.491 & 0.925 & NA & NA \\
				CSA.AMSE & 0.040 & 0.062 & 0.132 & 0.057 & 0.506 & 0.943 & 2.120 & 1.000 \\
				CSA.CV & 0.039 & 0.062 & 0.126 & 0.064 & 0.485 & 0.943 & 6.168 & 6.000 \\
				\multicolumn{3}{l}{\underline{$\pi_0:$ half-zero}} \\
				OLS & 0.018 & 0.091 & 0.063 & 0.092 & 0.247 & 0.828 & NA & NA \\
				TSLS & 0.041 & 0.070 & 0.126 & 0.070 & 0.467 & 0.930 & NA & NA \\
				DN.M & 1.357 & 0.034 & 0.228 & 0.081 & 1.642 & 0.965 & 8.525 & 11.000 \\
				KO.M & 0.044 & 0.071 & 0.129 & 0.068 & 0.473 & 0.932 & NA & NA \\
				CSA.AMSE & 0.043 & 0.067 & 0.135 & 0.069 & 0.497 & 0.940 & 4.085 & 1.000 \\
				CSA.CV & 0.043 & 0.066 & 0.133 & 0.065 & 0.480 & 0.940 & 6.235 & 6.000 \\
				\hline
			\end{tabular}
		}
	\end{center}
	\footnotesize
	\renewcommand{\baselineskip}{11pt}
	\textbf{Note:} See the Note below Table \ref{tb:m19-24-small} for details.
\end{table}

\begin{table}[ht]
	\caption{Simulation Results \\($N=100$, $K=20$, $\sig_{u\eps}=0.1$, $\rho_z=0.5$)} \label{tb:m13-18-small}

	\begin{center}
		\resizebox{\textwidth}{!}{
			
			\begin{tabular}{lcccccccc}
				\hline
				& MSE & Bias & MAD & Median Bias & Range & Coverage & Mean($\what{k}$) & Med($\what{k}$) \\
				\hline
				\multicolumn{9}{c}{\underline{$R_f^2=0.01$ (weak IV signal)}} \\
				\multicolumn{3}{l}{\underline{$\pi_0:$ flat}} \\
				OLS & 0.018 & 0.090 & 0.063 & 0.090 & 0.245 & 0.835 & NA & NA \\
				TSLS & 0.039 & 0.065 & 0.131 & 0.062 & 0.460 & 0.930 & NA & NA \\
				DN.M & 0.212 & -0.002 & 0.213 & -0.007 & 0.774 & 0.953 & 5.838 & 4.000 \\
				KO.M & 0.044 & 0.055 & 0.140 & 0.059 & 0.509 & 0.920 & NA & NA \\
				CSA.AMSE & 0.098 & -0.009 & 0.205 & -0.009 & 0.724 & 0.975 & 2.530 & 1.000 \\
				CSA.CV & 0.078 & 0.010 & 0.173 & 0.014 & 0.654 & 0.960 & 3.697 & 3.000 \\
				\multicolumn{3}{l}{\underline{$\pi_0:$ decreasing}} \\
				OLS & 0.020 & 0.099 & 0.066 & 0.100 & 0.262 & 0.823 & NA & NA \\
				TSLS & 0.061 & 0.097 & 0.146 & 0.101 & 0.578 & 0.932 & NA & NA \\
				DN.M & 6.252 & 0.132 & 0.398 & 0.090 & 1.898 & 0.973 & 3.970 & 2.000 \\
				KO.M & 0.066 & 0.094 & 0.162 & 0.091 & 0.622 & 0.932 & NA & NA \\
				CSA.AMSE & 0.161 & 0.079 & 0.237 & 0.053 & 0.964 & 0.973 & 4.197 & 1.000 \\
				CSA.CV & 0.182 & 0.073 & 0.241 & 0.076 & 1.015 & 0.975 & 2.740 & 1.000 \\
				\multicolumn{3}{l}{\underline{$\pi_0:$ half-zero}} \\
				OLS & 0.020 & 0.099 & 0.067 & 0.099 & 0.259 & 0.825 & NA & NA \\
				TSLS & 0.062 & 0.098 & 0.150 & 0.096 & 0.582 & 0.925 & NA & NA \\
				DN.M & 2.085 & 0.069 & 0.398 & 0.074 & 2.021 & 0.978 & 4.003 & 2.000 \\
				KO.M & 0.067 & 0.098 & 0.159 & 0.096 & 0.612 & 0.925 & NA & NA \\
				CSA.AMSE & 0.152 & 0.074 & 0.229 & 0.053 & 0.939 & 0.970 & 4.580 & 1.000 \\
				CSA.CV & 0.180 & 0.073 & 0.231 & 0.082 & 1.051 & 0.970 & 2.748 & 1.000 \\
				\multicolumn{9}{c}{\underline{$R_f^2=0.1$ (strong IV signal)}} \\
				\multicolumn{3}{l}{\underline{$\pi_0:$ flat}} \\
				OLS & 0.007 & 0.044 & 0.046 & 0.044 & 0.187 & 0.863 & NA & NA \\
				TSLS & 0.008 & 0.011 & 0.061 & 0.011 & 0.222 & 0.943 & NA & NA \\
				DN.M & 0.009 & 0.004 & 0.065 & 0.006 & 0.233 & 0.935 & 13.170 & 13.000 \\
				KO.M & 0.008 & 0.010 & 0.062 & 0.011 & 0.223 & 0.943 & NA & NA \\
				CSA.AMSE & 0.009 & -0.002 & 0.065 & -0.003 & 0.231 & 0.958 & 3.470 & 1.000 \\
				CSA.CV & 0.008 & -0.000 & 0.066 & -0.002 & 0.229 & 0.945 & 6.143 & 6.000 \\
				\multicolumn{3}{l}{\underline{$\pi_0:$ decreasing}} \\
				OLS & 0.018 & 0.089 & 0.062 & 0.090 & 0.249 & 0.835 & NA & NA \\
				TSLS & 0.039 & 0.063 & 0.130 & 0.065 & 0.457 & 0.938 & NA & NA \\
				DN.M & 0.147 & -0.007 & 0.203 & 0.010 & 0.735 & 0.958 & 5.435 & 4.000 \\
				KO.M & 0.044 & 0.052 & 0.133 & 0.046 & 0.499 & 0.925 & NA & NA \\
				CSA.AMSE & 0.101 & -0.008 & 0.207 & -0.003 & 0.743 & 0.980 & 2.163 & 1.000 \\
				CSA.CV & 0.075 & 0.009 & 0.167 & 0.014 & 0.642 & 0.968 & 3.980 & 3.000 \\
				\multicolumn{3}{l}{\underline{$\pi_0:$ half-zero}} \\
				OLS & 0.018 & 0.090 & 0.064 & 0.086 & 0.243 & 0.833 & NA & NA \\
				TSLS & 0.040 & 0.065 & 0.130 & 0.070 & 0.465 & 0.935 & NA & NA \\
				DN.M & 0.373 & 0.004 & 0.214 & 0.023 & 0.856 & 0.948 & 6.543 & 4.000 \\
				KO.M & 0.044 & 0.056 & 0.132 & 0.060 & 0.524 & 0.932 & NA & NA \\
				CSA.AMSE & 0.104 & -0.002 & 0.209 & -0.005 & 0.753 & 0.975 & 2.783 & 1.000 \\
				CSA.CV & 0.081 & 0.012 & 0.159 & 0.020 & 0.673 & 0.960 & 4.160 & 3.000 \\
				\hline
			\end{tabular}
		}
	\end{center}
	\footnotesize
	\renewcommand{\baselineskip}{11pt}
	\textbf{Note:} See the Note below Table \ref{tb:m19-24-small} for details.
\end{table}

\begin{table}[ht]
	\caption{Simulation Results \\($N=1000$, $K=30$, $\sig_{u\eps}=0.1$, $\rho_z=0$)}\label{tb:m1-6-large}

	\begin{center}
		\resizebox{\textwidth}{!}{
			
			\begin{tabular}{lcccccccc}
				\hline
				& MSE & Bias & MAD & Median Bias & Range & Coverage & Mean($\what{k}$) & Med($\what{k}$) \\
				\hline
				\multicolumn{9}{c}{\underline{$R_f^2=0.01$ (weak IV signal)}} \\
				\multicolumn{3}{l}{\underline{$\pi_0:$ flat}} \\
				OLS  & 0.011 & 0.100 & 0.020 & 0.099 & 0.077 & 0.105 & NA & NA \\
				TSLS & 0.029 & 0.072 & 0.111 & 0.070 & 0.390 & 0.940 & NA & NA \\
				DN.M & 1.081 & 0.039 & 0.299 & 0.047 & 1.648 & 0.975 & 7.685 & 3.000 \\
				KO.M & 0.035 & 0.074 & 0.126 & 0.074 & 0.440 & 0.930 & NA & NA \\
				CSA.AMSE & 0.028 & 0.072 & 0.114 & 0.065 & 0.376 & 0.940 & 6.690 & 1.000 \\
				CSA.CV& 0.029 & 0.073 & 0.114 & 0.071 & 0.392 & 0.945 & 6.995 & 7.000 \\
				\multicolumn{3}{l}{\underline{$\pi_0:$ decreasing}} \\
				OLS & 0.011 & 0.100 & 0.021 & 0.099 & 0.078 & 0.113 & NA & NA \\
				TSLS & 0.030 & 0.068 & 0.113 & 0.068 & 0.407 & 0.930 & NA & NA \\
				DN.M & 0.231 & 0.012 & 0.205 & 0.039 & 0.901 & 0.965 & 7.560 & 5.000 \\
				KO.M & 0.036 & 0.059 & 0.124 & 0.055 & 0.454 & 0.925 & NA & NA \\
				CSA.AMSE & 0.030 & 0.068 & 0.114 & 0.069 & 0.425 & 0.932 & 3.317 & 1.000 \\
				CSA.CV & 0.031 & 0.069 & 0.115 & 0.066 & 0.412 & 0.932 & 7.003 & 7.000 \\
				\multicolumn{3}{l}{\underline{$\pi_0:$ half-zero}} \\
				OLS & 0.011 & 0.100 & 0.021 & 0.099 & 0.076 & 0.107 & NA & NA \\
				TSLS & 0.029 & 0.067 & 0.110 & 0.063 & 0.384 & 0.943 & NA & NA \\
				DN.M & 2.307 & 0.100 & 0.325 & 0.049 & 2.183 & 0.990 & 7.985 & 2.000 \\
				KO.M & 0.036 & 0.066 & 0.117 & 0.057 & 0.451 & 0.927 & NA & NA \\
				CSA.AMSE & 0.029 & 0.066 & 0.110 & 0.061 & 0.381 & 0.945 & 7.147 & 1.000 \\
				CSA.CV & 0.029 & 0.068 & 0.108 & 0.064 & 0.390 & 0.940 & 7.095 & 7.000 \\
				\multicolumn{9}{c}{\underline{$R_f^2=0.1$ (strong IV signal)}} \\
				\multicolumn{3}{l}{\underline{$\pi_0:$ flat}} \\
				OLS & 0.009 & 0.091 & 0.020 & 0.091 & 0.073 & 0.122 & NA & NA \\
				TSLS & 0.007 & 0.021 & 0.055 & 0.019 & 0.203 & 0.958 & NA & NA \\
				DN.M & 0.007 & 0.021 & 0.054 & 0.017 & 0.200 & 0.960 & 28.940 & 30.000 \\
				KO.M & 0.007 & 0.021 & 0.055 & 0.019 & 0.203 & 0.958 & NA & NA \\
				CSA.AMSE & 0.007 & 0.020 & 0.054 & 0.016 & 0.203 & 0.963 & 1.000 & 1.000 \\
				CSA.CV & 0.007 & 0.021 & 0.057 & 0.017 & 0.205 & 0.960 & 22.727 & 23.000 \\
				\multicolumn{3}{l}{\underline{$\pi_0:$ decreasing}} \\
				OLS & 0.009 & 0.091 & 0.020 & 0.090 & 0.074 & 0.128 & NA & NA \\
				TSLS & 0.007 & 0.016 & 0.057 & 0.022 & 0.215 & 0.950 & NA & NA \\
				DN.M & 0.009 & 0.005 & 0.066 & 0.004 & 0.232 & 0.945 & 15.085 & 13.000 \\
				KO.M & 0.008 & 0.015 & 0.058 & 0.022 & 0.217 & 0.943 & NA & NA \\
				CSA.AMSE & 0.007 & 0.015 & 0.057 & 0.020 & 0.213 & 0.953 & 1.630 & 1.000 \\
				CSA.CV & 0.007 & 0.015 & 0.059 & 0.022 & 0.215 & 0.948 & 22.762 & 23.000 \\
				\multicolumn{3}{l}{\underline{$\pi_0:$ half-zero}} \\
				OLS & 0.009 & 0.090 & 0.020 & 0.088 & 0.073 & 0.130 & NA & NA \\
				TSLS & 0.007 & 0.016 & 0.058 & 0.016 & 0.216 & 0.948 & NA & NA \\
				DN.M & 0.008 & 0.010 & 0.060 & 0.010 & 0.224 & 0.945 & 22.858 & 22.000 \\
				KO.M & 0.007 & 0.016 & 0.058 & 0.016 & 0.219 & 0.950 & NA & NA \\
				CSA.AMSE & 0.008 & 0.016 & 0.059 & 0.015 & 0.220 & 0.953 & 1.000 & 1.000 \\
				CSA.CV & 0.007 & 0.015 & 0.058 & 0.016 & 0.216 & 0.948 & 22.578 & 23.000 \\
				\hline
			\end{tabular}
		}
	\end{center}
	\footnotesize
	\renewcommand{\baselineskip}{11pt}
	\textbf{Note:} See the Note below Table \ref{tb:m19-24-small} for details.
\end{table}


\begin{table}[ht]
	\caption{Simulation Results \\($N=1000$, $K=30$, $\sig_{u\eps}=0.1$, $\rho_z=0.5$)}\label{tb:m13-18-large}
	
	\begin{center}
		\resizebox{\textwidth}{!}{
			
			\begin{tabular}{lcccccccc}
				\hline
				& MSE & Bias & MAD & Median Bias & Range & Coverage & Mean($\what{k}$) & Med($\what{k}$) \\
				\hline
				\multicolumn{9}{c}{\underline{$R_f^2=0.01$ (weak IV signal)}} \\
				\multicolumn{3}{l}{\underline{$\pi_0:$ flat}} \\
				OLS & 0.008 & 0.087 & 0.020 & 0.086 & 0.073 & 0.145 & NA & NA \\
				TSLS & 0.005 & 0.012 & 0.048 & 0.015 & 0.172 & 0.955 & NA & NA \\
				DN.M & 0.006 & 0.004 & 0.055 & 0.004 & 0.185 & 0.943 & 18.198 & 18.000 \\
				KO.M & 0.005 & 0.011 & 0.049 & 0.014 & 0.172 & 0.950 & NA & NA \\
				CSA.AMSE & 0.006 & -0.002 & 0.052 & -0.003 & 0.196 & 0.953 & 3.228 & 1.000 \\
				CSA.CV & 0.006 & 0.001 & 0.051 & -0.001 & 0.187 & 0.948 & 7.665 & 7.000 \\
				\multicolumn{3}{l}{\underline{$\pi_0:$ decreasing}} \\
				OLS & 0.011 & 0.100 & 0.021 & 0.099 & 0.077 & 0.105 & NA & NA \\
				TSLS & 0.029 & 0.069 & 0.116 & 0.067 & 0.402 & 0.943 & NA & NA \\
				DN.M & 0.538 & 0.055 & 0.200 & 0.005 & 0.807 & 0.965 & 6.298 & 4.000 \\
				KO.M & 0.036 & 0.059 & 0.127 & 0.053 & 0.444 & 0.927 & NA & NA \\
				CSA.AMSE & 0.088 & 0.006 & 0.186 & 0.011 & 0.705 & 0.980 & 3.875 & 1.000 \\
				CSA.CV & 0.066 & 0.008 & 0.158 & 0.020 & 0.605 & 0.970 & 4.412 & 3.000 \\
				\multicolumn{3}{l}{\underline{$\pi_0:$ half-zero}} \\
				OLS & 0.011 & 0.100 & 0.022 & 0.098 & 0.078 & 0.107 & NA & NA \\
				TSLS & 0.029 & 0.067 & 0.107 & 0.061 & 0.399 & 0.945 & NA & NA \\
				DN.M & 0.222 & 0.014 & 0.220 & 0.027 & 0.937 & 0.968 & 7.085 & 4.000 \\
				KO.M & 0.035 & 0.060 & 0.114 & 0.059 & 0.449 & 0.932 & NA & NA \\
				CSA.AMSE & 0.086 & 0.001 & 0.182 & 0.008 & 0.690 & 0.985 & 4.407 & 1.000 \\
				CSA.CV & 0.066 & 0.012 & 0.156 & 0.024 & 0.607 & 0.965 & 4.575 & 3.000 \\
				\multicolumn{9}{c}{\underline{$R_f^2=0.1$ (strong IV signal)}} \\
				\multicolumn{3}{l}{\underline{$\pi_0:$ flat}} \\
				OLS & 0.002 & 0.037 & 0.013 & 0.037 & 0.050 & 0.510 & NA & NA \\
				TSLS & 0.001 & 0.001 & 0.016 & 0.000 & 0.058 & 0.948 & NA & NA \\
				DN.M & 0.001 & 0.001 & 0.016 & -0.000 & 0.058 & 0.948 & 28.093 & 29.000 \\
				KO.M & 0.001 & 0.001 & 0.016 & 0.000 & 0.058 & 0.948 & NA & NA \\
				CSA.AMSE & 0.001 & -0.001 & 0.015 & -0.001 & 0.060 & 0.955 & 1.587 & 1.000 \\
				CSA.CV & 0.001 & -0.000 & 0.016 & -0.001 & 0.058 & 0.953 & 12.693 & 12.000 \\
				\multicolumn{3}{l}{\underline{$\pi_0:$ decreasing}} \\
				OLS & 0.009 & 0.090 & 0.020 & 0.089 & 0.073 & 0.135 & NA & NA \\
				TSLS & 0.007 & 0.016 & 0.058 & 0.018 & 0.206 & 0.963 & NA & NA \\
				DN.M & 0.009 & 0.004 & 0.065 & 0.006 & 0.229 & 0.948 & 12.055 & 10.000 \\
				KO.M & 0.007 & 0.014 & 0.059 & 0.015 & 0.215 & 0.960 & NA & NA \\
				CSA.AMSE & 0.009 & -0.004 & 0.065 & -0.004 & 0.231 & 0.955 & 3.362 & 1.000 \\
				CSA.CV & 0.008 & 0.002 & 0.061 & 0.001 & 0.224 & 0.950 & 8.465 & 8.000 \\
				\multicolumn{3}{l}{\underline{$\pi_0:$ half-zero}} \\
				OLS & 0.009 & 0.090 & 0.020 & 0.089 & 0.074 & 0.130 & NA & NA \\
				TSLS & 0.007 & 0.015 & 0.055 & 0.015 & 0.200 & 0.948 & NA & NA \\
				DN.M & 0.007 & 0.009 & 0.054 & 0.007 & 0.217 & 0.958 & 20.840 & 20.000 \\
				KO.M & 0.007 & 0.015 & 0.055 & 0.015 & 0.200 & 0.948 & NA & NA \\
				CSA.AMSE & 0.010 & -0.004 & 0.065 & -0.006 & 0.248 & 0.955 & 1.755 & 1.000 \\
				CSA.CV & 0.008 & 0.005 & 0.057 & 0.002 & 0.218 & 0.945 & 10.525 & 10.000 \\

				\hline
			\end{tabular}
		}
	\end{center}
	\footnotesize
	\renewcommand{\baselineskip}{11pt}
	\textbf{Note:} See the Note below Table \ref{tb:m19-24-small} for details.
\end{table}

\begin{table}
	\caption{Comparison of CSA.AMSE for Different Random Draws, $R$}\label{tb:m19-small-R}
	\begin{center}
		\resizebox{\textwidth}{!}{
			\begin{tabular}{lcccccccc}
				\hline
				& MSE & Bias & MAD & Median Bias & Range & Coverage & Mean($\what{k}$) & Med($\what{k}$) \\
				\hline
				$R=all$ &  0.082 & 0.037 & 0.168 & 0.082 & 0.635 & 0.890 & 1.107 & 1.000 \\
				$R=1,000$ & 0.076 & 0.048 & 0.146 & 0.098 & 0.621 & 0.885 & 1.130 & 1.000 \\
				$R=500$ & 0.087 & 0.043 & 0.156 & 0.092 & 0.658 & 0.879 & 1.112 & 1.000 \\
				$R=250$ & 0.075 & 0.037 & 0.151 & 0.086 & 0.643 & 0.889 & 1.072 & 1.000 \\
				$R=100$ & 0.090 & 0.029 & 0.178 & 0.071 & 0.648 & 0.890 & 1.058 & 1.000 \\
				\hline
			\end{tabular}
		}
	\end{center}
	\footnotesize
	\renewcommand{\baselineskip}{11pt}
	\textbf{Note:} This simulation is based on the following model: $N=100$, $K=20$, $\sig_{u\eps}=0.9$ (high endogeneity), $\rho_z=0.5$ (moderate correlation among $z$), $R_f^2=0.01$ (weak IV signal), $\pi_0:$ flat. $R$ is the number of random sampling from the complete subset for each $k$. When $R=all$, the CSA projection matrix is calculated by using all complete subsets. See the note below Table \ref{tb:m19-24-small} for other details.
	
\end{table}

\clearpage
\bibliographystyle{chicago}
{
	\small
	\bibliography{LS_CSA}}

\end{document}